%% file: paper.tex
\documentclass[10pt,journal,nofonttune]{IEEEtran}

\PassOptionsToPackage{protrusion, expansion}{microtype} 

\input{tex/00_preamble.tex}

\begin{document}

\title{TAC+:  Optimizing Error-Bounded Lossy Compression for 3D AMR Simulations}

\author{
    Daoce Wang\orcidlink{0000-0002-4444-3634},
    Jesus Pulido\orcidlink{0000-0002-9010-7762},
    Pascal Grosset\orcidlink{0000-0003-2192-3843},
    Sian Jin\orcidlink{0009-0009-9250-0611},
    Jiannan Tian\orcidlink{0000-0003-1101-9148},\\
    Kai Zhao\orcidlink{0000-0001-5328-3962},
    James Ahrens\orcidlink{0000-0002-5388-3156},
    and Dingwen Tao\orcidlink{0000-0001-5422-4497}, \textit{Senior Member, IEEE}
\IEEEcompsocitemizethanks{
\IEEEcompsocthanksitem Daoce Wang, Jiannan Tian, and Dingwen Tao (corresponding author) are with Indiana University, Bloomington, IN 47405, USA.
\IEEEcompsocthanksitem Jesus Pulido, Pascal Grosset, and James Ahrens are with Los Alamos National Laboratory, Los Alamos, NM 87545, USA.
\IEEEcompsocthanksitem Sian Jin is with Temple University, Philadelphia, PA 19122, USA.
\IEEEcompsocthanksitem Kai Zhao is with Florida State University, Tallahassee, FL 32306, USA.
}}

\markboth{IEEE TRANSACTIONS ON PARALLEL AND DISTRIBUTED SYSTEMS, 2024}{Wang \MakeLowercase{\textit{et al.}}}

\maketitle

\input{abstract}


\newcommand{\kdtree}{$k$-d tree}

\setlength{\textfloatsep}{6pt}

\input{tex/01_introduction}

\input{tex/02_background}
\input{tex/03_design}

\input{tex/04_evaluation}

\input{tex/06_conclusion}
\input{tex/99_acknowledge}

\bibliographystyle{IEEEtran}
\bibliography{refs.bib}

\begin{IEEEbiography}[{\includegraphics[width=1in,height=1.25in,clip,keepaspectratio]{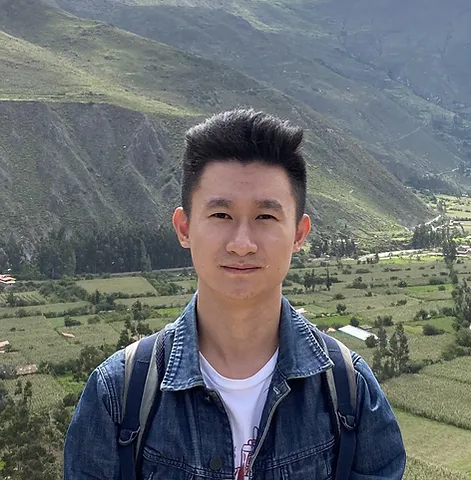}}]{Daoce Wang} is a PhD student in Intelligent Systems Engineering at Indiana University Bloomington. He received his bachelor's degree in Computer Science from University of Electronic Science and Technology of China in 2018 and his master's degree in Computer Science from University of Florida in 2020. His research interests include scientific data reduction, AMR simulations, and parallel algorithms. Email: daocwang@iu.edu.
\end{IEEEbiography}

\begin{IEEEbiography}[{\includegraphics[width=1in,height=1.25in,clip,keepaspectratio]{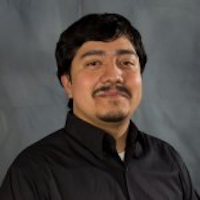}}]{Jesus Pulido} is a research scientist in the Data Science at Scale Team at Los Alamos National Laboratory. Pulido received his Ph.D. in Computer Science at University of California, Davis in 2019. Pulido specializes in data analysis, data reduction, visualization, high performance computing, wavelets and multi-resolution methods. He has experience in applications of image sensors, astronomy, turbulence and cosmology. Email: pulido@lanl.gov. 
\end{IEEEbiography}

\begin{IEEEbiography}[{\includegraphics[width=1in,height=1.25in,clip,keepaspectratio]{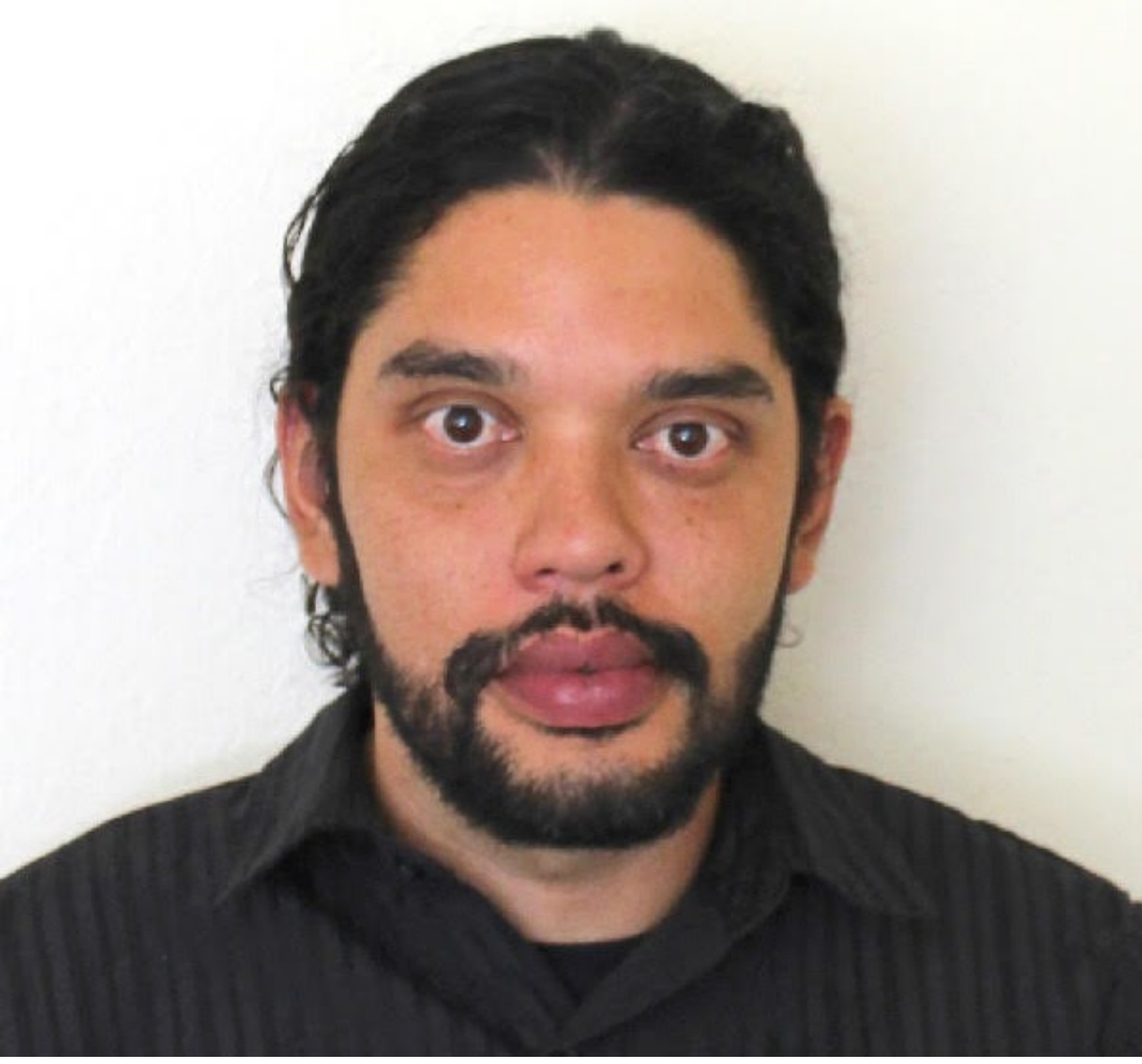}}]{Pascal Grosset} is a scientist in the Data Science at Scale team at Los Alamos National Laboratory. His primary research interests are large-scale data analysis and visualization, and data reduction. He received his Ph.D. in Computing: Graphics and Visualization from the University of Utah in 2016 where his research focused on large scale visualization. Email: pascalgrosset@lanl.gov.
\end{IEEEbiography}

\begin{IEEEbiography}[{\includegraphics[width=1in,height=1.25in,clip,keepaspectratio]{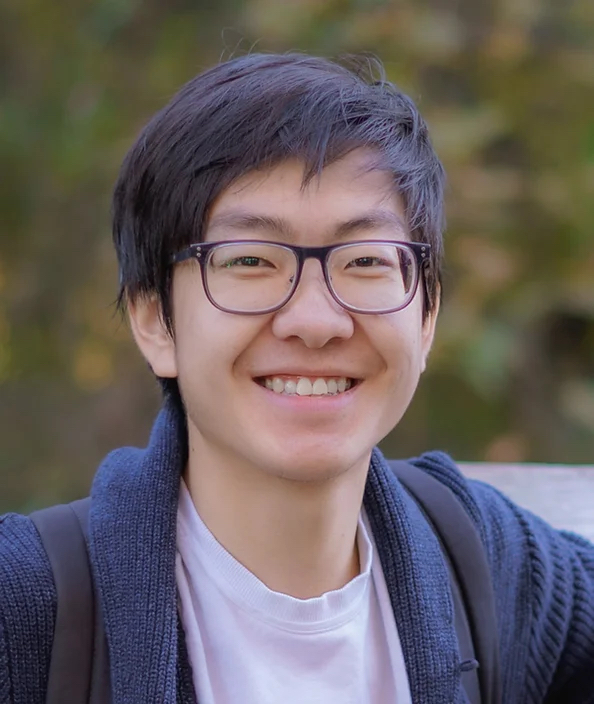}}]{Sian Jin}
is an assistant professor at Temple University. He received his Ph.D. in Computer Engineering from Indiana University in 2023 and B.S. in Physics from Beijing Normal University in 2018. His research interests include high performance computing, data compression, neural networks, and parallel computing. He has published several papers in major journals and international conferences including the SC, PPoPP, VLDB, EuroSys, HPDC, IPDPS, and ICDE. Email: sianjin@iu.edu.
\end{IEEEbiography}

\begin{IEEEbiography}[{\includegraphics[width=1in,height=1.25in,clip,keepaspectratio]{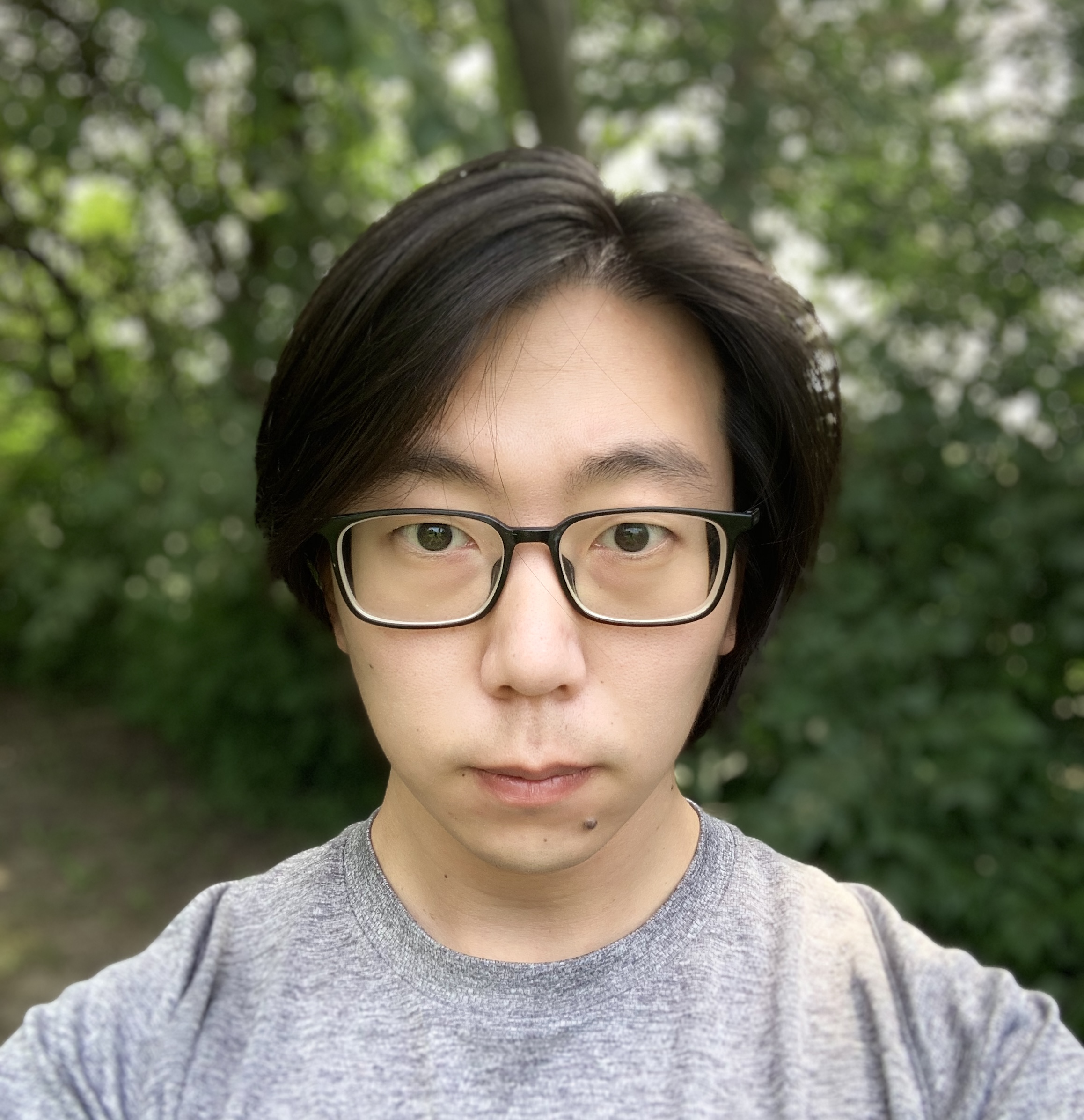}}]{Jiannan Tian}
is a PhD candidate in Intelligent Systems Engineering at Indiana University Bloomington. His research interests include lossy compression for scientific data and error analysis, and GPU-centric computing.
His ongoing project including developing GPU-accelerated compression algorithm and system design optimization of lossy compression. Email: jti1@iu.edu.
\end{IEEEbiography}

\begin{IEEEbiography}[{\includegraphics[width=1in,height=1.25in,clip,keepaspectratio]{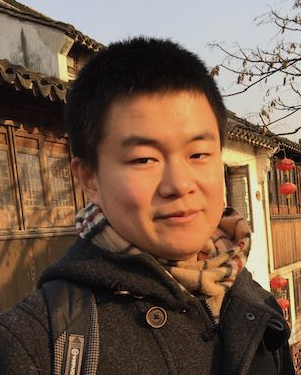}}]{Kai Zhao}
is an assistant professor of computer science at Florida State University. He received his Ph.D. in computer science from University of California, Riverside in 2022. He received his bachelor's degree from Peking University in 2014. His research interests include high-performance computing, scientific data management and reduction, and resilient machine learning. Email: kzhao@cs.fsu.edu.
\end{IEEEbiography}

\begin{IEEEbiography}[{\includegraphics[width=1in,height=1.25in,clip,keepaspectratio]{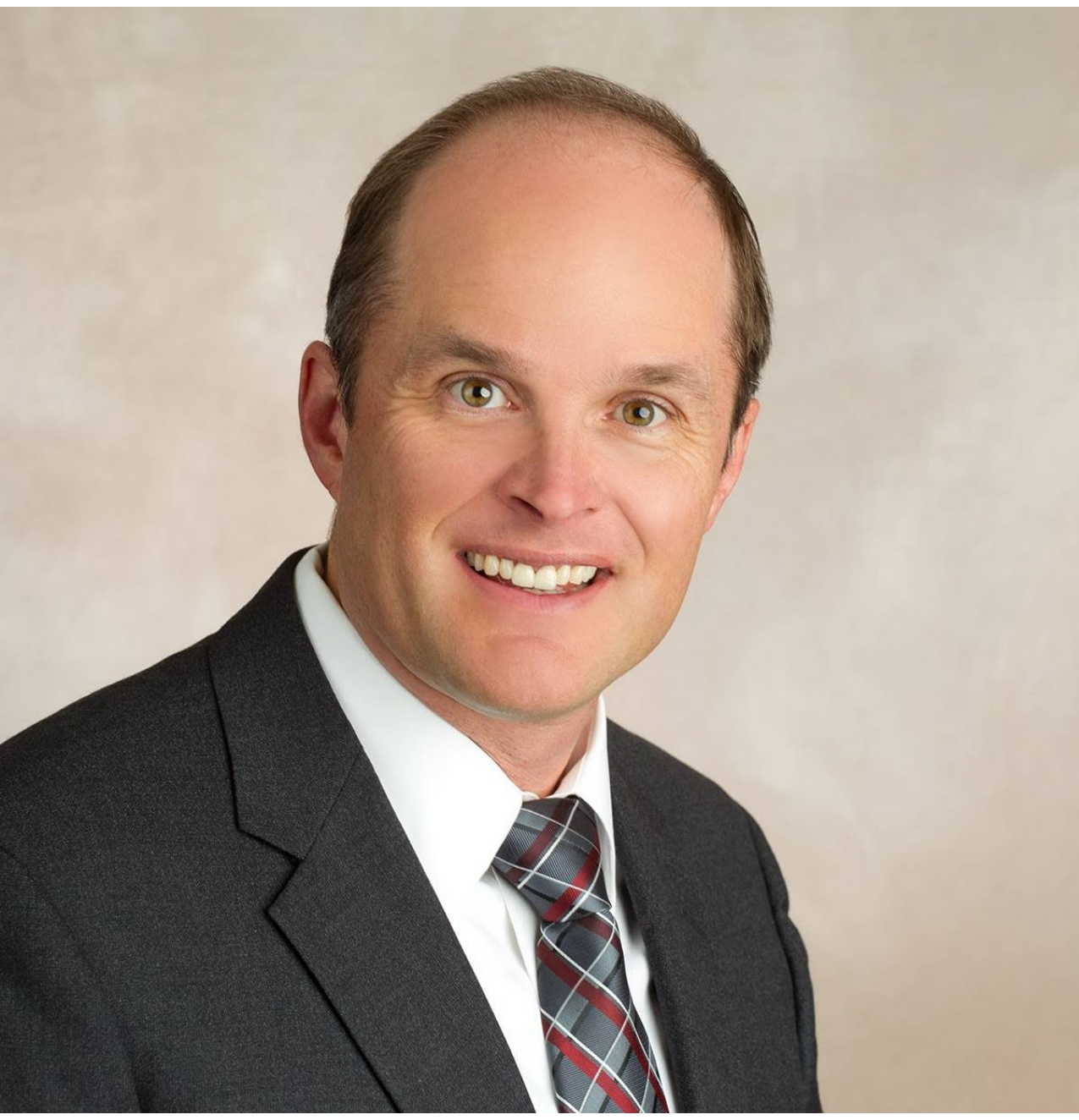}}]{James P. Ahrens} received the BS degree in computer science from the University of Massachusetts at Amherst in 1989, and the PhD degree in computer science from the University of Washington at Seattle in 1996. He is a senior scientist at Los Alamos National Laboratory. His primary research interests include visualization, computer graphics, data science, and parallel systems. He is author of more than 100 peer reviewed papers and the founder/design lead of ParaView, an open-source visualization tool designed to handle extremely large data. ParaView is broadly used for scientific visualization and is in use at supercomputing and scientific centers worldwide. He is the chair of the IEEE Computer Society Technical Committee on Visualization and Graphics (TCVG). Email: ahrens@lanl.gov.
\end{IEEEbiography}

\begin{IEEEbiography}[{\includegraphics[width=1in,height=1.25in,clip,keepaspectratio]{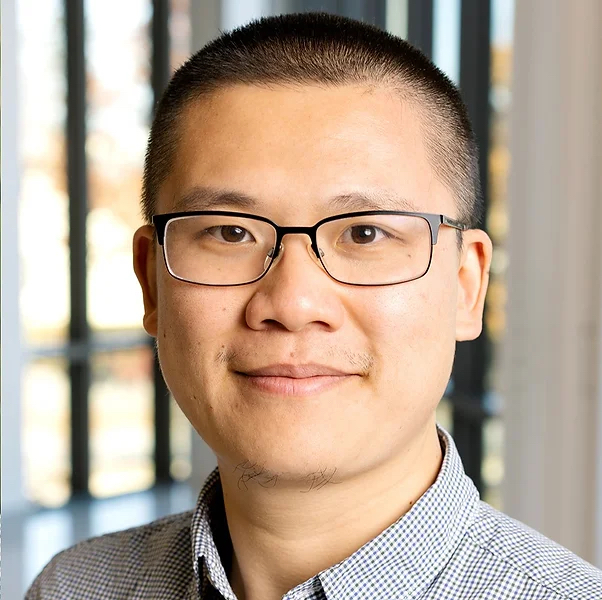}}]{Dingwen Tao} is an associate professor at Indiana University Bloomington, where he directs the High-Performance Data Analytics and Computing Lab. He received his Ph.D. in Computer Science from University of California, Riverside in 2018 and B.S. in Mathematics from University of Science and Technology of China in 2013. 
He is the recipient of various awards including NSF CAREER Award (2023), Amazon Research Award (2022), Meta Research Award (2022), R\&D100 Awards Winner (2021), IEEE Computer Society TCHPC Early Career Researchers Award for Excellence in HPC (2020), NSF CRII Award (2020), and IEEE CLUSTER Best Paper Award (2018).
He is serving as an Associate Editor of IEEE Transactions on Parallel and Distributed Systems. 
He is a senior member of IEEE and ACM. Email: ditao@iu.edu.
\end{IEEEbiography}

\end{document}

%% file: tex/00_preamble.tex
\usepackage{graphicx}

\usepackage{amsmath, amsthm}
    
\usepackage{graphicx, xcolor} 
\usepackage{hyperref, url}

\usepackage{adjustbox}

\usepackage{booktabs, multirow}
\usepackage{caption}
\usepackage{subcaption}
\usepackage{enumitem}
\usepackage{lipsum}
\usepackage{setspace}
\usepackage{array}
\usepackage{comment}
\usepackage{orcidlink}
\usepackage[noadjust]{cite}

\usepackage[ruled, linesnumbered]{algorithm2e}
\SetKwComment{Comment}{/* }{ */}
\usepackage{algpseudocode}
\algnewcommand{\algorithmicand}{\textbf{ and }}
\algnewcommand{\AND}{\algorithmicand}
\algnewcommand{\algorithmicor}{\textbf{ or }}
\algnewcommand{\OR}{\algorithmicor}

\SetCommentSty{mycommfont}

\usepackage{footnote}
\makesavenoteenv{tabular}
\makesavenoteenv{table}

\usepackage[zerostyle=d]{newtxtt}

%% file: abstract.tex
\begin{abstract}
Today's scientific simulations require significant data volume reduction because of the enormous amounts of data produced and the limited I/O bandwidth and storage space. Error-bounded lossy compression has been considered one of the most effective solutions to the above problem. However, little work has been done to improve error-bounded lossy compression for Adaptive Mesh Refinement (AMR) simulation data. Unlike the previous work that only leverages 1D compression, in this work, we propose an approach (\textsc{TAC}) to leverage high-dimensional SZ compression for each refinement level of AMR data. To remove the data redundancy across different levels, we propose several pre-process strategies and adaptively use them based on the data features. We further optimize \textsc{TAC} to \textsc{TAC+} by improving the lossless encoding stage of SZ compression to handle many small AMR data blocks after the pre-processing efficiently. \textcolor{black}{Experiments on 10 AMR datasets from three real-world large-scale AMR simulations} demonstrate that \textsc{TAC+} can improve the compression ratio by up to 4.9$\times$ under the same data distortion, compared to the state-of-the-art method. In addition, we leverage the flexibility of our approach to tune the error bound for each level, which achieves much lower data distortion on two application-specific metrics. 
\end{abstract}

%% file: tex/01_introduction.tex
\section{Introduction}
\label{sec:introduction}
\IEEEPARstart{T}he increase in supercomputer performance over the past decades has been insufficient to solve many challenging modeling and simulation problems. For example, the complexity of solving evolutionary partial differential equations scales as $\Omega(n^4)$, where $n$ is the number of mesh points per dimension. Thus, the performance improvement of about three orders of magnitudes over the past 30 years has meant just a 5.6$\times$ gain in spatio-temporal resolution~\cite{burstedde2008towards}. To address this issue, many high-performance computing (HPC) simulation packages~\cite{dubey2014survey} (such as AMReX~\cite{zhang2019amrex} and Athena++~\cite{stone2020athena++}) use Adaptive Mesh Refinement (AMR)---which applies computation to selective regions of most interest---to increase resolution. Compared to the method where a high resolution is applied everywhere, the AMR method greatly reduces the computational complexity and storage overhead; thus, it is one of the most widely used frameworks for many HPC applications~\cite{almgren2013nyx, runnels2021massively, whitman2018simulation, sverdrup2018highly}. 

Although AMR can save storage space to some extent, AMR applications running on supercomputers still generate large amounts of data, bringing challenges to data transmission and storage. 
For example, one Nyx simulation~\cite{nyx} with a resolution of $4096^3$ (i.e., $0.5 \times 2048^3$ mesh points in the coarse level and $0.5 \times 4096^3$ in the fine level ) can generate up to 1.8 TB of data for a single snapshot; a total of 1.8 PB of disk storage is needed assuming running the simulation 5 times with 200 snapshots dumped per simulation.
Therefore, reducing data size is necessary to lower the storage overhead and I/O cost and improve the overall application performance for running large-scale AMR simulations on supercomputers. 

A straightforward way to address this issue is to use data compression. However, traditional lossless compression techniques such as GZIP~\cite{gzip} and Zstandard~\cite{zstd} can only provide a compression ratio by up to 2$\times$ for scientific data~\cite{son2014data}. On the other hand, a new generation of lossy compressors that can provide strict error control (called ``error-bounded'' lossy compression) has been developed, such as SZ~\cite{sz16,sz17,sz18}, ZFP~\cite{zfp}, MGARD~\cite{ainsworth2017mgard,liang2021mgard+}, and TTHRESH~\cite{ballester2019tthresh}. Using those error-bounded lossy compressors, scientists can achieve relatively high compression ratios while minimizing the quality loss of reconstructed data and post-analysis, as seen in 
~\cite{cappello2019use, jin2020understanding, grosset2020foresight, lu2018understanding, baker2014methodology, baker2017toward, gok2018pastri, wu2019full}. 

\textcolor{black}{However, a gap exists between data compression and AMR data. The root of this disparity lies in the hierarchical nature of AMR data, where the entire dataset possesses varying resolutions/levels. Data at each level is sparse because it only covers a portion of the domain. Yet, current scientific compressors exclusively support the compression of non-sparse data with uniform resolution. Consequently, AMR data must be preprocessed prior to compression.}

\textcolor{black}{A straightforward pre-process would be to flatten the high-dimensional AMR data from different levels into a 1D array for compression. However, this approach would cause the data to lose a significant amount of spatial information, which is critical for compression performance optimization.}

\textcolor{black}{To leverage high-dimension compression, a common approach is to generate uniform-resolution data by upsampling the coarse-level data and merging it with the finest-level data. However, this method introduces redundant information, which significantly reduces the compression ratio. This degradation is especially pronounced when the upsampling rate is high or when multiple coarse levels need to be upsampled.
This presents us with a dilemma in compressing AMR data: we are forced to choose between losing spatial locality (by compressing in 1D) or introducing redundant information (by using upsampled 3D data).}

Only a few existing contributions have investigated advanced error-bounded lossy compression for AMR applications and datasets.
Recently,  
Luo~\textit{et~al.} introduced zMesh~\cite{zMesh}, a technique that pre-processs the data by grouping data points that are mapped to the same or adjacent geometric coordinates such that the dataset is smoother and more compressible. However, since zMesh maps data points from different AMR levels to adjacent geometric coordinates and generates a 1D array, it still cannot adopt 3D compression which most HPC simulations use. 
Moreover, zMesh is designed for patch-based AMR data\footnote{The patch-based AMR data redundantly saves the data block to be refined at the next finer level in the current coarse level (see Section~\ref{sec:diffAMR}).} 
and the reorganization approach proposed by zMesh cannot improve the data smoothness appropriately (see Section~\ref{sec:evaluation}).

To solve these issues, we propose \textsc{TAC} that removes the redundant data in coarser level(s) and employs 3D lossy compression for each level.
We note that each level may contain many empty/zero regions, where data points are saved in other levels, which may significantly decrease the data smoothness
and hence reduce the compression ratio. 
To this end, \textsc{TAC} either removes these empty regions using adaptive partition strategies or partially pads them with appropriate values, based on the density of empty regions.

Another challenge is that the partition strategies can generate many (e.g., 3,000+) small data blocks, whereas the SZ compressor performs poorly on small data sets because of the Huffman encoding cost (will be detailed in Section~\ref{sec:sle}).
TAC's solution to the heavy Huffman encoding cost is to linearize/merge the small blocks and then pass these merged blocks to SZ.  This approach can reduce the cost of the Huffman encoding.
However, \textsc{TAC} still faces a critical limitation: most of the merged small blocks are not adjacent in the original dataset, leading to rapid changes in the data values between these non-neighboring blocks, which can negatively impact the accuracy of SZ's predictor.

To address the limitations, we further optimize \textsc{TAC} to \textsc{TAC+} by designing a Shared Huffman Encoding (SHE) approach for the SZ compressor. 
This approach allows individual predictions for each small block while being encoded using a single shared Huffman tree, which can improve the prediction accuracy and compression ratio accordingly. 

The main contributions are summarized as follows. 
\begin{itemize}[topsep=0pt,partopsep=1ex,parsep=0pt]
    \item We propose to leverage 3D SZ compression to compress each level of an AMR dataset separately. We propose a hybrid compression approach based on the following three pre-process strategies and data characteristics.
    \item We propose an optimized sparse tensor representation to efficiently partition data and remove empty regions for sparse AMR data. 
    \item We propose an enhanced \kdtree~approach to reduce the time overhead of removing empty regions.
    \item We propose a padding approach to improve the smoothness and compressibility of dense AMR data.
    \item 
    We employ the SHE approach in the SZ compressor to reduce the high time and storage costs of compressing multiple small blocks in order to achieve better compression on AMR data after the partition.
    \item We tune the error bound for each AMR level
    to further improve the compression quality in terms of two application-specific post-analysis metrics.
    \item Experiments show that, compared to the state-of-the-art approach zMesh, our proposed AMR compression can improve the compression ratio by up to 4.9$\times$ under the same data distortion on the tested datasets. 
\end{itemize}

\textcolor{black}{We evaluate our proposed compression method on ten datasets from three real-world AMR applications: Nyx \cite{nyx}, WarpX \cite{warpx}, and IAMR \cite{IAMR}.}
We compare our method with four baselines including zMesh using generic metrics such as compression ratio and peak signal-to-noise ratio (PSNR) and application-specific metrics such as power spectrum and halo finder. Our code is available at \url{https://github.com/hipdac-lab/HPDC22-TAC}.

The remaining paper is organized as follows. In Section~\ref{sec:background}, we present background information about error-bounded lossy compression, AMR method, \kdtree, and related work on AMR data compression.
In Section~\ref{sec:design}, we describe our proposed pre-process strategies, SHE approach, and hybrid compression.
In Section~\ref{sec:evaluation}, we show the experimental results on different AMR datasets.
In Section~\ref{sec:conclusion}, we conclude our work and discuss the future work.

%% file: tex/02_background.tex
\section{Background and Related Work} 
\label{sec:background}

\subsection{Lossy Compression for Scientific Data} \label{sec:back-comps}
There are two main categories for data compression: lossless and lossy compression. Compared to lossless compression, lossy compression can offer a much higher compression ratio by trading a little bit of accuracy~\cite{jin2024concealing}. There are some well-developed lossy compressors for images and videos such as JPEG~\cite{wallace1992jpeg} and MPEG~\cite{le1991mpeg}, but they do not have a good performance on the scientific data because they are mainly designed for integers rather than floating points. 

In recent years there is a new generation of lossy compressors that are designed for scientific data, such as SZ~\cite{sz16, sz17, sz18}, ZFP~\cite{zfp}, MGARD~\cite{ainsworth2017mgard}, and TTHRESH~\cite{ballester2019tthresh}.
These lossy compressors provide parameters that allow users to finely control the information loss introduced by lossy compression. 
Unlike traditional lossy compressors such as JPEG~\cite{wallace1992jpeg} for images (in integers), SZ, ZFP, MGARD, and TTHRESH are designed to compress floating-point data and can provide a strict error-controlling scheme based on the user's requirements.
Generally, lossy compressors provide multiple compression modes, such as error-bounding mode and fixed-rate mode.
Error-bounding mode requires users to set an error type, such as the point-wise absolute error bound and point-wise relative error bound, and an error bound level (e.g., $10^{-3}$). The compressor ensures that the differences between the original data and the reconstructed data do not exceed the user-set error bound level.

In this work, we focus on the SZ lossy compression (2021 R\&D 100 Award Winner) because SZ typically provides a higher compression ratio than ZFP~\cite{lu2018understanding,zhao2021optimizing} and higher speeds than MGARD~\cite{zhao2021optimizing,liang2021error} and TTHRESH~\cite{ballester2019tthresh}. 
SZ is a prediction-based error-bounded lossy compressor for scientific data. 
It has three main steps: (1) predict each data point's value based on different prediction methods; (2) quantize the difference between the real value and predicted value based on the user-set error bound; and (3) apply a customized Huffman coding and lossless compression.

\textcolor{black}{The SZ framework comprises a series of algorithms tailored to various user and application needs. In this study, we primarily focus on its two principal algorithms: the compression algorithm that utilizes the Lorenzo and linear regression predictors, denoted as ``Lor/Reg'' \cite{sz18}, and the compression algorithm based on the spline interpolation approach, denoted as ``Interp'' \cite{zhao2021optimizing}. Specifically, the Lor/Reg method begins by truncating the entire input data into smaller blocks. It then applies either the Lorenzo predictor or the high-dimensional linear regression to each block separately. In contrast, the Interp algorithm carries out global interpolation across all three dimensions of the complete dataset. The Lor/Reg and Interp algorithms differ significantly, resulting in varied performance and features when compressing AMR data. A more in-depth discussion on this can be found in Sections~\ref{sec:sle},~\ref{sec:hybrid},~and\ref{sec:algovs}.}

\subsection{AMR Method and AMR Data}
\label{sec:amr}

\begin{figure}[t]
     \centering
     \begin{subfigure}[t]{0.28\linewidth}
         \centering
         \includegraphics[width=\linewidth]{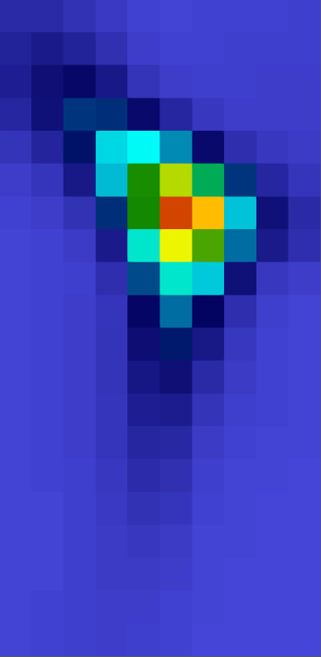} 
        \vspace{-3mm}
     \end{subfigure}
     \begin{subfigure}[t]{0.28\linewidth}
         \centering
         \includegraphics[width=\linewidth]{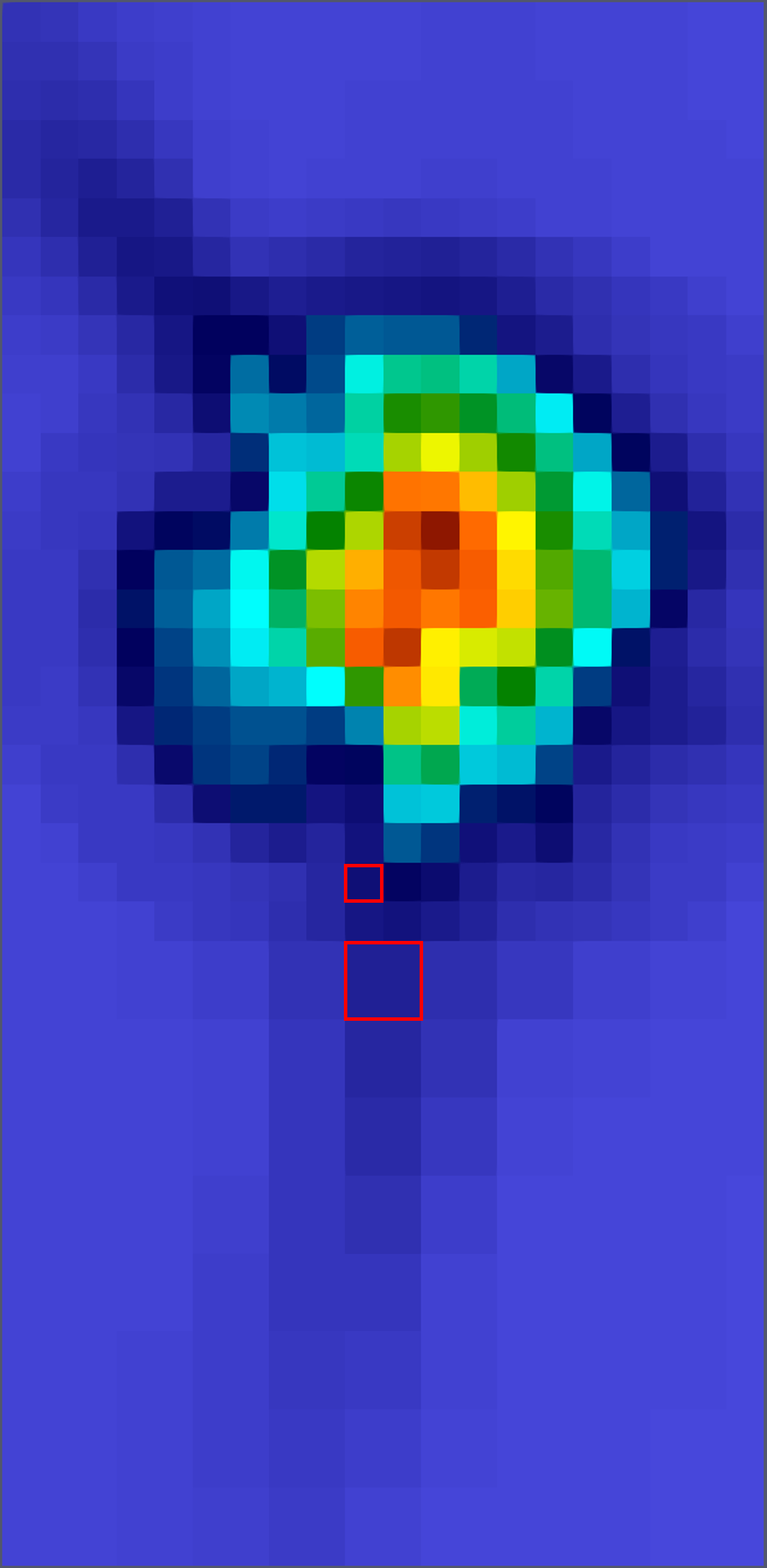}
        \vspace{-3mm}
     \end{subfigure}
      \begin{subfigure}[t]{0.28\linewidth}
         \centering
         \includegraphics[width=\linewidth]{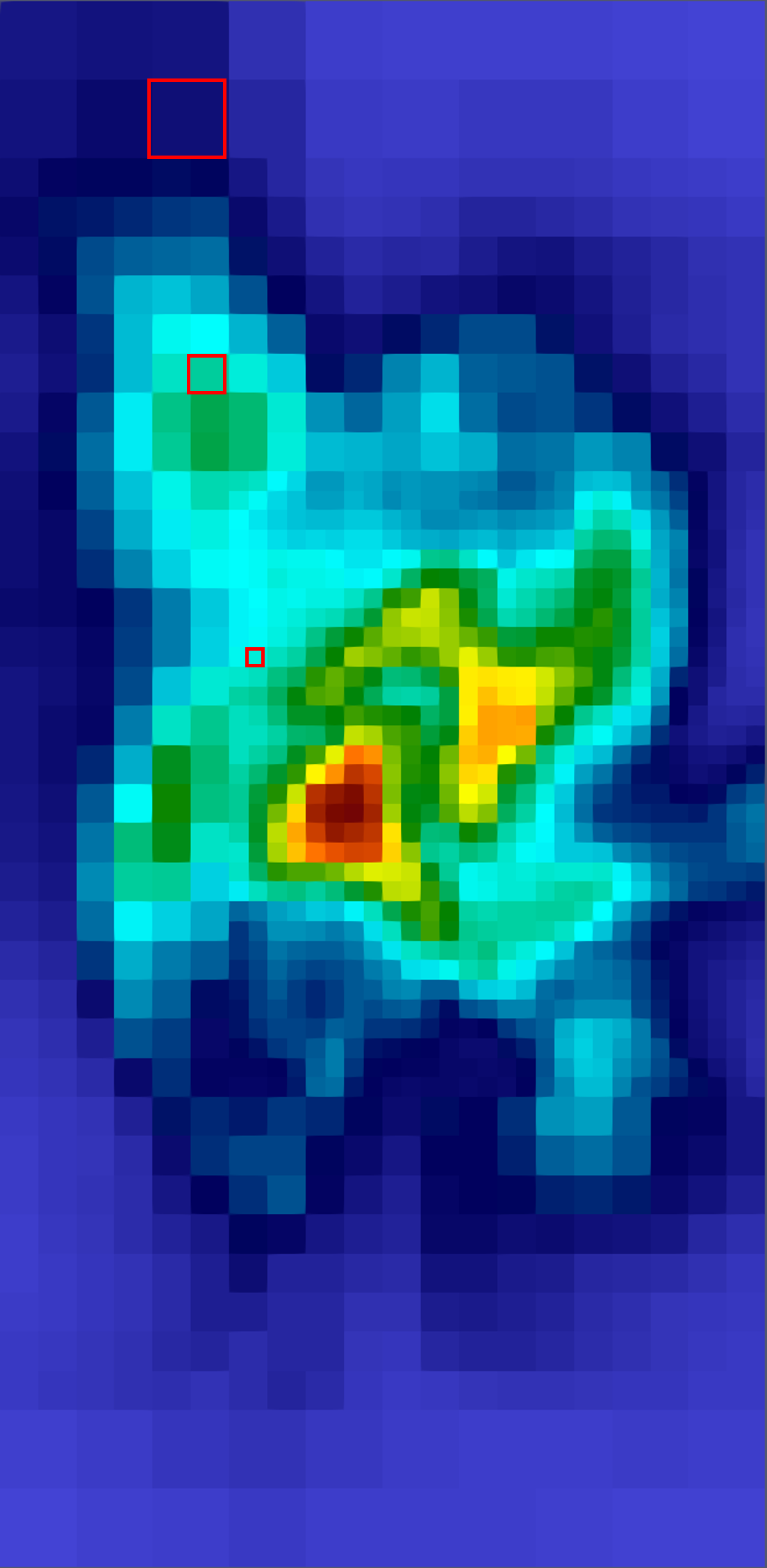}
        \vspace{-3mm}
     \end{subfigure}
        \vspace{-2mm}
        \caption[t]{Visualization (one zoom-in 2D slice) of three key timesteps generated from an AMR-based cosmology simulation. The grid structure changes with the universe's evolution. The red boxes indicate different resolutions within one AMR level.}
        \vspace{-2mm}
        \label{fig:visamr}
\end{figure}
AMR is a method adapting the accuracy of a solution 
by using a non-uniform grid to increase computational and storage savings while still achieving the desired accuracy. AMR applications change the mesh or spatial resolution based on the level of refinement needed by the simulation and use \textit{finer mesh in the regions with more importance/interest} and \textit{coarser mesh in the regions with less importance/interest}. Figure \ref{fig:visamr} shows that the mesh will be refined when 
the value meets the refinement criteria, e.g., refining a block when its norm of the gradients or maximum value is larger than a threshold. 

\begin{figure}[h]
    \centering 
    \includegraphics[width=0.99\columnwidth]{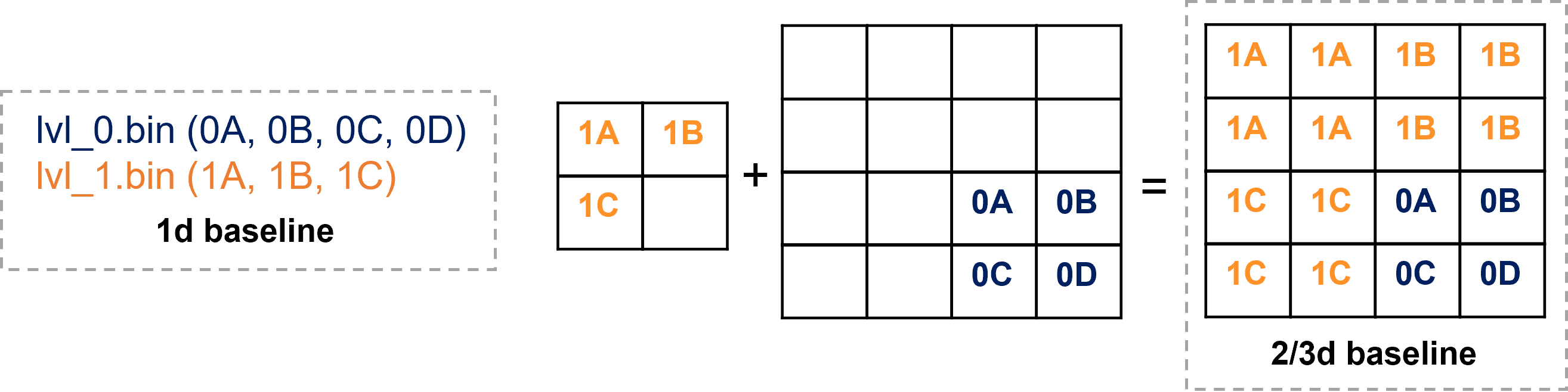}
     \vspace{-2mm}
    \caption{A typical example of AMR data storage and usage.}
    \vspace{-2mm}
    \label{fig:base_ex}
\end{figure} 

Clearly, the data generated by an AMR application are hierarchical data with different resolutions. The data of each AMR level are usually stored separately (e.g., in a 1D array). For example, Figure~\ref{fig:base_ex} (left) shows a simple example of two-level AMR data; ``0'' means high resolution (the fine level) and ``1'' for low resolution (the coarse level).
When the AMR data are needed for post analysis or visualization, users will typically convert the data from different levels to a uniform resolution. In the previous example, we will up-sample the data at the coarse level and combine it with the data at the fine level, as shown in Figure~\ref{fig:base_ex} (right).

\subsection{Tree-based and Patch-based AMR Data}
\label{sec:diffAMR}
There are two types of techniques to represent AMR data: patch-based AMR and tree-based AMR~\cite{cpuRay}. 
The main difference between them is that the patch-based AMR technique generates AMR data with redundancy across different levels. 
In other words, the patch-based AMR data structure redundantly saves data blocks to be refined at the next level in the current level, simplifying the computation in the refinement process.
By comparison, the tree-based AMR technique organizes the grids on the tree leaves, so there is no redundant data across different levels. But tree-based AMR data is more complex for post analysis and visualization compared to patch-based AMR data~\cite{patchVsTree,daoceamrvis23}.

In this work, we focus on a state-of-the-art patch-based AMR framework AMReX. Note that since the redundant coarser-level data in the patch-based AMR will not often be used in post-analysis, we discard them during compression to improve the compression ratio.

\subsection{Existing AMR Data Compression}
\label{sec:baseline}
\textit{1D AMR Compression:} 
The main challenge for AMR data compression is that the AMR data is comprehensive and hierarchical with different resolutions.
A naive approach is to compress the 1D data of each AMR level separately. However, this approach loses most of the topological/spatial information, which is critical for data compression.
zMesh~\cite{zMesh} is a state-of-the-art AMR data compression based on the 1D approach.
Different from the naive 1D approach, zMesh
re-organizes the 1D data based on each point's coordinate in the 2D layout; in other words, zMesh puts the points neighbored in the 2D layout closer in the 1D array.
It can increase the data smoothness/compressibility to benefit the following 1D compression such as SZ on patch-based AMR data with redundancy across different AMR levels. 
However, zMesh does not leverage high-dimensional compression, while many previous studies~\cite{sz17,zhao2020significantly,amric} proved that leveraging more dimensional information (e.g., spatial/temporal information) can significantly improve the compression performance.
Moreover, it only focuses on 2D AMR data. Our work aims to leverage high-dimensional data compression and supports 3D AMR data.

\textit{High-dimensional AMR Compression:}
Similar to the idea described in Section~\ref{sec:amr}, a straightforward way to leverage 3D compression on 3D AMR data is to compress different levels together by up-sampling coarse levels. 
However, this approach must
handle extra redundant data generated by the up-sampling process. 
As shown in Figure~\ref{fig:base_ex}, \textit{1A}, \textit{1B}, and \textit{1C} are redundant points in the compression. 
Note that the storage overhead of these redundant points will be higher when more data are in the coarse levels or the up-sampling rate is higher, especially for 3D AMR data. 
This is because we only need to duplicate one point from the coarse level 4 times for 2D AMR data but 8 times for 3D AMR data, with an up-sampling rate of 2. 
Another limitation of this approach is that it cannot apply different compression configurations (e.g., error bound) to different AMR levels. This is because after up-sampling all data points will have the same importance.
However, the purpose of using the AMR method is to set different interests to different AMR levels, so the error bound for each AMR level can be chosen adaptively. 

\subsection{$k$-D Tree for Particle Data Compression} 

\kdtree{}~\cite{kd1975} is a binary tree in which every node represents a certain space. Without loss of generality, for the 3D case, every non-leaf node in a \kdtree{} splits the space into two parts by a 2D plane associated with one of the three dimensions. 
The left subspace is associated with the left child of the node, while the right subspace is associated with the right child.
\kdtree{} is commonly used in particle data compression~\cite{Hoangldav21, kd-1, kd-2} to locate each particle and remove empty regions. Specifically, a \kdtree{} keeps dividing the space in between along one dimension until the space is empty or contains only one particle. 
We will optimize the classic \kdtree{} and use it to remove empty regions and increase the compressibility for each AMR level (to be detailed in Section~\ref{sec:kd}).

%% file: tex/03_design.tex
\section{Our Proposed Design}
\label{sec:design}

\begin{figure}[b]
    \includegraphics[width=0.95\columnwidth]{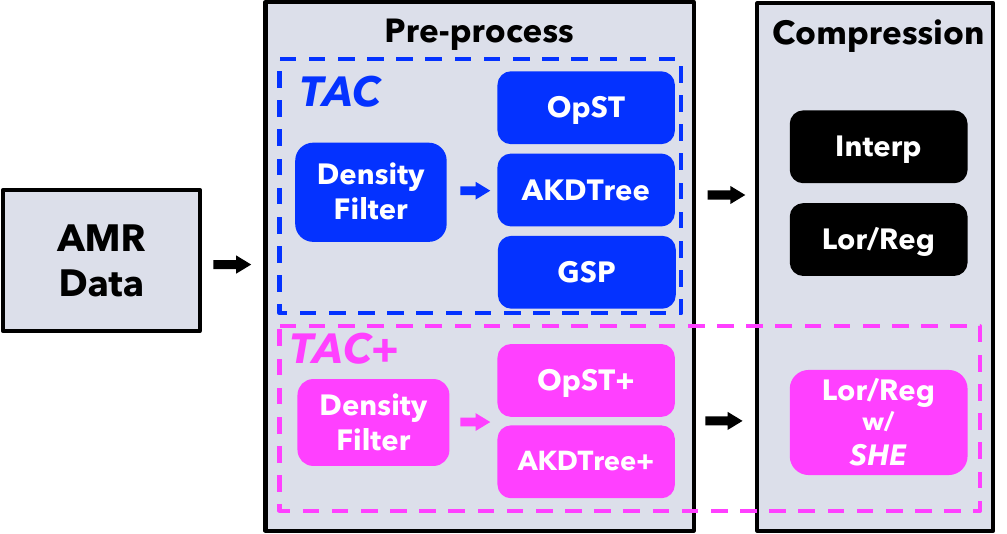}
    \caption{Workflow overview of our proposed \textsc{TAC} and \textsc{TAC+}.}
    \label{fig:flow}
\end{figure} 

\textcolor{black}{In this section, we introduce a compression framework tailored for AMR data, utilizing high-dimensional SZ lossy compression for each AMR level. Our initial proposal, \textsc{TAC}, integrates three pre-process strategies to address the challenges posed by irregular data distributions. Extending \textsc{TAC}, we present \textsc{TAC+}. In \textsc{TAC+}, we refine two pre-process methods from the original TAC and introduce Shared Huffman Encoding (SHE). We then integrate SHE with the Lor/Reg algorithm to bolster compression efficacy for AMR data. Furthermore, we propose adaptive approaches for \textsc{TAC}/\textsc{TAC+}, enabling the selection of the most suitable pre-process method based on each level's data density.}

\subsection{Ghost-Shell Padding for High-density Data}\label{sec:gsp}

\begin{figure}[t]
     \centering
     \begin{subfigure}[t]{0.49\linewidth}
         \centering
         \includegraphics[width=\linewidth]{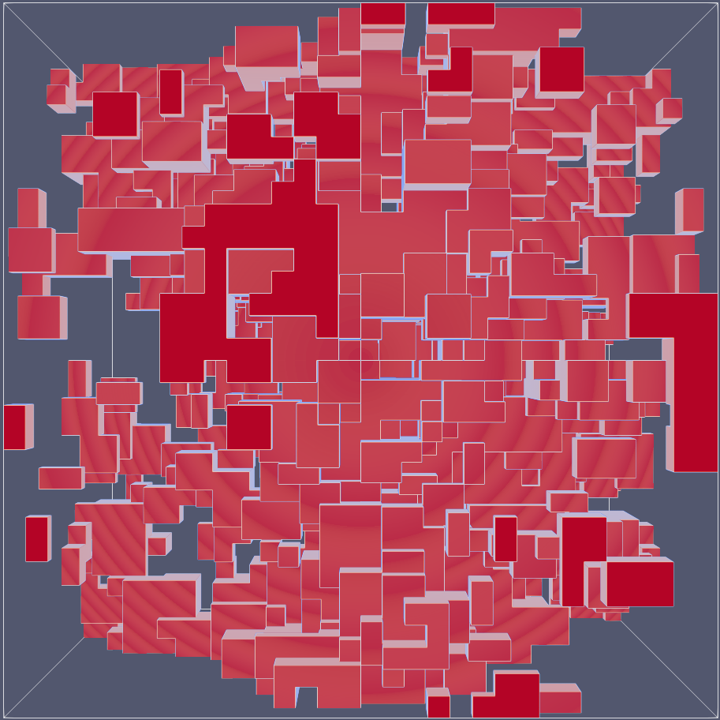}  
         \caption[t]{z10 fine level}
         \label{fig:dis_fine}
     \end{subfigure}
     \begin{subfigure}[t]{0.49\linewidth}
         \centering
         \includegraphics[width=\linewidth]{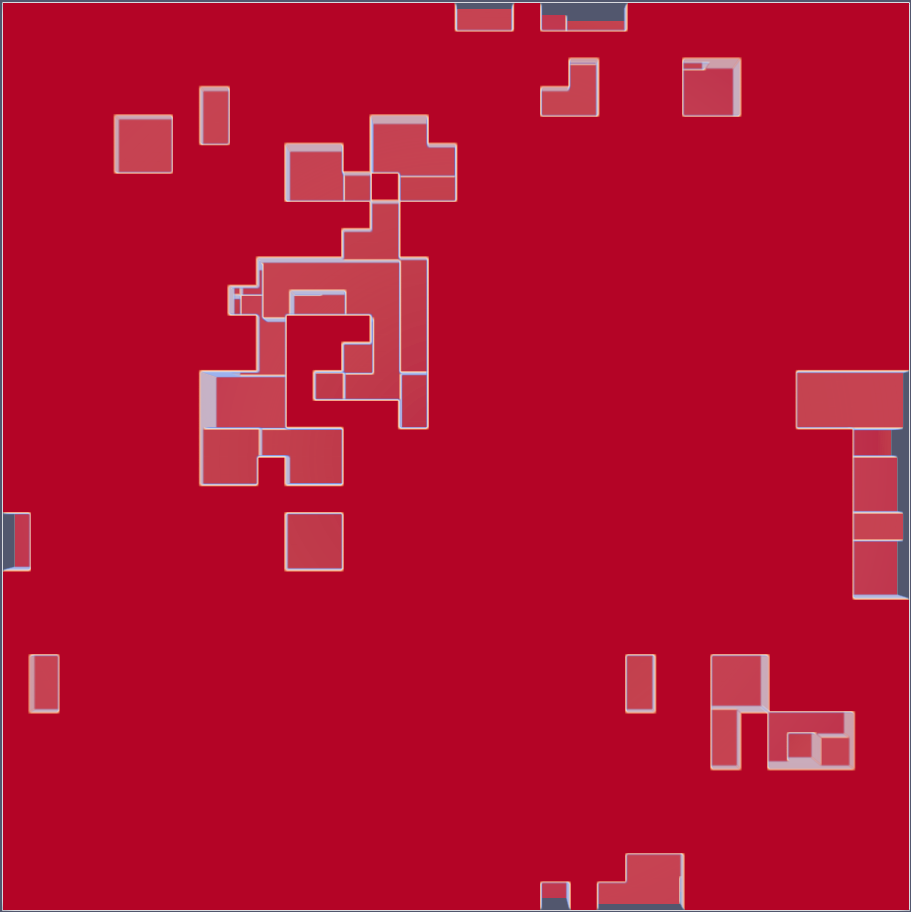}
         \caption{z10 coarse level}
         \label{fig:dis_coarse}
     \end{subfigure}
        \caption[t]{Visualization of data distributions of an example AMR data ``z10'', where z = redshift. Non-empty regions are shown in red. 
        }
       \vspace{-2mm}
        \label{fig:dis}
\end{figure}

To compress the AMR data in 3D, besides the aforementioned 3D baseline, we can also compress each level separately in 3D. 
In that way, however, the data will be split into multiple levels, 
and each level will have many empty regions and an irregular data distribution, as shown in Figure~\ref{fig:dis}. A naive solution to handle the irregular 3D data is to fill the empty regions with zeros and pass a large 3D block to the compressor.
Although the padded zeros will increase the size of data for compression, for high-density data such as z10's coarse level shown in  Figure~\ref{fig:dis_coarse} (i.e., about 77\% density), the size overhead will be small.

However, these padded zeros can also greatly reduce the performance of compression, especially for prediction-based lossy compression such as SZ, because these zeros can significantly affect the prediction accuracy of SZ, resulting in high compression errors on the boundaries, as shown in Figure~\ref{fig:zero_err}.  
More specifically, as mentioned in Section~\ref{sec:opst}, SZ uses each point's neighboring points' values to predict its value. 
Thus, for those boundary points that are adjacent to padded zeros, SZ will involve zero(s) in the prediction, while the actual values of these empty regions are typically non-zeros (saved in other AMR levels), which will seriously mislead the prediction. 

\begin{figure}[t]
    \centering 
    \includegraphics[width=0.90\columnwidth]{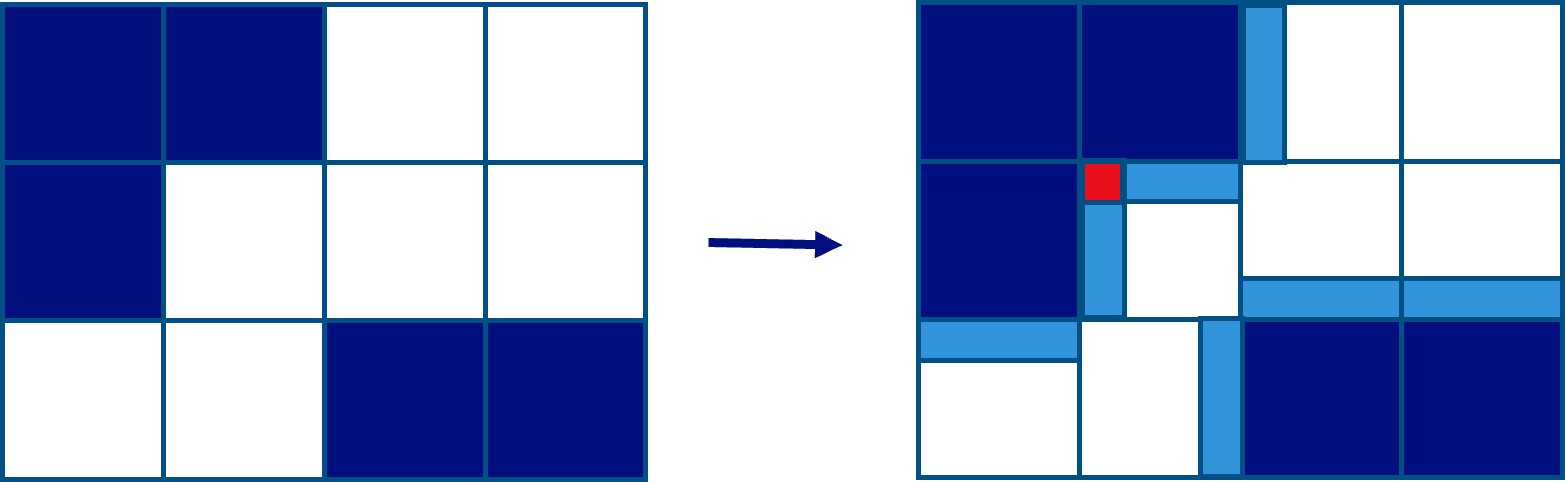}
    
    \caption{A 2D example of GSP approach. Non-empty blocks are in navy blue; padded blocks are in light blue/red; padded blocks based on more than one non-empty neighbor are in red. 
    }
   \vspace{-2mm}
    \label{fig:gsp-e}
\end{figure}

\begin{figure}[b]
     \centering
     \begin{subfigure}[t]{0.49\linewidth}
         \centering
         \includegraphics[width=\linewidth]{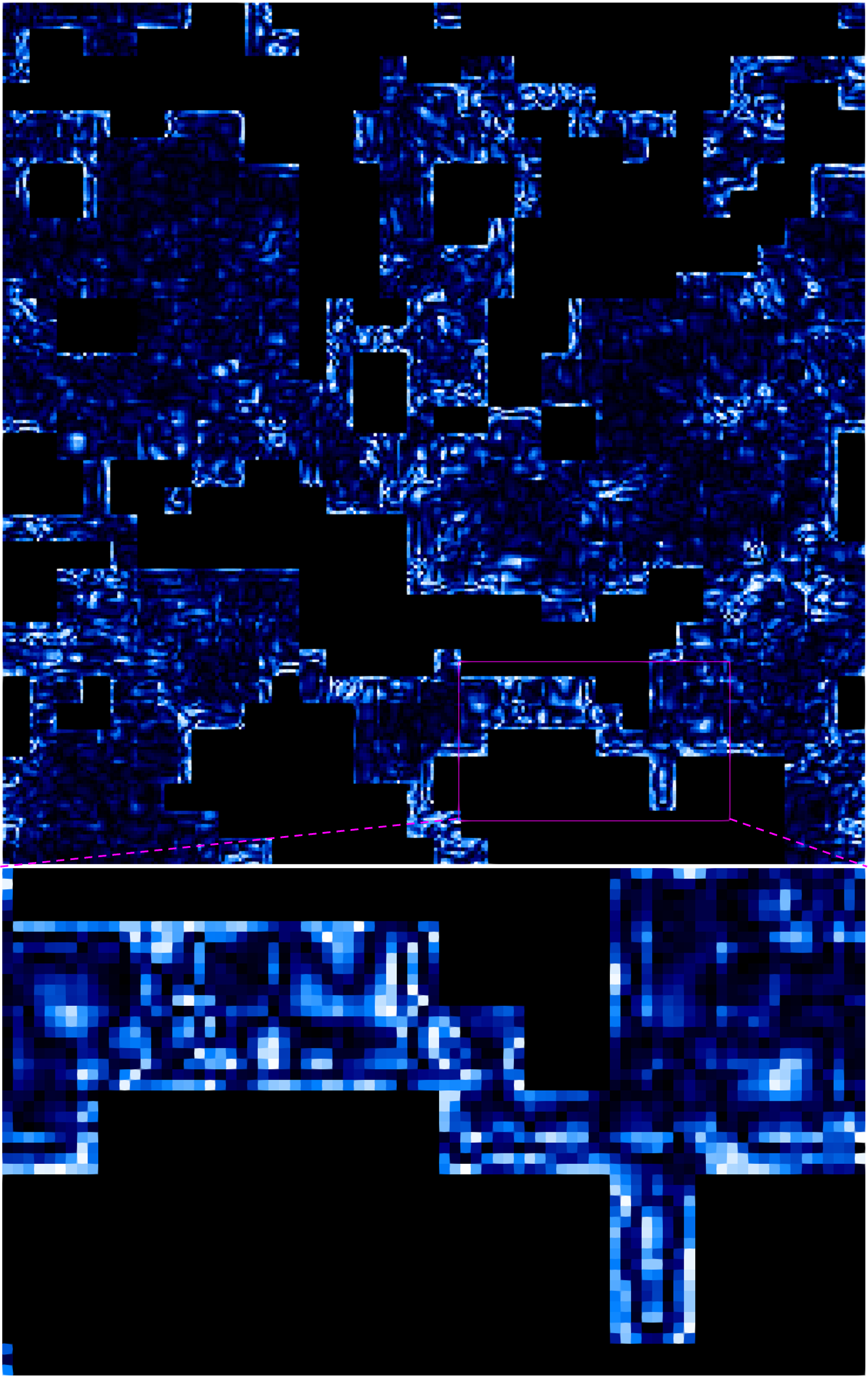}  
         \caption[t]{ZF (CR=156.7, PSNR=32.8dB)}
         \label{fig:zero_err}
     \end{subfigure}
     \begin{subfigure}[t]{0.49\linewidth}
         \centering
         \includegraphics[width=\linewidth]{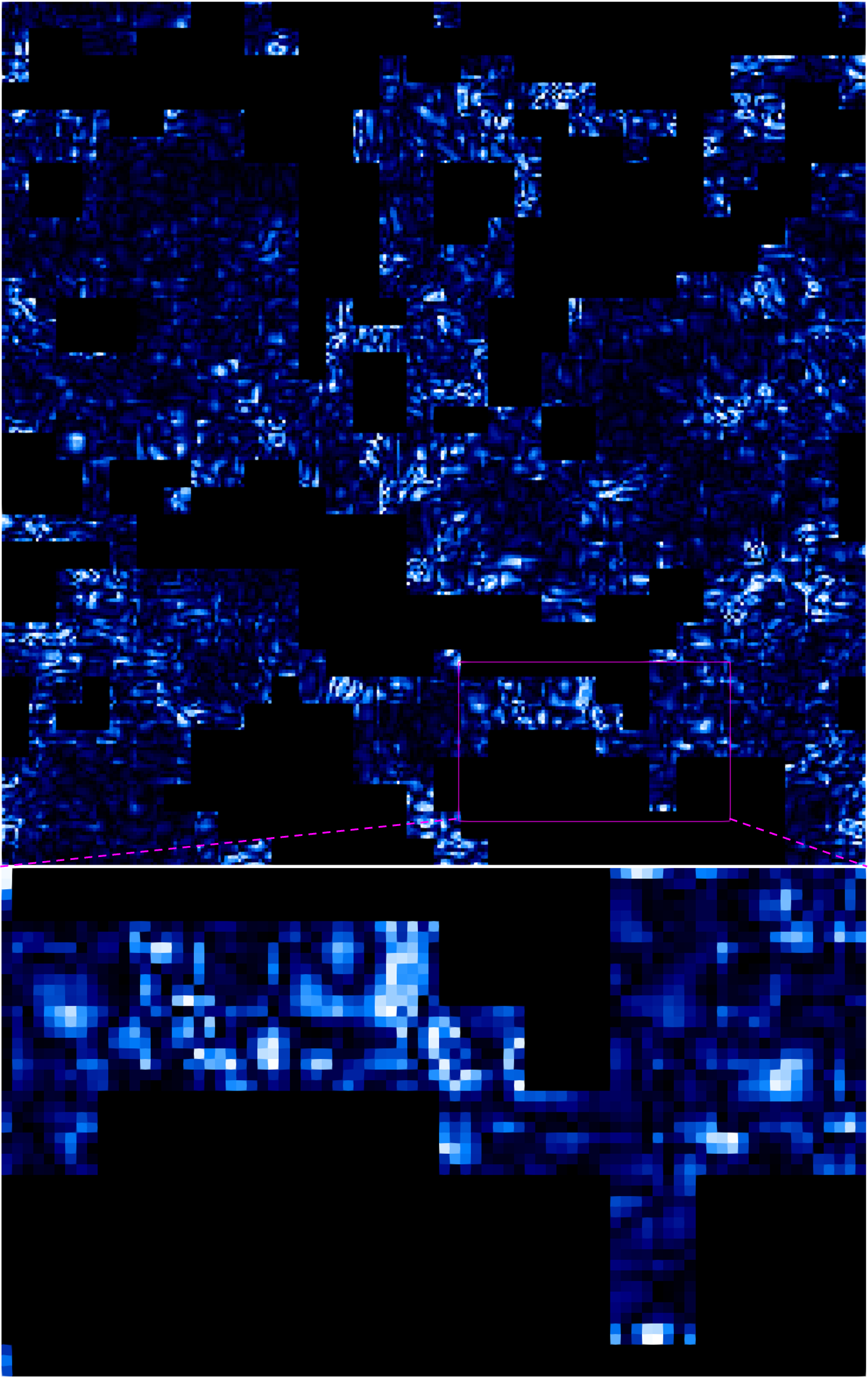}
         \caption{GSP (CR=161.3, PSNR=33.5dB)}
         \label{fig:gsp_err}
     \end{subfigure}
     \vspace{-2mm}
        \caption[t]
        {Visual comparison (one slice) of compression errors of two approaches using SZ based on Nyx's ``baryon density'' field (i.e., z10's coarse level, 77\% density). Brighter means higher compression error. The error bound is the relative error bound of $6.7\times10^{-3}$.}
        \label{fig:gsp_zero}
\end{figure}

\begin{algorithm}[t]
\caption{\footnotesize \textcolor{black}{Proposed Ghost Shell Padding Method}}\label{alg:gsp}  
\ttfamily\footnotesize
\KwIn {Data, $m$}
\KwOut {Data after padding}
\For{each unit block $b_i$}{
    $m = min(unitBlockSize/2, 4)$\;
  \If{$b_i$ is empty \AND $b_i$ has non-empty neighbor }{
    \For{each non-empty neighbor $n_j$}{
        pad slice = avg (first $m$ slices of $n_j$ next to $b_i$)\;
      \uIf{overlap edge}{
      $pad = {pad}/{2}$\;
      }
      \uElseIf{overlap corner}{
      $pad = {pad}/{3}$\;
      }
      \Else{
      continue\;
      }
      add an $m$-layers pad slice to $b_i$ next to $n_j$\;
    }
  }
}
return padded Data
\end{algorithm}

To eliminate the above issue of padding zeroes, 
we propose to use a ghost-shell padding strategy (\textbf{GSP}) to diffuse neighboring values to a padding layer.
Figure~\ref{fig:gsp-e} illustrates the high-level idea, and the detailed algorithm is described in Algorithm~\ref{alg:gsp}. 
Specifically, we first partition the data into unit blocks and then pad each empty unit block by using the average of its non-empty neighbors' boundary data values.

Note that some empty unit blocks have more than one non-empty neighbor such as the red box shown in Figure~\ref{fig:gsp-e}. For these blocks, we will use the average value of all its neighbors for padding. 
Correspondingly, we will remove these padded values in the decompression based on the saved padding information. 
Note that since the padding process is only for non-empty blocks, this metadata overhead is almost negligible for high-density data (e.g., 0.1\%).

After padding, each boundary point will be predicted using the average of all the boundary data in the unit block(s) to which it belongs or is neighbored. 
As shown in Figure~\ref{fig:gsp_zero}, compared to the zero-filling (ZF) approach, GSP can significantly reduce the overall compression error, especially for the boundary data. 
Moreover, the GSP approach can provide a similar compression ratio to the ZF approach on this high-density data and hence a better rate-distortion. 

\textcolor{black}{Note that besides using the average values, we can also use values that aid the compression predictor in making the most accurate predictions for padding. However, this requires running the predictor an extra time to determine the best values, making the process more time-consuming. Furthermore, as illustrated in Figure 6, average padding already significantly reduces the compression error on boundary when compared to zero padding. Thus, we chose to use the average values to avoid sacrificing too much performance.}

\subsection{Optimized Sparse Tensor Representation for Low-density Data} 
\label{sec:opst}
When most of the regions in the data are empty (e.g., about 77\% of the data is empty in Figure~\ref{fig:dis_fine}), the large amount of padded data would greatly increase the size of data for compression, resulting in a low compression ratio.

\begin{figure}[h]
    \centering 
    \vspace{-2mm}
    \includegraphics[width=0.8\columnwidth]{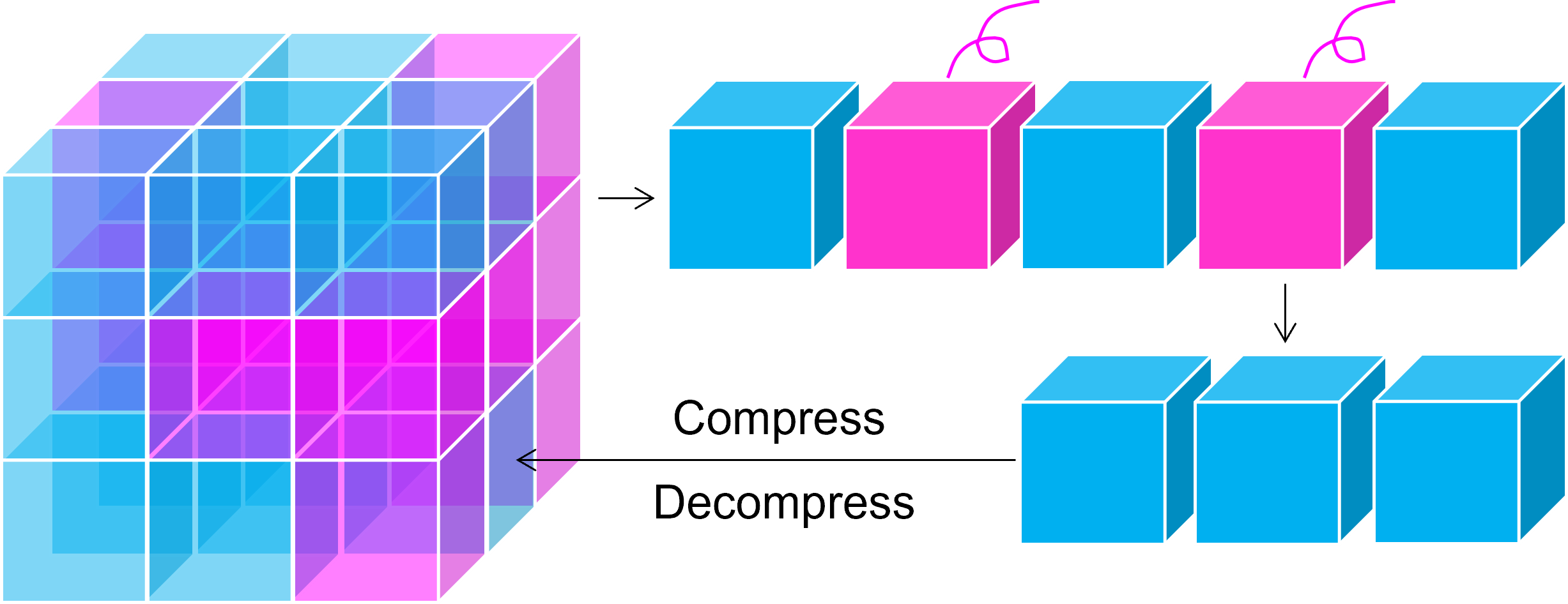}
    \caption{
    Workflow of the naive sparse tensor (NaST) method (empty regions marked in pink and non-empty regions marked in blue).}
    \vspace{-2mm}
    \label{fig:st-flow}
\end{figure}

To solve this issue, we propose to use a naive sparse-tensor-based approach (called \textbf{NaST}) to remove the empty regions, as shown in Figure~\ref{fig:st-flow}. NaST includes four main steps in the compression process: (1) partition the 3D data into multiple unit blocks, (2) remove the empty blocks, (3) linearize the remaining 3D blocks into a 4D array, and (4) pass the 4D array to the compressor. Note that in the decompression process, we will put the unit blocks from the decompressed 4D array back into the original data.

However, in order to completely remove the empty regions to form a sparse representation, the unit block size needs to be relatively small compared to the input data size (e.g., $16^3$ vs. $512^3$).
This results in a reduced spatial locality because of the partitioning.
Also, a high proportion of data will be on the block boundary. Given that the linearized unit blocks may not be contiguous in the original dataset, their boundaries aren't smooth, making it challenging for compressors like SZ to predict values accurately."
As a result, the NaST method without optimizing the unit block size would have low compression performance.

\begin{figure}[t]
    \centering 
    \vspace{-2mm}
    \includegraphics[width=\columnwidth]{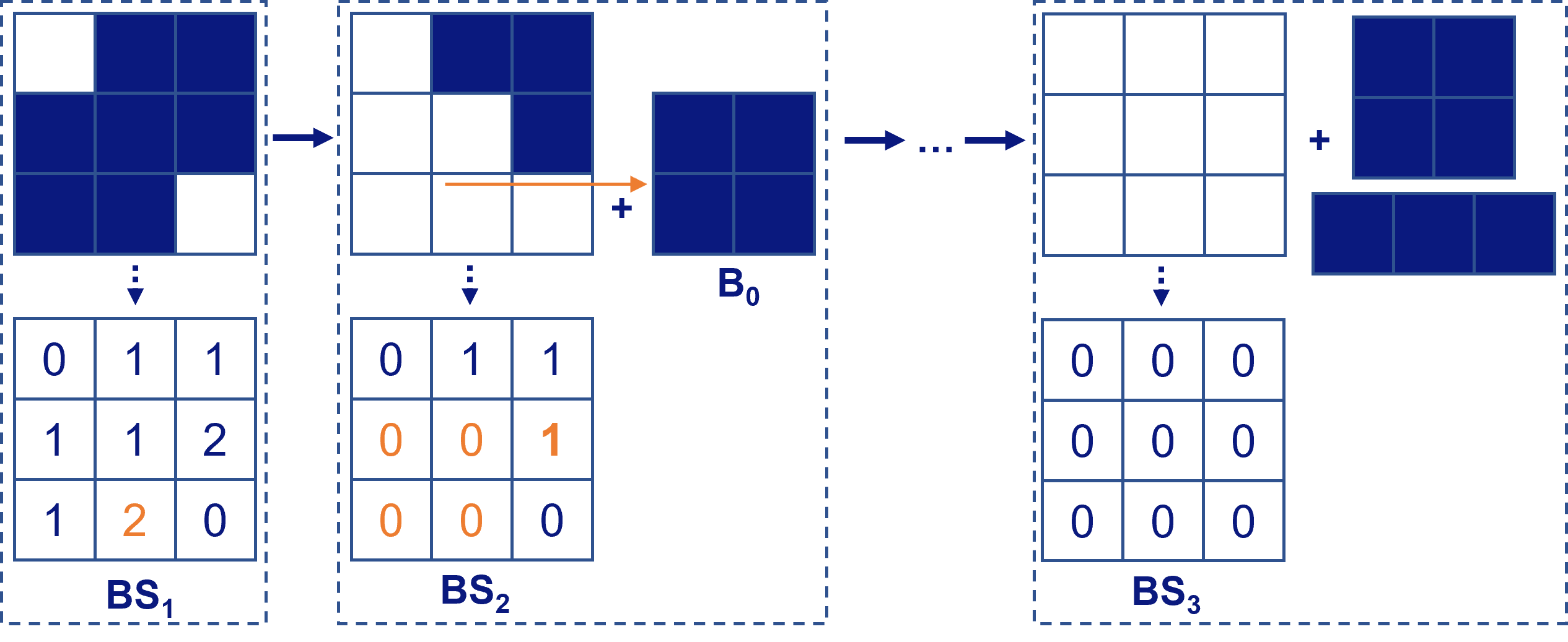}
     \vspace{-4mm}
    \caption{A 2D example of our proposed OpST approach. The sub-blocks are extracted according to our optimized sizes saved in $BS$. E.g., a 2-by-2 sub-block $B_0$ is extracted according to $BS_1[2][1]$.}
    \vspace{-2mm}
    \label{fig:opst-e}
\end{figure}

To address the above problems, we propose an optimized sparse tensor representation (called \textbf{OpST}) to effectively remove the empty regions as well as maintain a relatively large unit block size so as to increase the spatial locality and reduce the portion of boundary data.
A detailed description of our algorithm can be found in Algorithm~\ref{alg:opst}.
We use a 2D example to demonstrate our approach, as illustrated in Figure~\ref{fig:opst-e}. 
Specifically, (1) we partition the data into many small unit blocks.
(2) For each unit block, we use the dynamic programming method to initiate an array $BS$ to save the dimension/size of the maximum square whose bottom-right corner is that unit block (line 6, which will be discussed in the next paragraph). 
(3) We extract the sub-blocks (composed of multiple unit blocks) from the original data according to the sizes saved in $BS$ (lines 13). 
(4) Since the original data will be changed after the extraction, we need to partially update $BS$ based on \textit{maxSide} (lines 14, will be discussed later).  
We loop (3) and (4) from the bottom-right corner to the top-left corner until the original data is empty.
(5) After extracting all the sub-blocks, we put them into multiple 3D arrays (to be compressed) based on their sizes. 
Note that the sub-blocks with the same size will be merged into the same array for easy compression.

\begin{algorithm}[h]
\caption{\footnotesize Proposed Optimized Sparse Tensor Method}\label{alg:opst}
\ttfamily\footnotesize
\KwIn {Sparse 3D data S}
\KwOut {multiple 4D array $D_n$}
\For{each unit block $b(x,y,z)$}{
  \If{$b(x,y,z)$ is non-empty}{
    \eIf{x is 0 \OR y is 0 \OR z is 0}{
        $BS(x,y,z) = 1$
      }{
        $BS(x,y,z) = \min( BS(x-1, y, z),~BS(x,y-1,z), ~BS(x,y,z-1), ~BS(x-1,y-1,z), ~BS(x,y-1,z-1), ~BS(x-1,y,z-1), ~BS(x-1,y-1,z-1)) + 1$
        \tcc*{BS(x,y,z) is the dimension size of the maximum cube whose bottom right rear corner is the unit block with index (x,y,z) in the original data}
        $maxSide = \max(maxSide, ~BS(x,y,z))$
      }
  }
}
\For{each unit block $b(x,y,z)$}{
    \If{$BS(x,y,z) \geq 1$}{
        $size = BS(x,y,z)$
        $D_{size} \leftarrow S((x-size:x) * blkSize,~ (y-size:y) * blkSize,~ (z-size:z)* blkSize)$
        \tcc*{put the sub-block to the according to 4D array}
        $b(x-size:x,~y-size:y,~z-size:z) \leftarrow empty$
        $BS(x-size:x,~y-size:y,~z-size:z) = 0$
        $BS = updateBs(BS,~x,~y,~z,~ maxSide )$
     }
}
return $D_n$
\end{algorithm}

When initializing the $BS$ in step (2), we start with the $b'[i][j]$ with $i = 0$ or $j = 0$ (i.e., on the top-left edge), where $b'[\cdot][\cdot]$ are the unit blocks: if $b'[i][j]$ is empty, we will set $BS[i][j]$ to 0 otherwise 1. 
For the remaining unit blocks, if it is empty, $BS[i][j]$ will be 0; otherwise, $BS[i][j]$ will be set to 1 plus the minimum value among its three neighboring blocks (i.e., upper block, left block, and upper-left block).
In other words, we have $BS[i][j] = 1 + \min(BS[i][j-1], BS[i-1][j], BS[i-1][j-1])$ for the 2D case. For example, $BS_1[2][1]$ is 2 because all its upper-left neighbors are 1 (as shown in Figure~\ref{fig:opst-e}). However, both $BS_1[1][1]$ and $BS_2[1][2]$ can only reach 1 because one of their neighbors is set to 0, having no chance to form a sub-block with the size of 2. 
\textcolor{black}{Then, for step 3, we perform extraction based on the BS. For instance, as illustrated in Figure~\ref{fig:opst-e}, according to $BS_1[2][1]$, we extract a 2-by-2 sub-block, $B_0$, with its bottom right corner at $BS_1[2][1]$.}
Moreover, as mentioned in step (4), we need to update $BS$ after each extraction.
Specifically, for each sub-block we extract, we have to set its corresponding values in $BS$ to zeros. 
For instance, as shown in Figure~\ref{fig:opst-e}, after we extract a 2-by-2 sub-block $B_0$ at $BS_1[2][1]$, we need to set $BS_2[1][0]$, $BS_2[1][1]$, $BS_2[2][0]$, and $BS_2[2][1]$ to zeros.
In addition, we also need to recalculate a part of $BS$ (line 17 in Algorithm~\ref{alg:opst}) because the extraction could influence other $BS$ values. 
For example, we need to recalculate $BS_2[1][2]$ (marked in bold orange) after extracting $B_0$.
Note that this update is a partial update as the $BS$ values to be updated will be bounded by \textit{maxSide} which is the dimension size of the largest cube in the dataset (line 7). 

Similar to NaST, in decompression, we will put the sub-blocks back to reconstruct the data based on the saved coordinates. Note that after our optimization, each sub-block size will be relatively large (e.g., $96^3$ vs the original data size of $512^3$), the overhead of saving the coordinates of all the sub-blocks will be negligible (e.g., 0.1\%). 

\begin{figure}[t]
     \centering
     \begin{subfigure}[t]{0.49\linewidth}
         \centering
         \includegraphics[width=\linewidth]{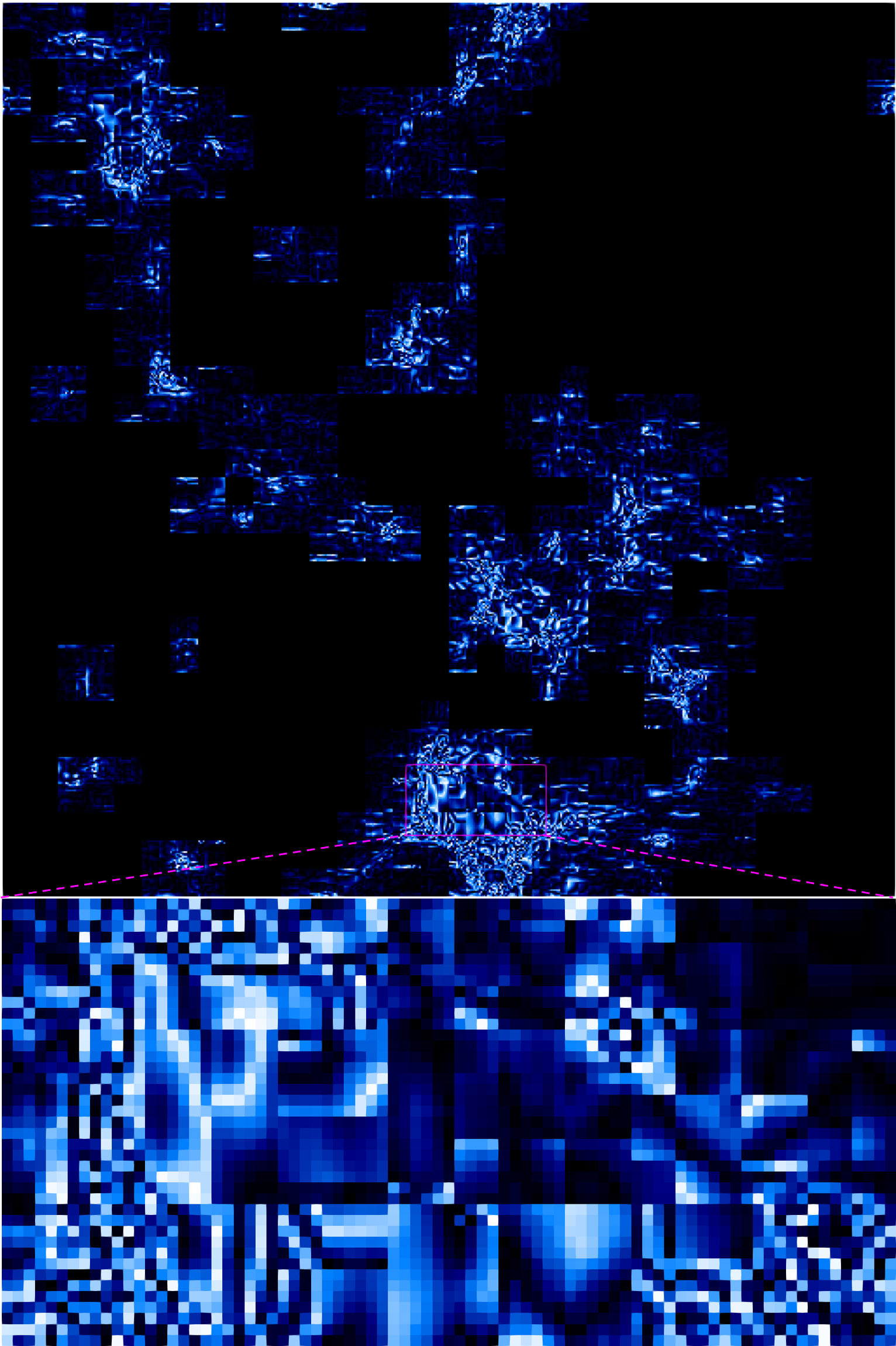}  
         \caption[t]{NaST(CR=245, PSNR=77.5dB)}
         \label{fig:st-err}
     \end{subfigure}
     \begin{subfigure}[t]{0.49\linewidth}
         \centering
         \includegraphics[width=\linewidth]{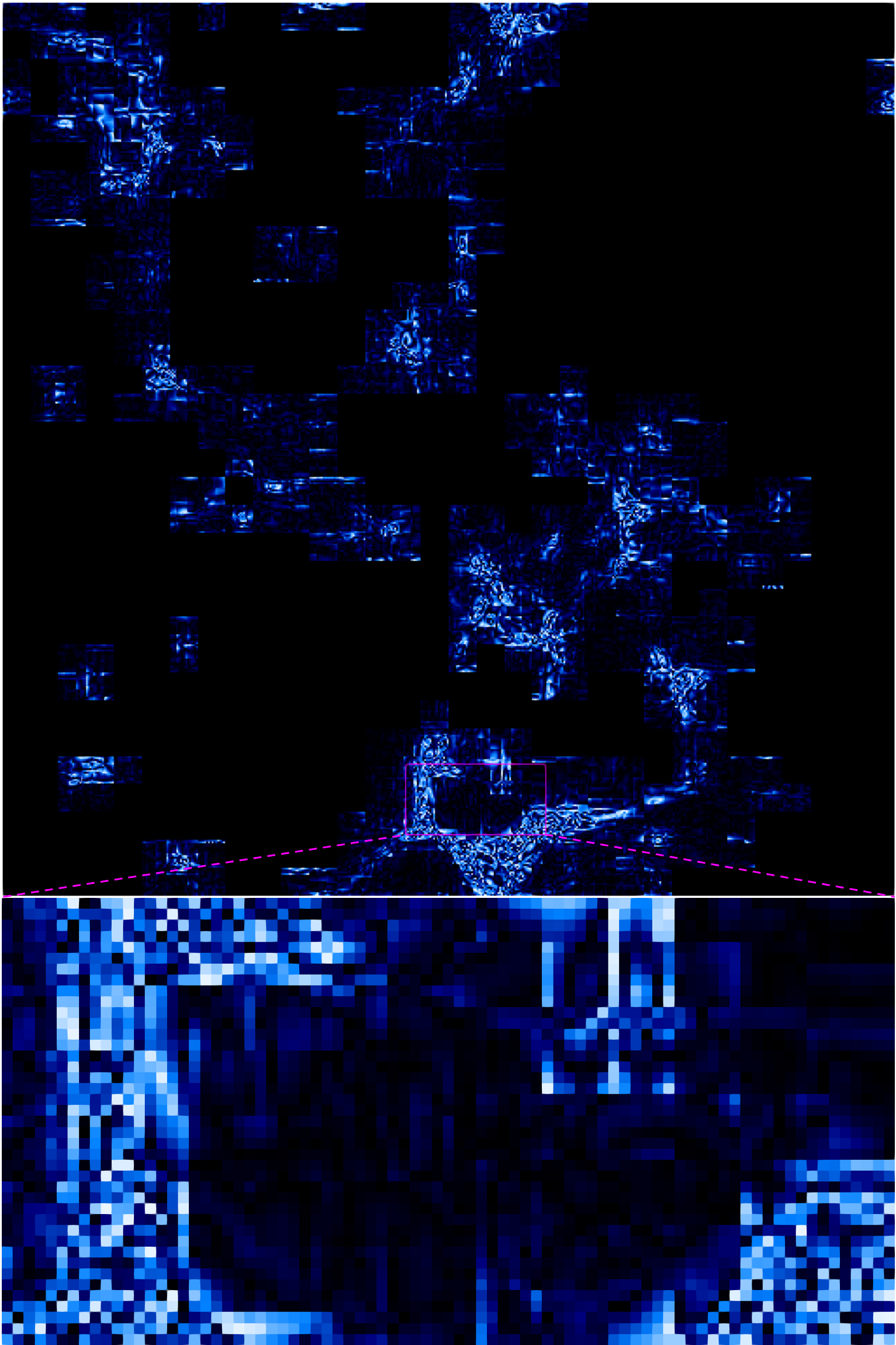}
         \caption{OpST(CR=248, PSNR=78.0dB)}
         \label{fig:opst-err}
     \end{subfigure}
     \vspace{-2mm}
        \caption[t]{Visual comparison (one slice) of compression errors of two approaches using SZ based on Nyx's ``baryon density'' field (i.e., z10's fine level, 23\% density). Brighter means higher compression error. The error bound is the relative error bound of $7.2\times10^{-4}$.}
        \vspace{-2mm}
        \label{fig:st_opst}
\end{figure}

Finally, we show a visual comparison of the compression quality between NaST and OpST in Figure~\ref{fig:st_opst}.
Note that both use SZ with the same error bound. Brighter means more errors. 
We can observe that compared to the NaST method, OpST can significantly reduce the overall compression error, especially for the data points on the boundary.
It is worth noting that even with a lower error, our OpST can still provide a higher compression ratio than NaST.
This is because our proposed optimization will generate larger sub-blocks, which provide more information for prediction-based lossy compressors such as SZ to achieve better rate-distortion.
A detailed evaluation will be shown in Section~\ref{sec:evaluation}.

\subsection{Adaptive $k$-D Tree for Medium-density Data}
\label{sec:kd}

The OpST approach proposed for low-density data, however, has a high computation overhead, especially when the data is relatively dense. 
This is because, on one hand, OpST needs to update \textit{BS} based on \textit{maxSide} for each extraction of a sub-block, while the larger the \textit{maxSide}, the more values in \textit{BS} that need to be updated; on the other hand, \textit{maxSide} is the dimension size of the largest non-empty cube in the dataset, which is highly related to the density of the dataset.
Thus, the time complexity of OpST can be expressed as $O(N^2 \cdot d)$, where $N$ is the unit block number and $d$ is the density. 
Note that here density describes how dense the data is. 
For example, the density of 77\% means that 23\% of the data is empty.
Clearly, when the density of an AMR level is relatively high, using OpST will be relatively time-consuming. 

\begin{figure}[t]
    \centering 
    \includegraphics[width=0.95\columnwidth]{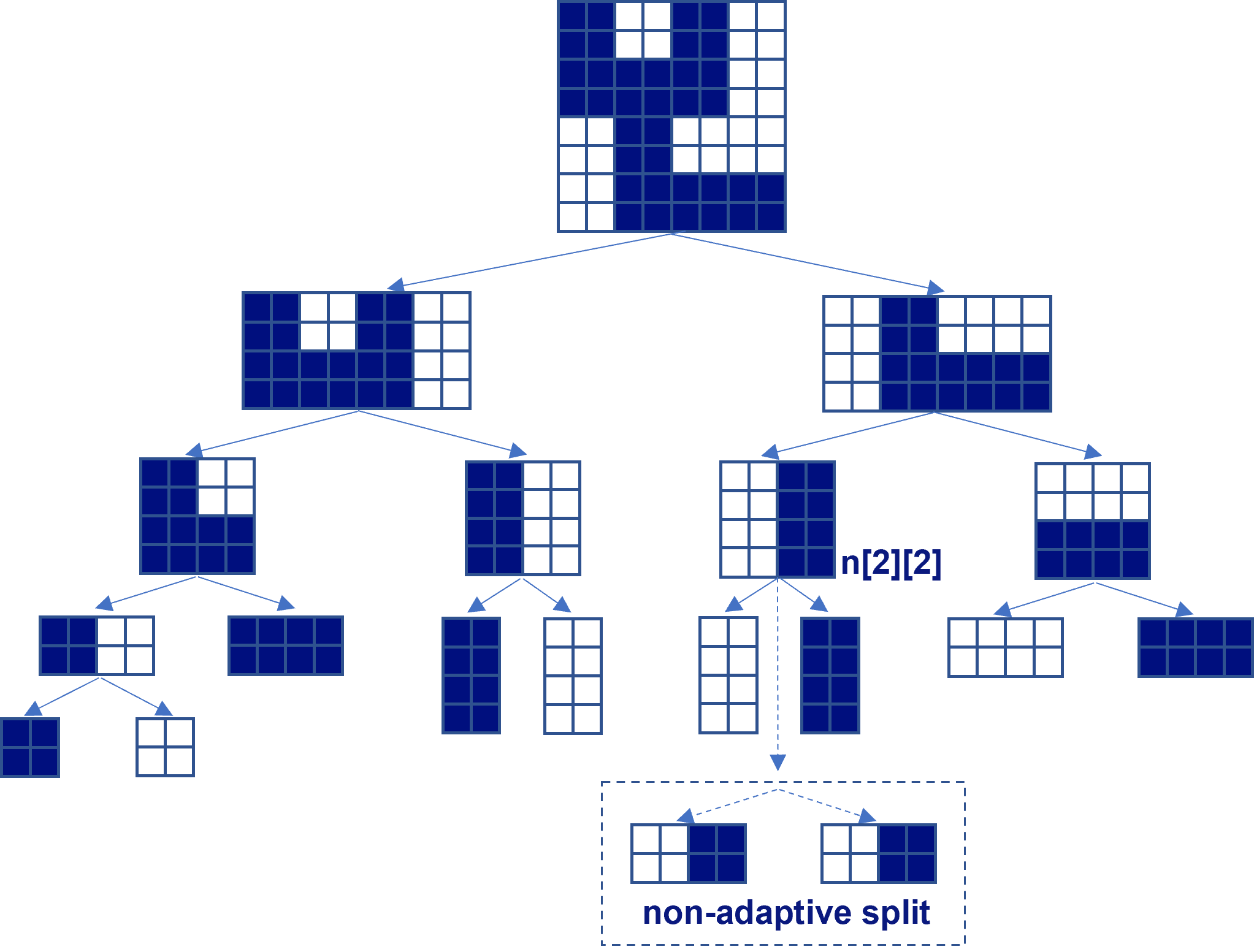}
    \caption{2D example of adaptive $k$-D tree. Sub-block will be adaptively split to effectively remove empty regions and get bigger full sub-blocks.}
    \vspace{-4mm}
    \label{fig:kde}
\end{figure} 

\begin{algorithm}[h]
\textcolor{black}{\caption{\footnotesize Dynamic $k$-D Tree}}\label{alg:kd}
\ttfamily\footnotesize
\KwIn {data block $d$, counts information}
\KwOut {$k$-d tree}
node.count $\leftarrow$ counts information\;
\eIf{$d$ is empty \OR $d$ is full}{
    continue \tcc*{stop splitting}
}{
    \eIf{\textcolor{black}{max(x,y,z)/min(x,y,z) > 2 \AND pre-split is not over}}{
        \textcolor{black}{split from the max dimension to get $d_1$, $d_2$\;
        node.left\ \ = AKDTree ($d_1$)\;
        node.right = AKDTree ($d_2$)\;}
    }{
        \uIf{$d$ is a cube}{
        split d equally into 8 oct-blocks: $s_1,\cdots,s_8$\;
        get the counts $c_1,...c_8$ for $s_1,\cdots,s_8$\;
        find the maxDiff partition $d_1$,$d_2$\;
        node.left\ \ = AKDTree ($d_1$, four $c_i$ of $d_1$)\;
        node.right = AKDTree ($d_2$, four $c_i$ of $d_2$)\;
        }
        \uElseIf{$d$ is a flat cuboid}{
          get the counts $c_1,\cdots,c_4$ from counts information\;
          find the maxDiff partition $d_1$, $d_2$\;
          node.left\ \ = AKDTree ($d_1$, two $c_i$ of $d_1$)\;
          node.right = AKDTree ($d_2$, two $c_i$ of $d_2$)\;
        }
        \uElseIf{$d$ is a slim cuboid}{
          get the counts $c_1,c_2$ from counts information\;
          split $d$ along the largest dimension to get $d_1$,$d_2$\;
          node.left\ \ = AKDTree ($d_1$, $c_1$)\;
          node.right = AKDTree ($d_2$, $c_2$)\;
        }
    }
}
return node;
\end{algorithm}

To address the above high overhead issue of OpST, we propose an adaptive \kdtree, called \textbf{AKDTree}, to remove empty regions and extract sub-blocks (containing multiple unit blocks). AKDTree has a lower time complexity of $O(\frac{1}{3} N \cdot \log N)$ (will be discussed later). Figure~\ref{fig:kde} shows a simple 2D example. 
Specifically,
(1) we partition the data into small unit blocks. 
(2) We use a tree to hierarchically represent the whole data. Each node in the tree is associated with a sub-block of the data. 
Moreover, each node stores the number of non-empty unit blocks in the sub-block associated with the node. 
(3) For each node, we split its associated sub-block from the middle along one dimension to form two sub-blocks for its two children. Note that while keeping the 3D feature of data, we select one dimension that can maximize the difference of the numbers of non-empty unit blocks of the two children (will be discussed in the next two paragraphs).
(4) We keep splitting a node until it has no empty unit block or itself is empty. 
(5) Once finishing the construction of the tree, we collect all the leaf nodes and send them to the compressor. Note that a non-empty leaf node does not have any empty unit block; otherwise, it will keep splitting. Thus, a leaf node must be an empty or full node, as shown in Figure~\ref{fig:kde}. 
More detail is described in Algorithm~\ref{alg:kd}.

As mentioned in step (3), We are distributing the non-empty unit blocks unevenly to two children for each node. This is done to maximize the number of leaf nodes with large sub-block sizes. \textcolor{black}{Large data blocks contain more spatial information, which can enhance compression}.
If we keep splitting sub-blocks in a fixed way, for instance, first split along the $x$-axis, second split along the $y$-axis, third split along the $x$-axis, fourth split along the $y$-axis, and so on, 
we will get a 2-by-2 sub-block for the node $n[2][2]$ as shown in the dashed box, while its largest possible sub-block could be 4 by 2 as shown in Figure~\ref{fig:kde}.

\begin{figure}[h]
    \centering 
    \includegraphics[width=0.85\columnwidth]{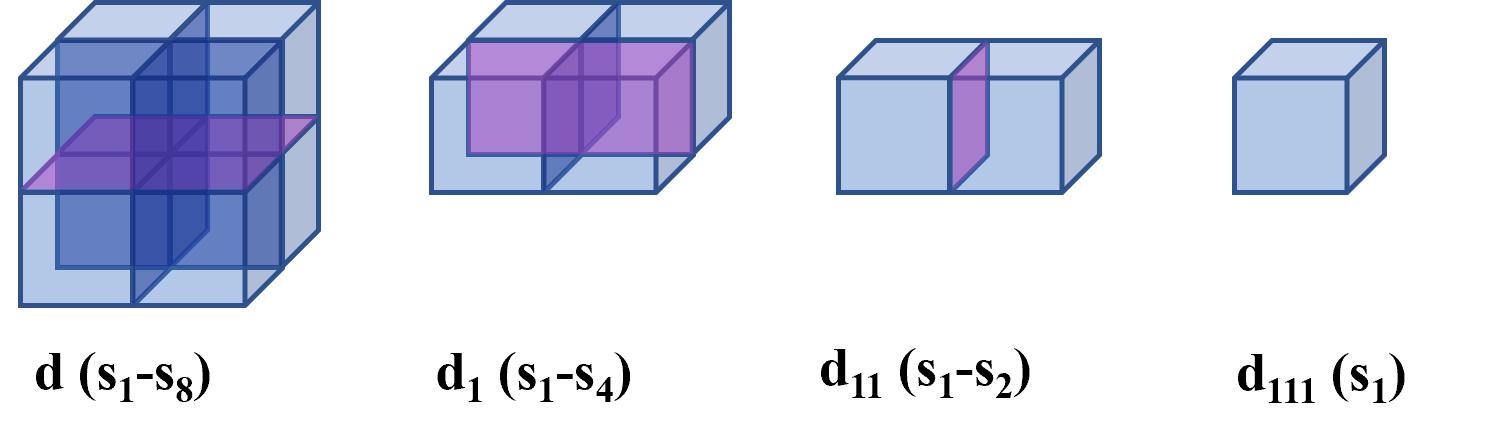}
    \vspace{-2mm}
    \caption{Example of adaptive splitting, different shapes will have a different number of choices for splitting.}
   \vspace{-4mm}
    \label{fig:8sp}
\end{figure} 

To select one of the dimensions to unevenly distribute its non-empty unit blocks to the two children, we now present our dynamic splitting approach.

\textcolor{black}{To maintain the high-dimensional characteristics of the dataset, it's essential to prioritize splitting along the largest dimension, especially if it is significantly greater than the others. For instance, with a dataset sized $8\times8\times64$, it's preferable to split along the z-axis rather than the x or y axes. Splitting along the x or y would risk flattening a 3D dataset down to 2D. Consequently, our initial approach is to divide the dataset along the dominant dimension until:}
\textcolor{black}{\begin{equation}
   max(x,y,z)/min(x,y,z) < 2
\end{equation}}
\textcolor{black}{Then, we categorize nodes into three different types: ``cube'' nodes, ``flat'' nodes, and ``slim'' nodes, whose dimension ratios are x:y:z, 2x:2y:z (or 2x:y:2z or x:2y:2z), 2x:y:z (or x:y:2z or x:2y:z), respectively. Here, x, y, and z represent the dimensions of the data block after pre-split.}

First of all, for the cube node $d$, we first divide it into eight oct-blocks, i.e., $s_1$, $s_2$, $\cdots$, $s_8$ (as shown in Figure~\ref{fig:8sp}), each sized $\frac{x}{2}\times\frac{y}{2}\times\frac{z}{2}$. x, y, and z are the dimension sizes of the block after pre-split.
Then, we can get the counts of non-empty unit blocks of the eight oct-blocks, i.e., $c_1$, $c_2$, $\cdots$, $c_8$. 
After that, we will decide along which dimension to split the cube node $d$ based on the counts. Specifically, we can calculate the following three differences:
\begin{align*}
    \text{diff}_x &= |c_1+c_3+c_5+c_7-c_2-c_4-c_6-c_8|,\\
    \text{diff}_y &= |c_1+c_2+c_5+c_6-c_3-c_4-c_7-c_8|,\\
    \text{diff}_z &= |c_1+c_2+c_3+c_4-c_5-c_6-c_7-c_8|.
\end{align*}

Finally, we compare these three values and choose the dimension with the maximum difference to split. For example, if the maximum difference is $\text{diff}_z$, we will split $d$ along the z-axis (i.e., the pink 2D plane shown in Figure~\ref{fig:8sp}) and get two flat nodes $d_1$ and $d_2$.
For the flat nodes such as $d_1$, we can reuse $c_1$, $\cdots$, $c_4$ to decide whether to split $d_1$ along the x-axis or y-axis by choosing the larger one among the following two differences. 
\begin{align*}
    \text{diff}_x = |c_1+c_3-c_2-c_4|, \; \text{diff}_y = |c_1+c_2-c_3-c_4|.
\end{align*}

For the slim nodes such as $d_{11}$, 
we simply split it along the x-axis to get two cube nodes $s_1$ and $s_2$. 
This process (i.e., cube nodes$\to$flat nodes$\to$slim nodes) in step (3) will be looped until the node becomes a leaf node (empty or full). 

Note that based on the above description, the counting process is required for every three nodes in each three paths (i.e., only for the ``cube'' nodes). 
Thanks to this dynamic splitting approach, we can lower the time complexity of the AKDTree algorithm to $O\big(\frac{1}{3} \cdot N \cdot \log N\big)$, where $N$ is the number of unit blocks while extracting as many relatively large sub-blocks without empty unit block as possible. 

In addition, after the dynamic splitting, we will have a series of sub-blocks with the same size but in different directions (e.g., 2x:2y:z, 2x:y:2z, x:2y:2z). We will align the sub-blocks with the same size based on their splitting dimensions (instead of in-memory transposing them), merge them into an array, and compress multiple merged arrays together. 

\begin{figure*}[ht]
\centering
\begin{subfigure}{0.32\linewidth}\centering
    \includegraphics[width=0.99\linewidth]{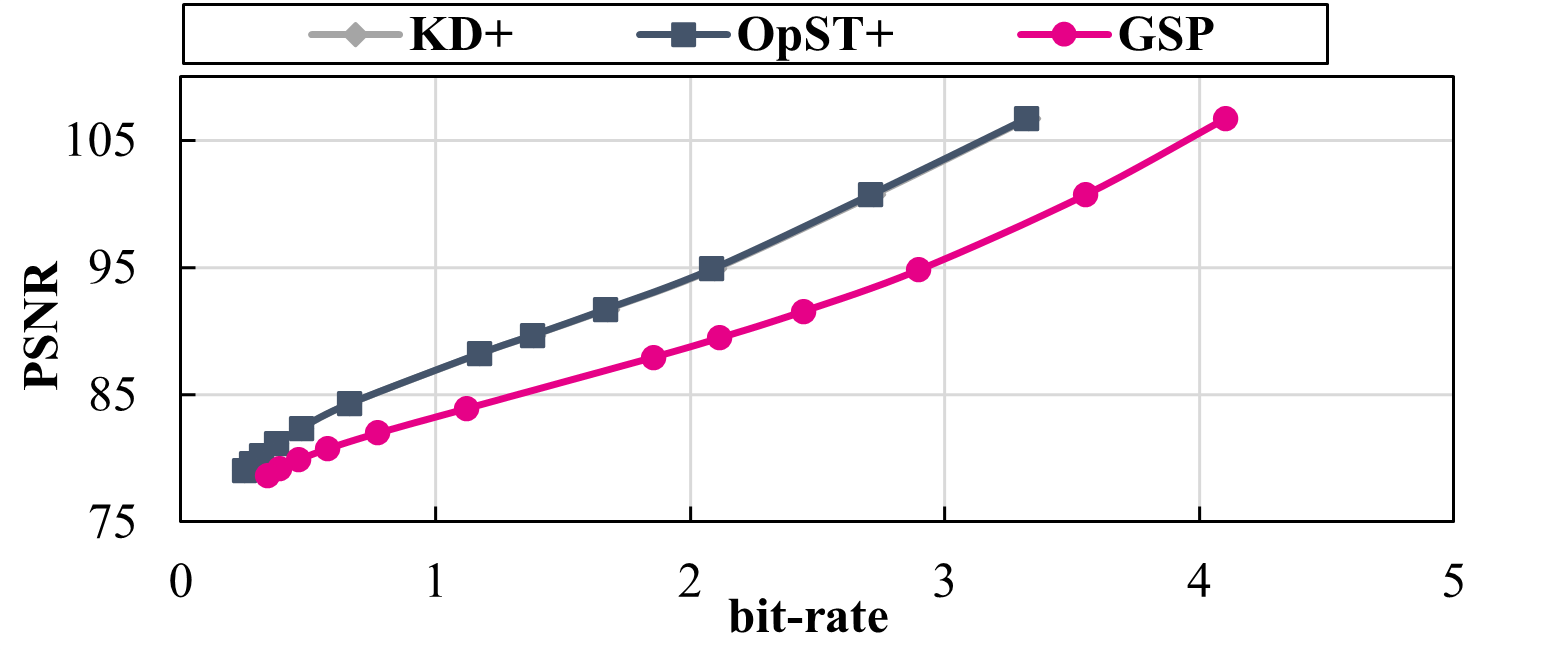}
    \vspace{-3mm}
	\caption{\footnotesize\centering d = 23}\label{fig:tt-1}
\end{subfigure}
\begin{subfigure}{0.32\linewidth}\centering
    \includegraphics[width=0.99\linewidth]{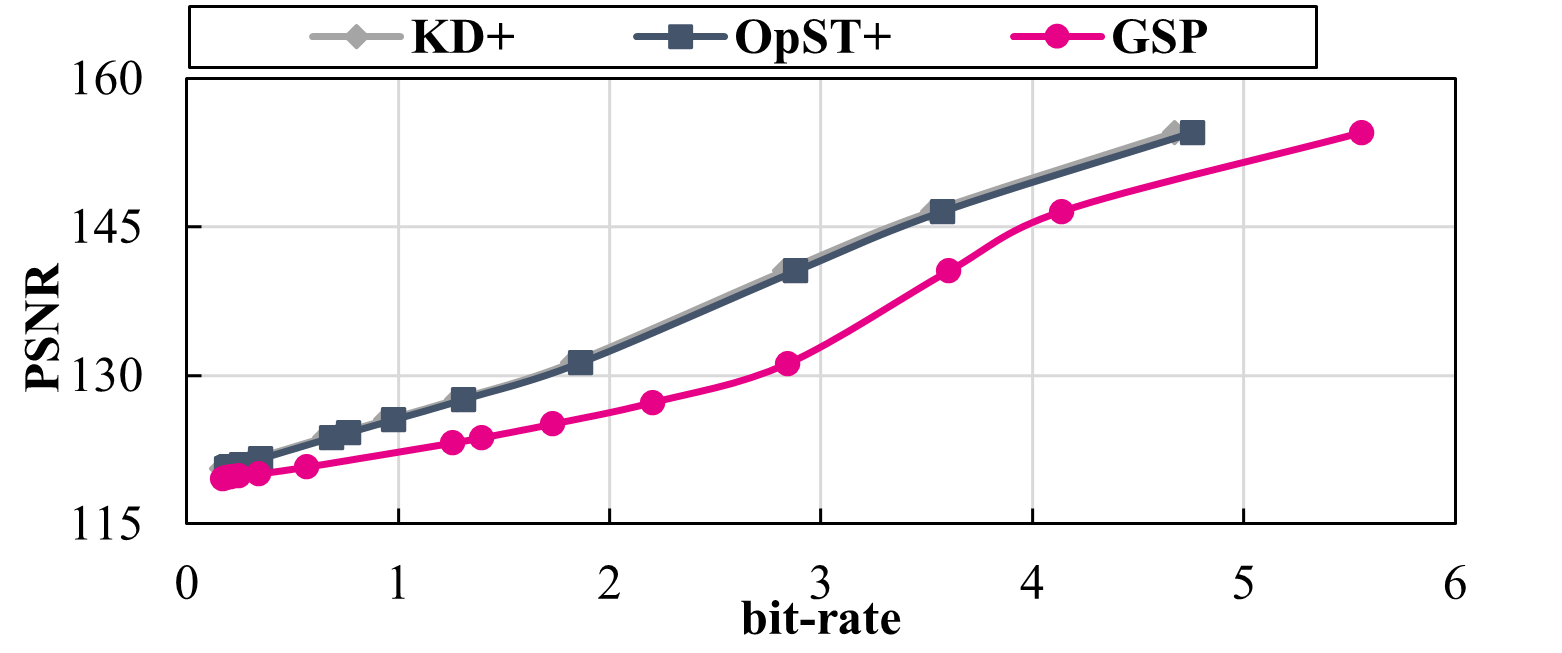}
    \vspace{-3mm}
	\caption{\footnotesize\centering d = 42}\label{fig:tt-4}
\end{subfigure}
\begin{subfigure}{0.32\linewidth}\centering
    \includegraphics[width=0.99\linewidth]{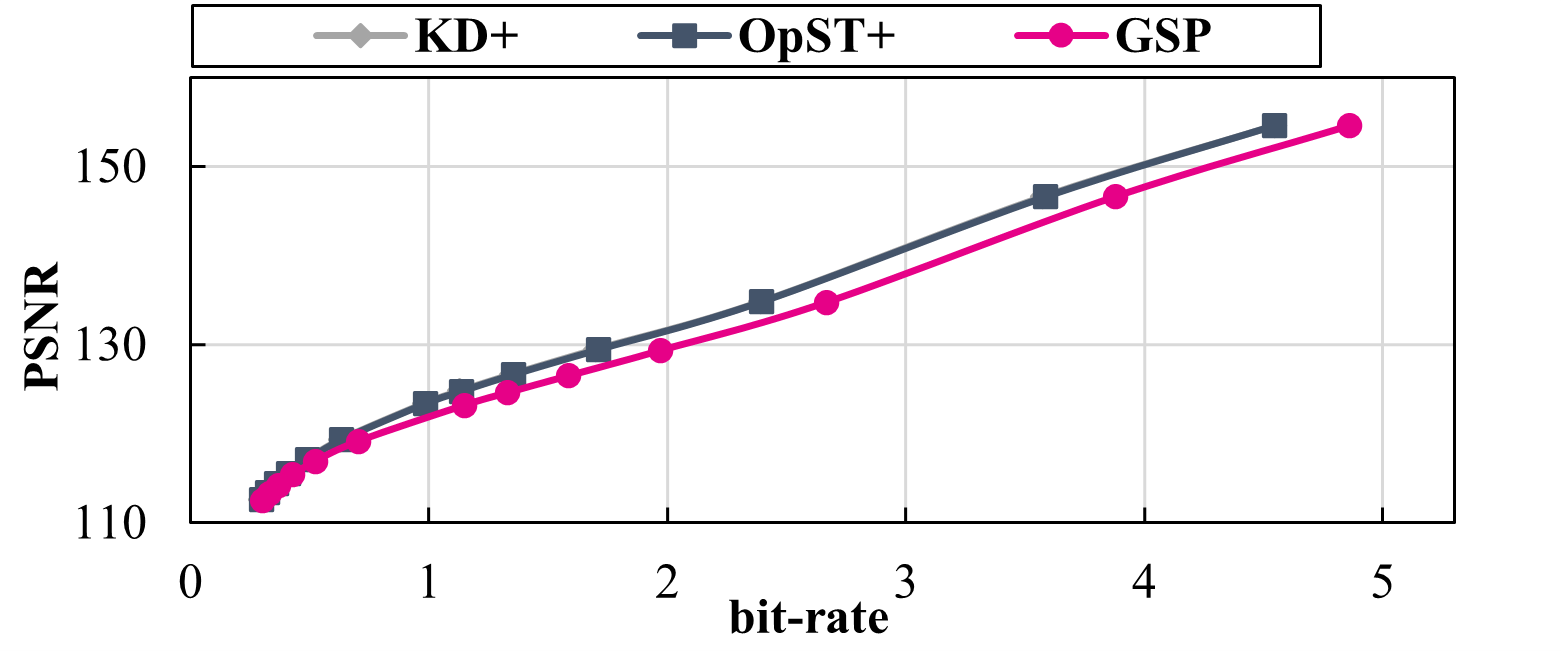}
    \vspace{-3mm}
	\caption{\footnotesize\centering d = 58}\label{fig:tt-2}
\end{subfigure}
\begin{subfigure}{0.32\linewidth}\centering
    \includegraphics[width=0.99\linewidth]{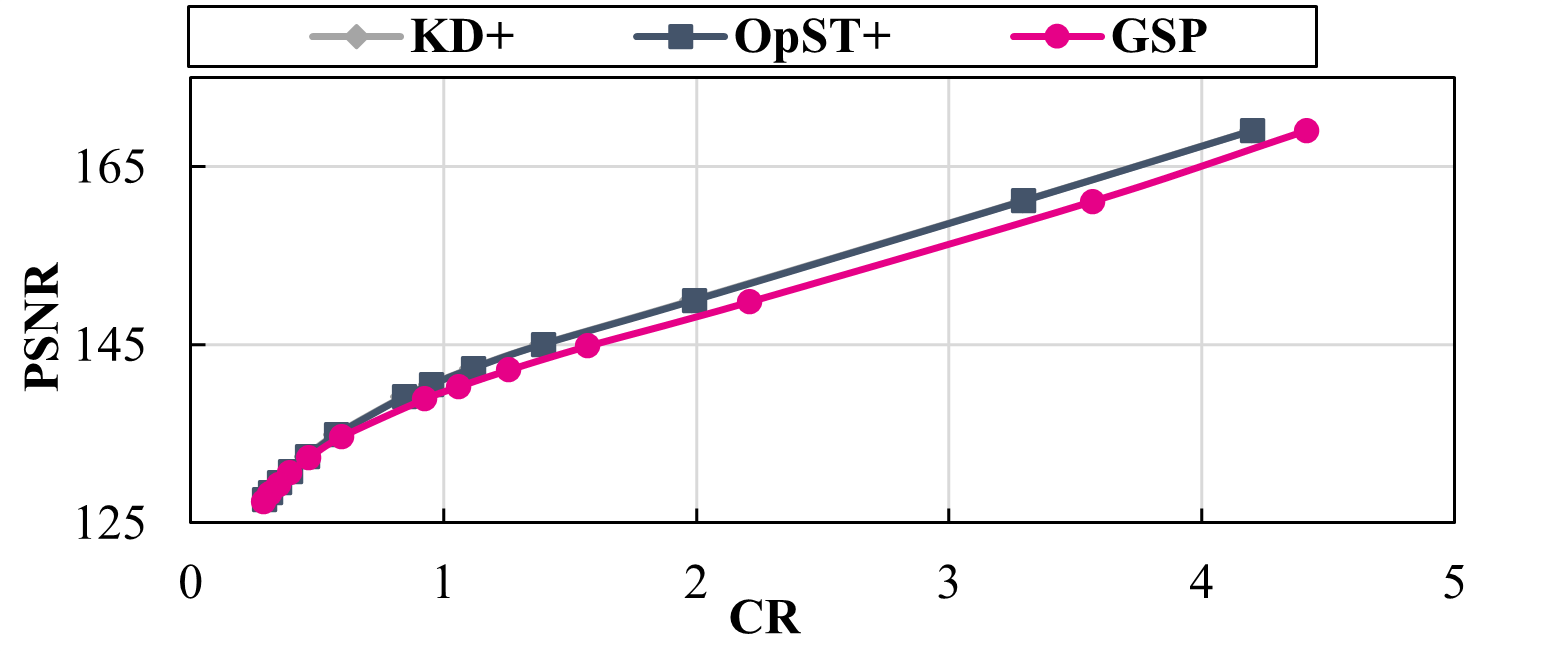}
    \vspace{-3mm}
	\caption{\footnotesize\centering d = 63}\label{fig:tt-3}
\end{subfigure}
\begin{subfigure}{0.32\linewidth}\centering
    \includegraphics[width=0.99\linewidth]{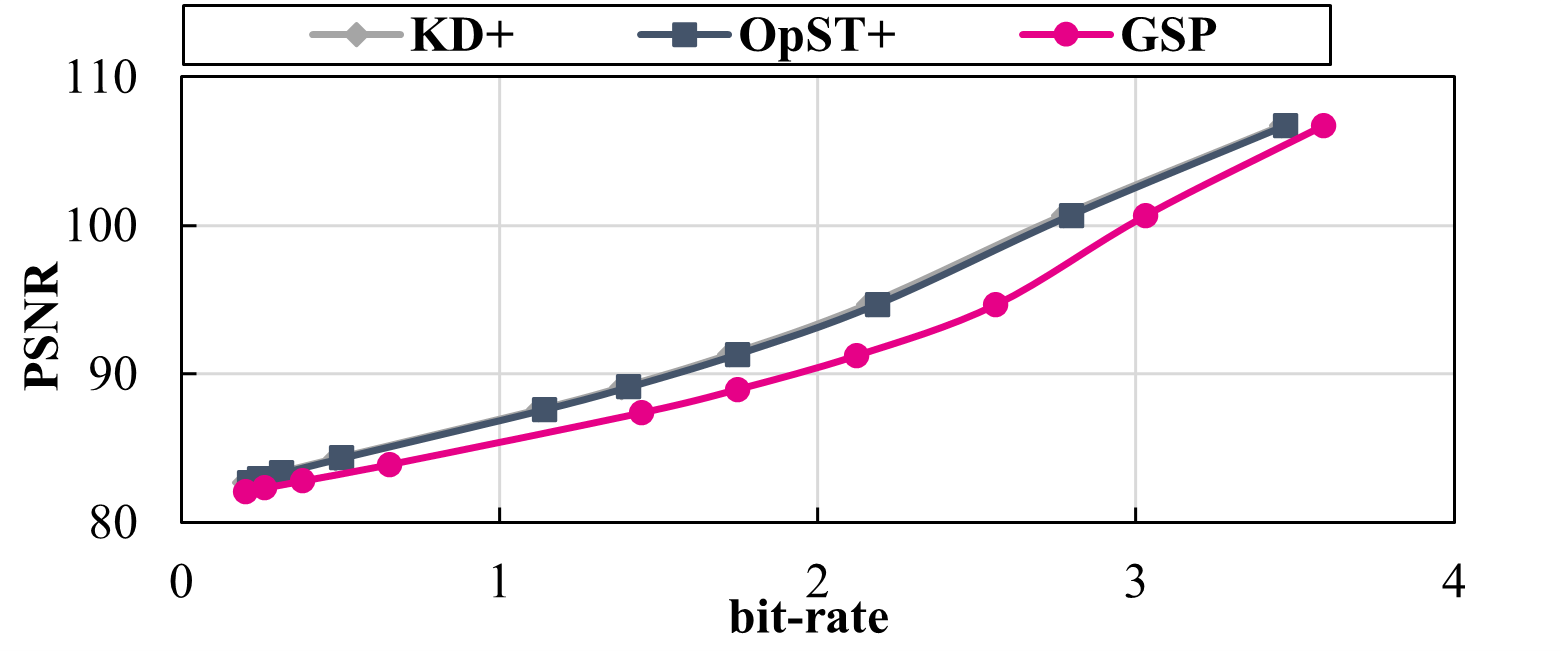}
    \vspace{-3mm}
	\caption{\footnotesize\centering d = 77}\label{fig:tt-5}
\end{subfigure}
\begin{subfigure}{0.32\linewidth}\centering
    \includegraphics[width=0.99\linewidth]{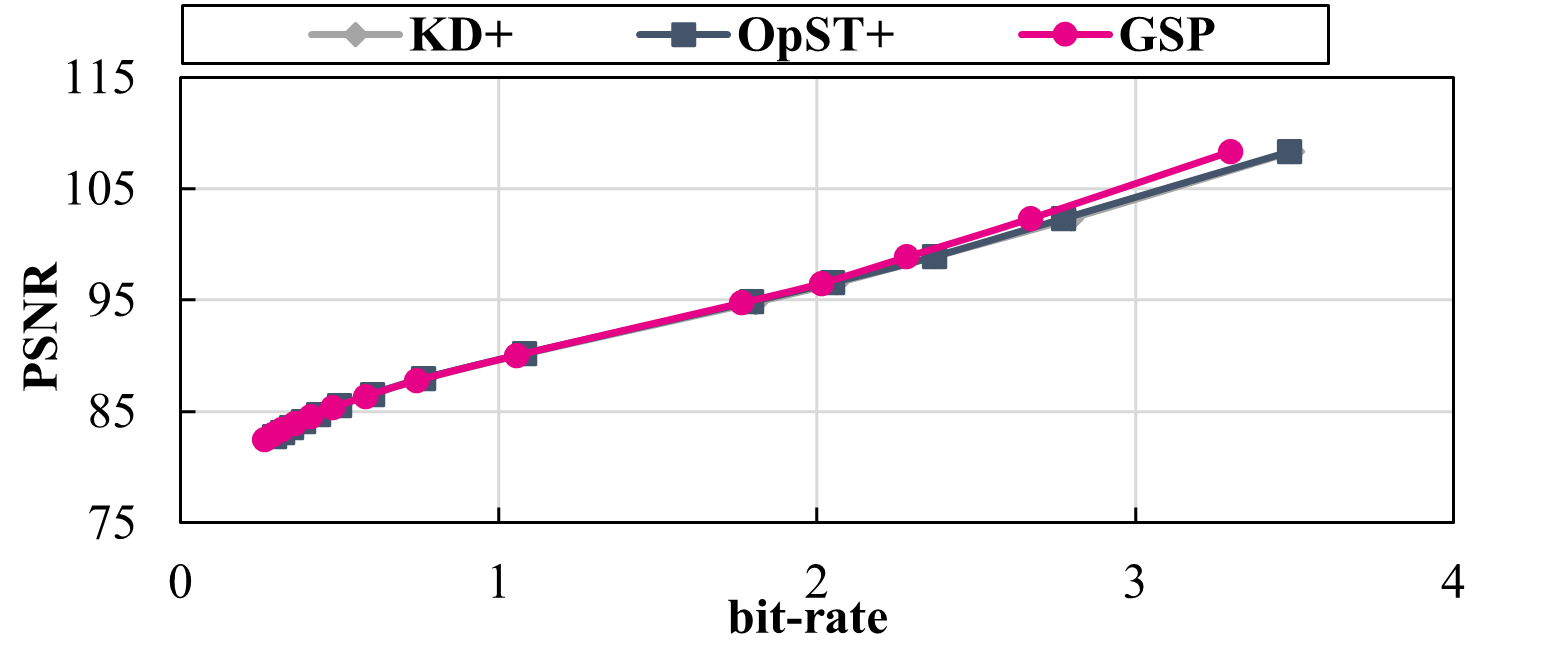}
    \vspace{-3mm}
	\caption{\footnotesize\centering d = 99}\label{fig:tt-6}
\end{subfigure}
 \vspace{-2mm}
\caption{Compression performance comparison of GSP, OpST+, and AKDTree+ on 6 datasets with different densities using Lor/Reg algorithm.}
\label{fig:tt}
\vspace{-2mm}
\end{figure*}

\begin{figure*}[h]
\centering
\begin{subfigure}{0.32\linewidth}\centering
    \includegraphics[width=0.99\linewidth]{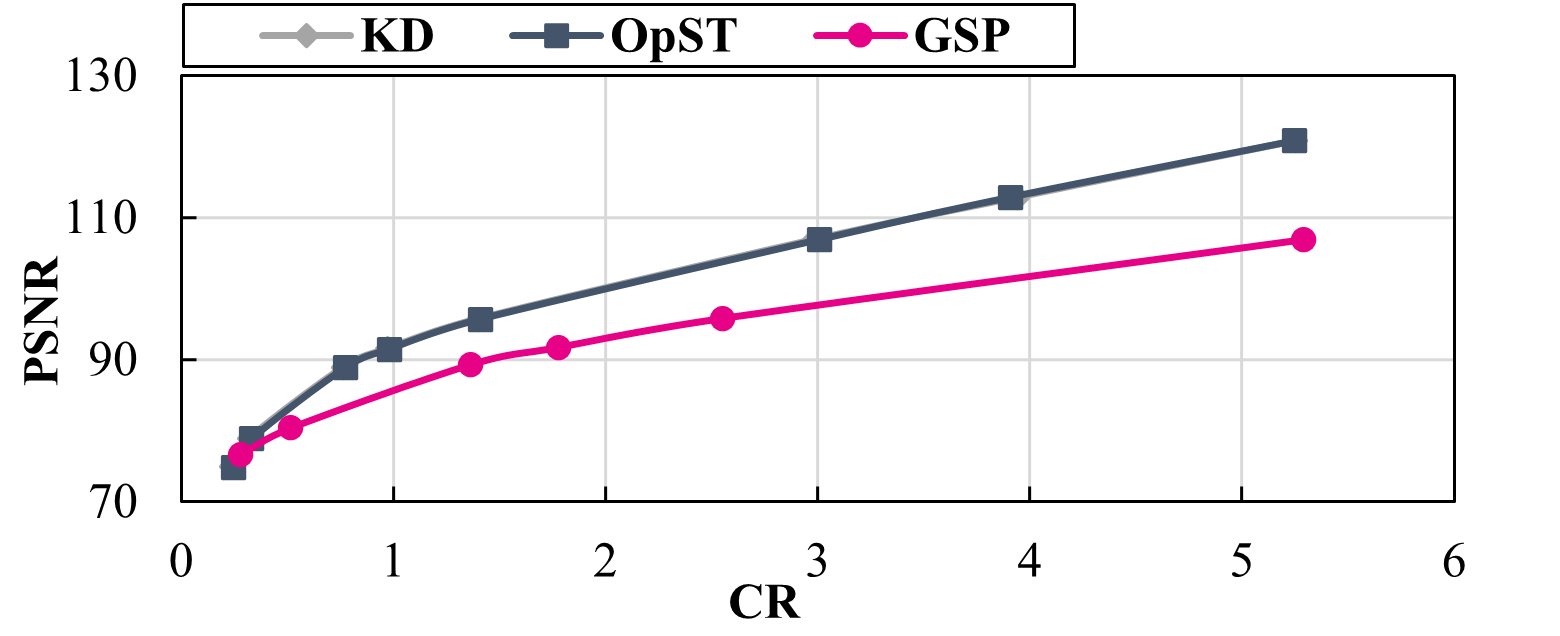}
    \vspace{-4mm}
	\caption{\footnotesize\centering d = 23}\label{fig:tt2-1}
\end{subfigure}
\begin{subfigure}{0.32\linewidth}\centering
    \includegraphics[width=0.99\linewidth]{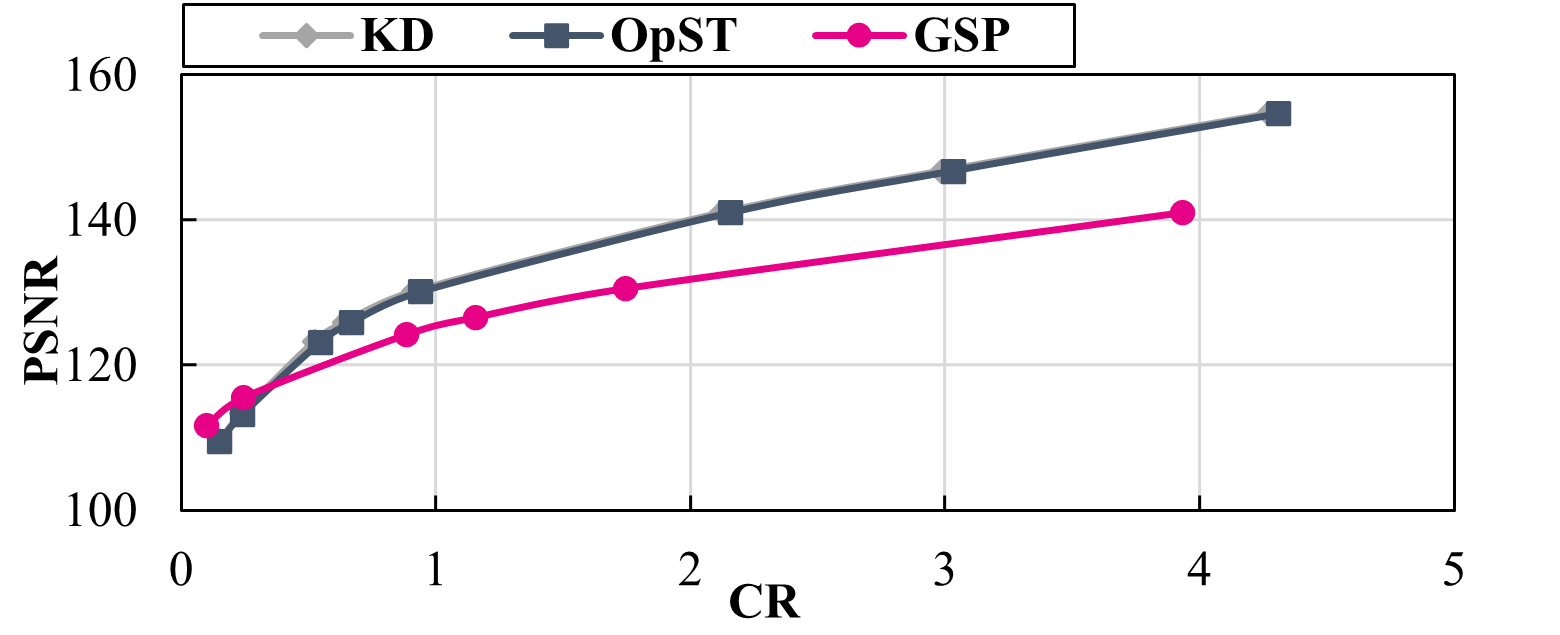}
    \vspace{-4mm}
	\caption{\footnotesize\centering d = 42}\label{fig:tt2-2}
\end{subfigure}
\begin{subfigure}{0.32\linewidth}\centering
    \includegraphics[width=0.99\linewidth]{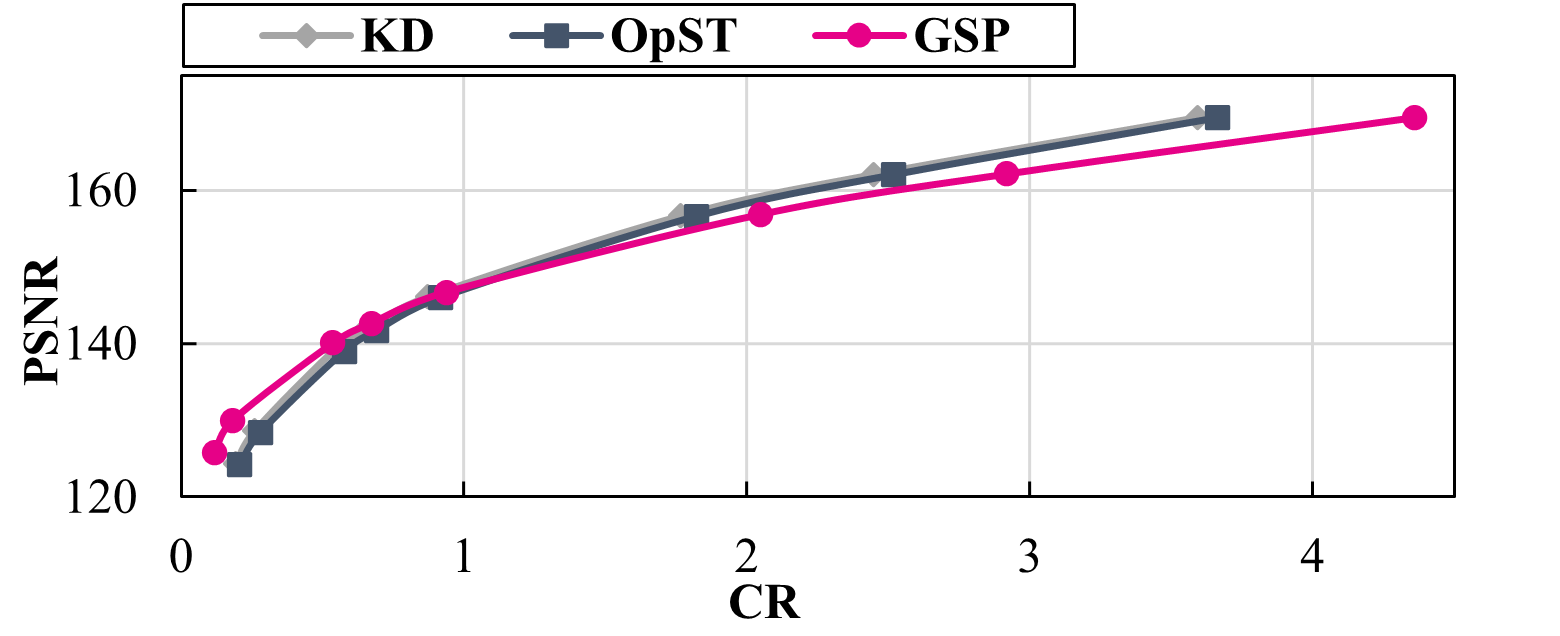}
    \vspace{-4mm}
	\caption{\footnotesize\centering d = 63}\label{fig:tt2-3}
\end{subfigure}
\begin{subfigure}{0.32\linewidth}\centering
    \includegraphics[width=0.99\linewidth]{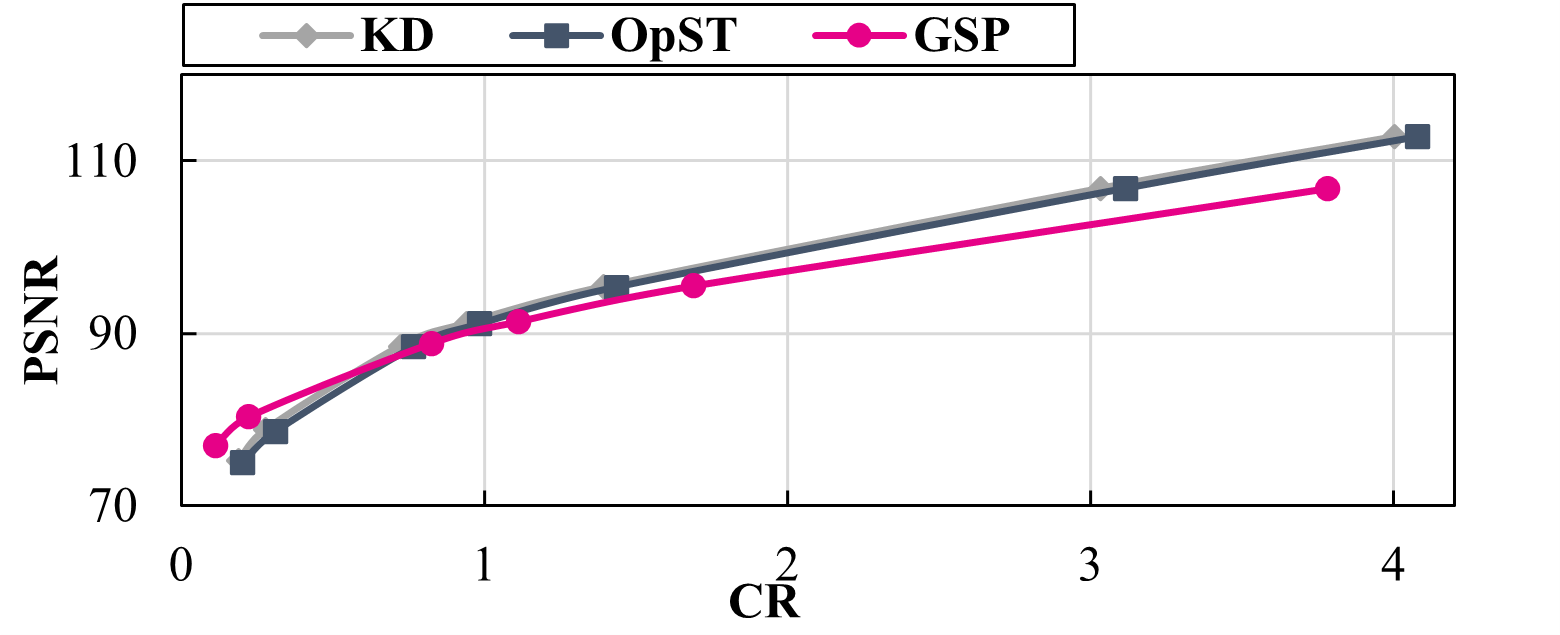}
    \vspace{-4mm}
	\caption{\footnotesize\centering \textbf{d = 77}}\label{fig:tt2-4}
\end{subfigure}
\begin{subfigure}{0.32\linewidth}\centering
    \includegraphics[width=0.99\linewidth]{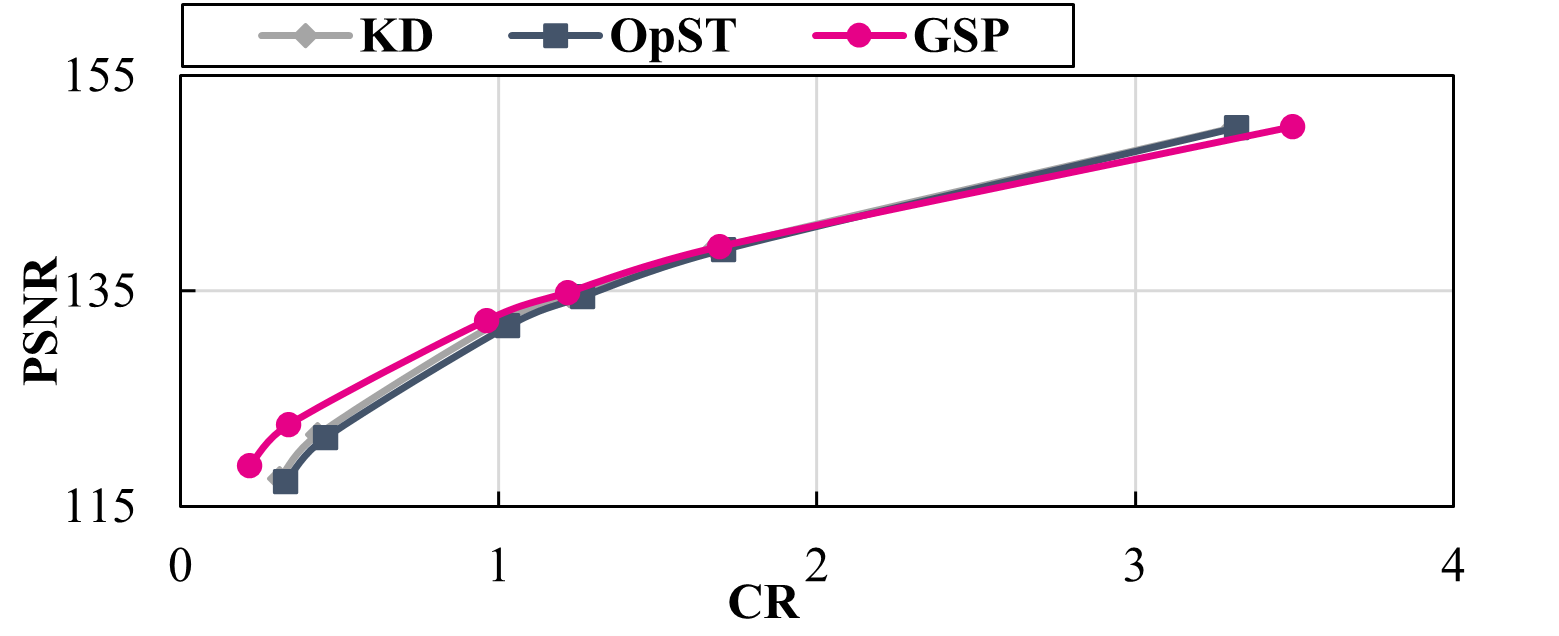}
    \vspace{-4mm}
	\caption{\footnotesize\centering \textbf{d = 85}}\label{fig:tt2-5}
\end{subfigure}
\begin{subfigure}{0.32\linewidth}\centering
    \includegraphics[width=0.99\linewidth]{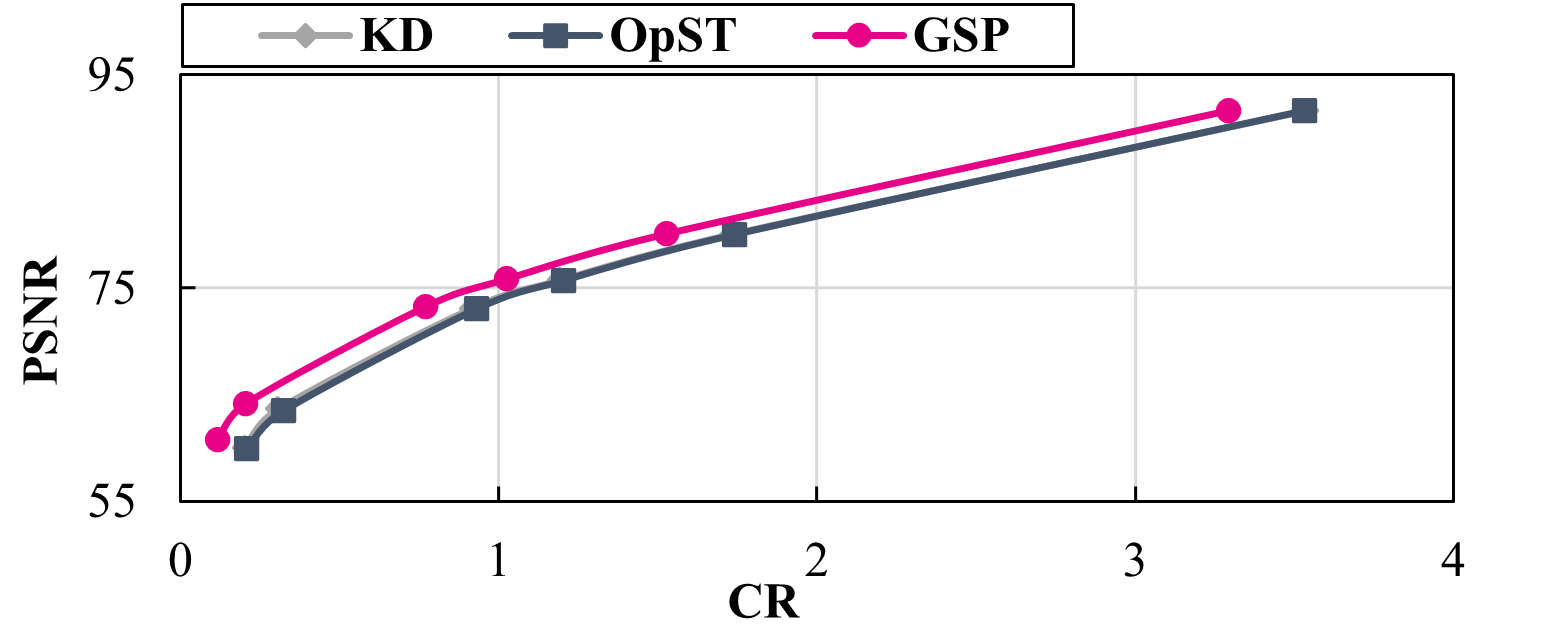}
    \vspace{-4mm}
	\caption{\footnotesize\centering d = 99}\label{fig:tt2-6}
\end{subfigure}
 \vspace{-2mm}
\caption{\textcolor{black}{Compression performance comparison of GSP, OpST, and AKDTree on 6 datasets with different densities using Interp algorithm.}}
\label{fig:tt2}
\vspace{-6mm}
\end{figure*}

\subsection{Shared Huffman encoding} \label{sec:sle}

As mentioned in Section \ref{sec:introduction}, \ref{sec:kd}, and \ref{sec:opst}, the OpST and AKDtree methods collect and linearize the data blocks with the same shape into a 4D array and send it to SZ. 
However, as highlighted in section~\ref{sec:opst}, these blocks may not be contiguous in the original dataset. This leads to a lack of locality/smoothness at the boundaries between non-neighbored data blocks, which can still compromise prediction accuracy.
\textcolor{black}{This adverse effect is particularly pronounced for SZ‘s Lor/Reg algorithm. As it is a local prediction algorithm that depends solely on local data, its performance plummets at block boundaries where local smoothness is absent.}

A potential solution could be compressing each data block individually using SZ. However, this approach would result in low Huffman encoding efficiency because the entire dataset would be divided into many small blocks, requiring SZ to build a separate Huffman tree for each of these small blocks.
In other words, the original SZ method either requires predicting and encoding small blocks together (by forcing them merged into 4D arrays), leading to low prediction accuracy, or predicting and encoding each small block separately, which results in high Huffman encoding overhead.
Furthermore, even if the data blocks with the same shape can be compressed together, the data blocks with different shapes still need to be compressed separately, resulting in low Huffman encoding performance and a high time cost of launching SZ multiple times.

\begin{algorithm}[t]
\caption{\footnotesize SZ compression with SHE}\label{alg:comsle}
\ttfamily\footnotesize
\KwIn {multiple data block $D_{1...n}$}
\KwOut {compressed data $S$}
quantCode $Q$, regreCoeff $R$, compressed data $S$\;

\For{each data block $D_i$}{ 
  quantCode block $q_i$, regreCoeff block $r_i$\;
  $q_i$, $r_i$ = SZ.compress($D_i$)\;
  $Q$.append($q_i$)\;
  $R$.append($r_i$)\;
}
$S \leftarrow HuffmanEncode(Q)$\;
$S \leftarrow HuffmanEncode(R)$\tcc*{end compression}
return S;

\end{algorithm}

\begin{figure}[t]
    \centering 
    \includegraphics[width=0.75\columnwidth]{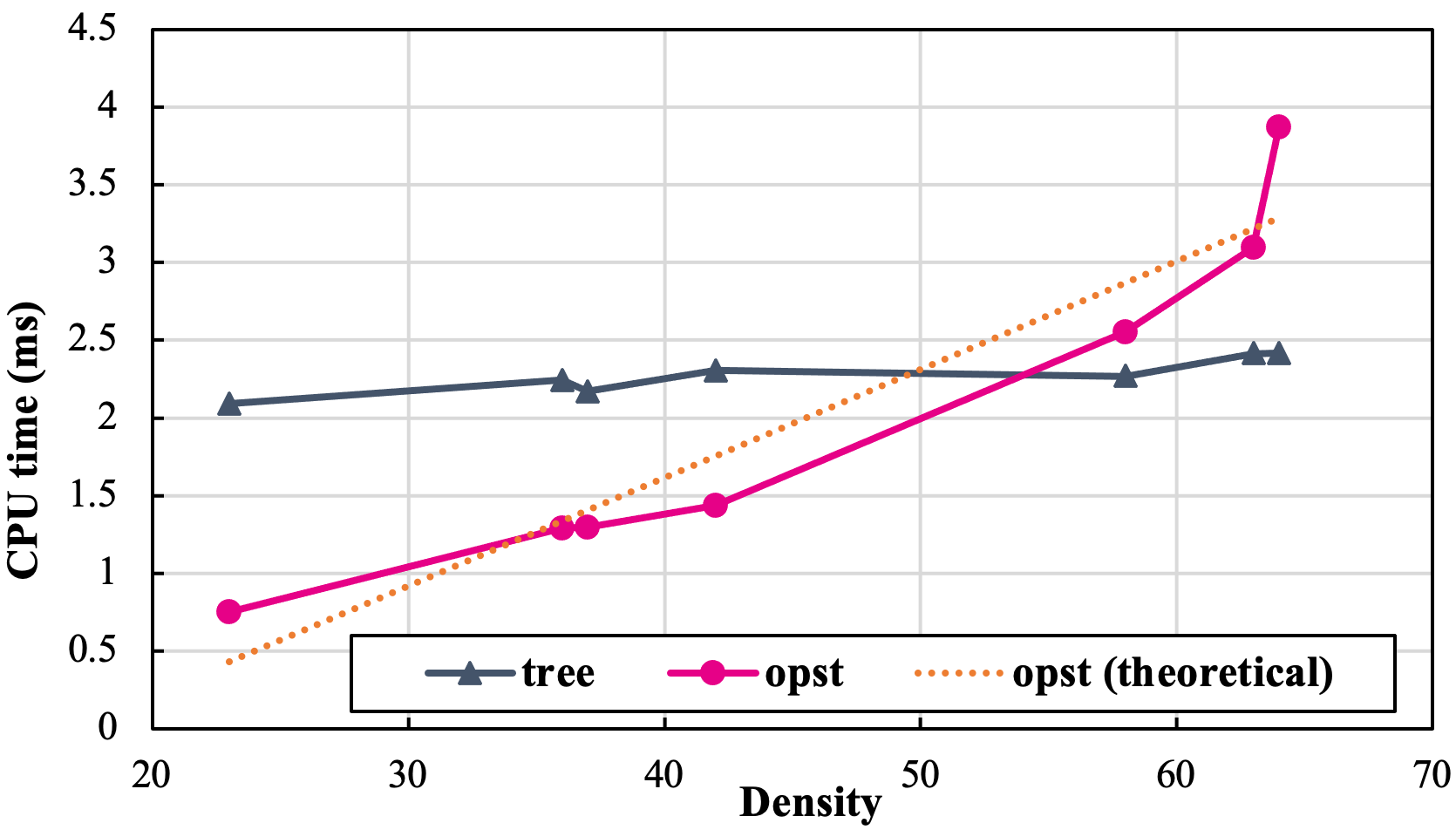}
    \vspace{-2mm}
    \caption{Time overhead comparison of OpST and AKDTree on different datasets with different densities.}
    \vspace{-2mm}
    \label{fig:time}
\end{figure} 

\begin{figure}[t]
     \centering
     \begin{subfigure}[t]{0.49\linewidth}
         \centering
         \includegraphics[width=\linewidth]{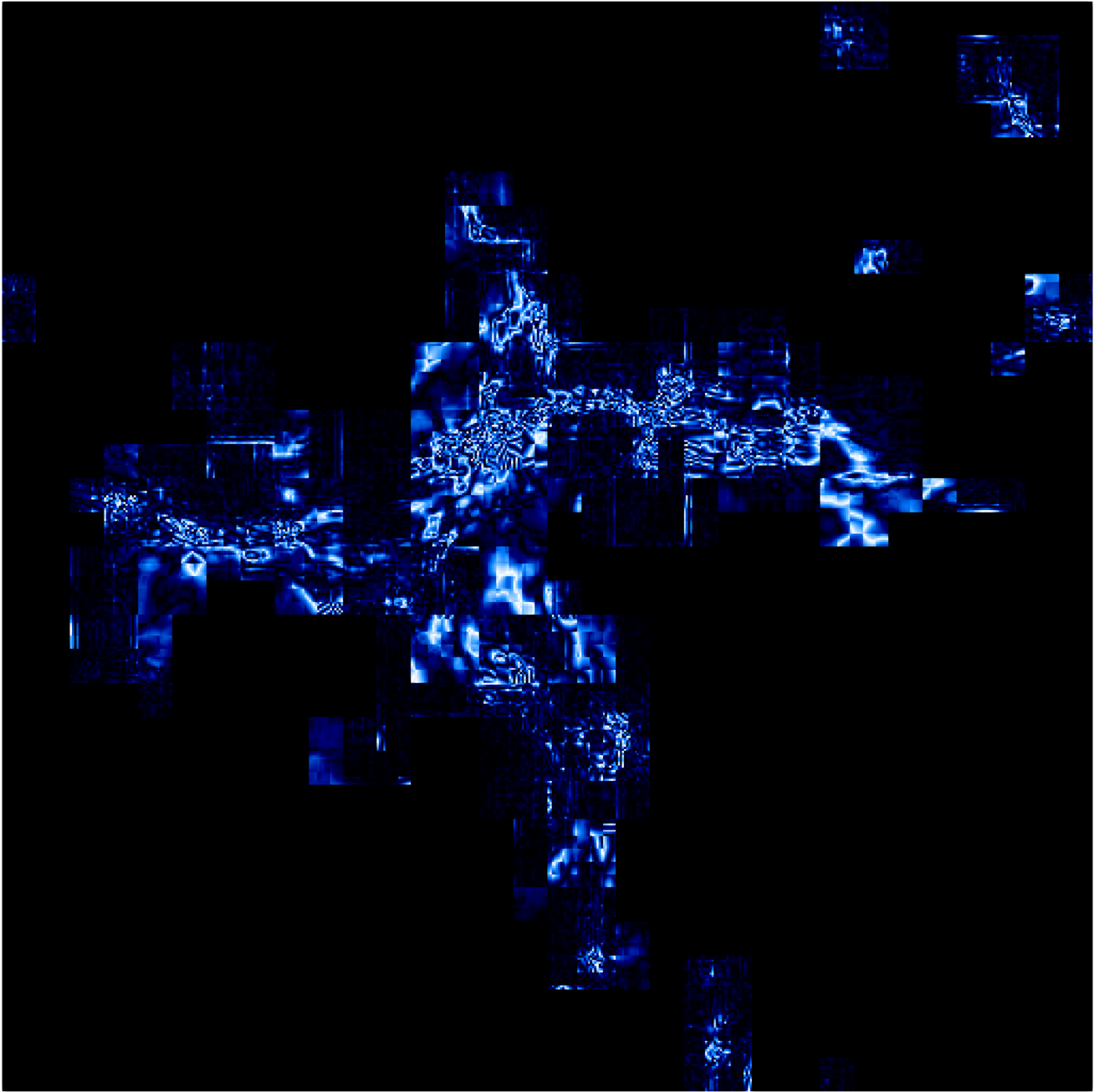}  
         \caption[t]{AKDtree\\(CR=222, PSNR=78.5dB)}
         \label{fig:kd-err}
     \end{subfigure}
     \begin{subfigure}[t]{0.49\linewidth}
         \centering
         \includegraphics[width=\linewidth]{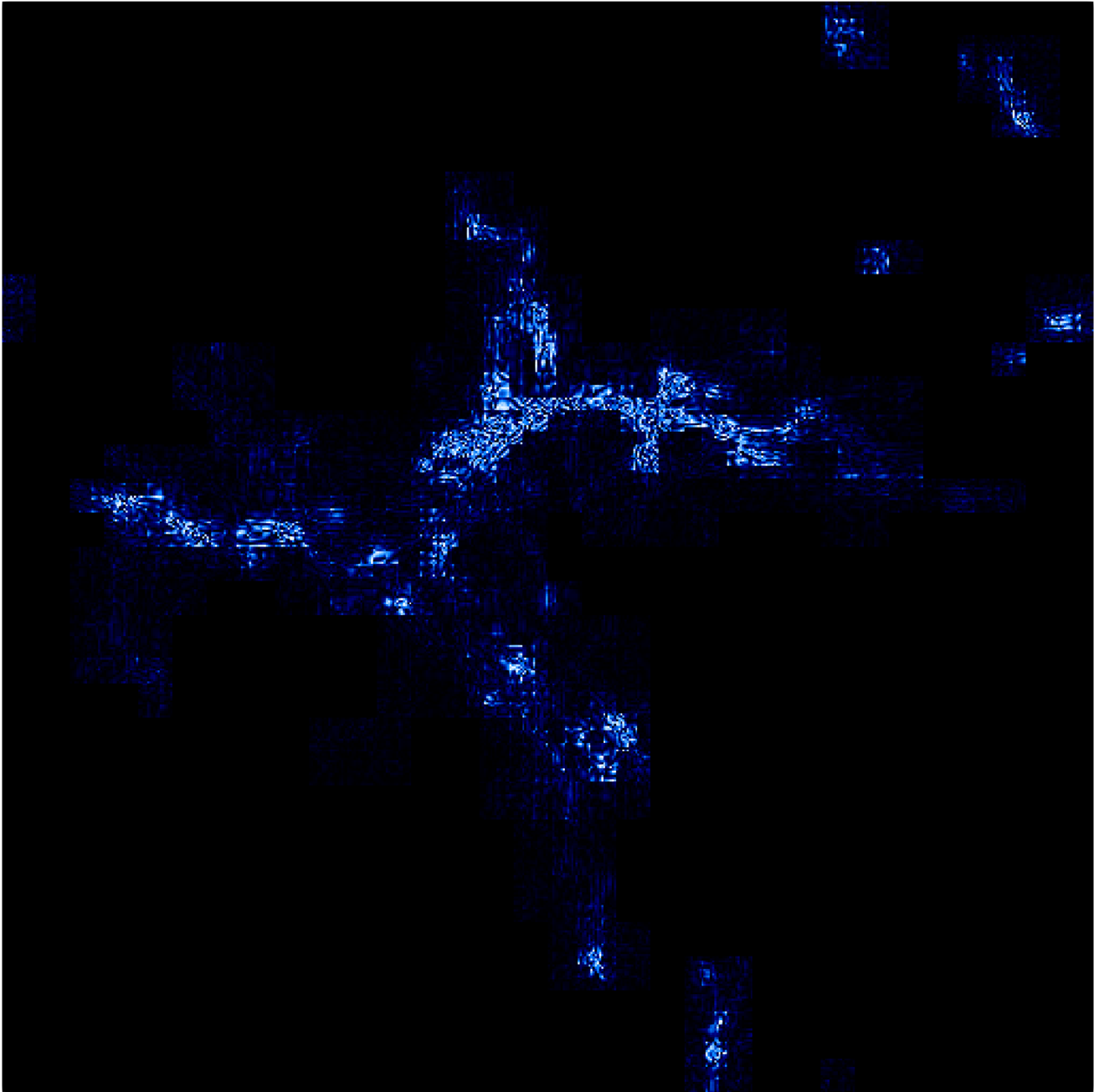}
         \caption{AKDtree + SHE\\(CR=231, PSNR=79.6dB)}
         \label{fig:sle-err}
     \end{subfigure}
     \vspace{-2mm}
        \caption[t]{Visual comparison (one slice) of compression errors of two approaches using SZ's Lor/Reg prediction on Nyx's ``baryon density'' field (i.e., z10's fine level, 23\% density). Brighter means higher compression error. The error bound is the relative error bound of $4.8\times10^{-4}$.}
        \vspace{-2mm}
        \label{fig:kd_sle}
\end{figure}

\begin{figure}[t]
    \centering \includegraphics[width=0.85\columnwidth]{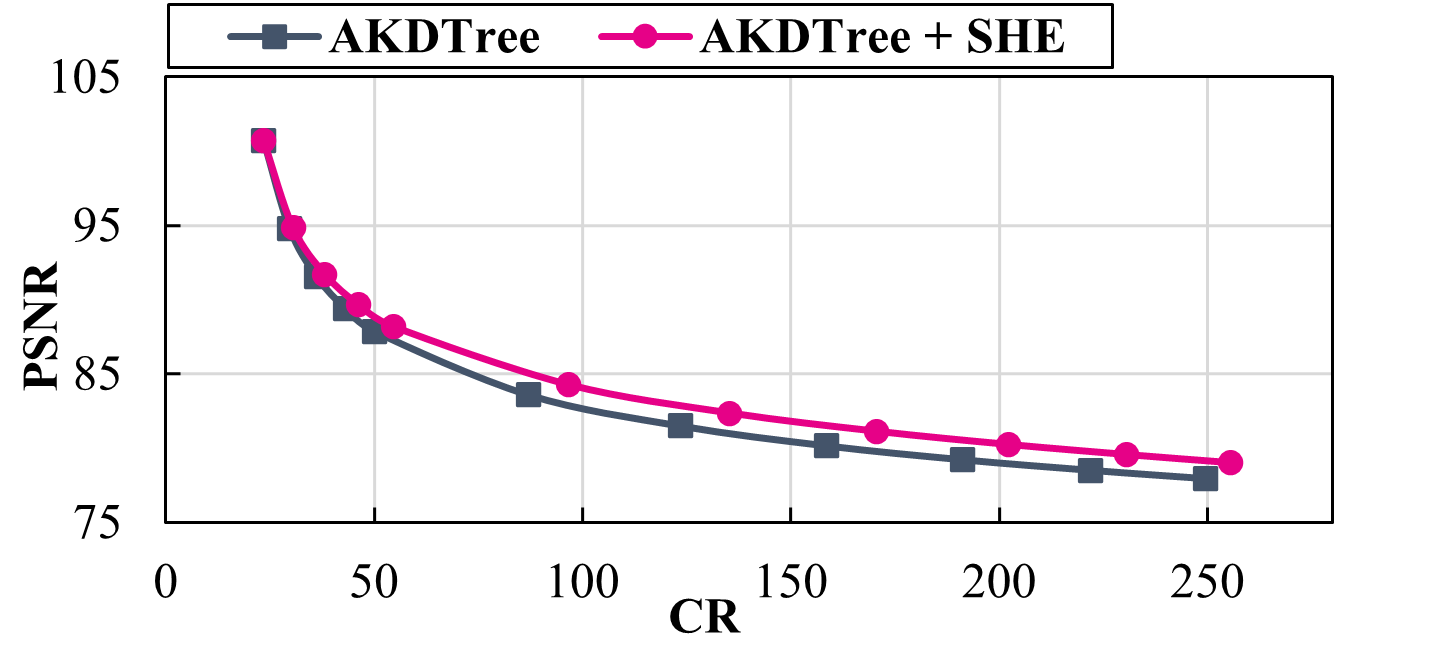}
    \vspace{-2mm}
    \caption{Comparison 
    of original AKDTree and AKDTree with SHE on Nyx's ``baryon density'' field (i.e., z10's fine level, 23\% density).}
    \vspace{-2mm}
    \label{fig:sle}
\end{figure}

To this end, for OpST and AKDtree, we propose a shared Huffman encoding technique to predict data blocks separately while encoding them together using a single shared Huffman tree. The detail is shown in Algorithm~\ref{alg:comsle}. Each data block is first predicted and quantized separately. Then, the quantization codes and regression coefficients of each data block are aggregated to build a shared Huffman tree and encoded at one time.
This approach can significantly improve the prediction performance of SZ's Lor/Reg prediction without introducing high time overhead to the encoding of SZ.
As shown in Figure~\ref{fig:kd_sle}, compared to the original AKDtree, AKDtree with SHE can significantly reduce the overall compression error, especially for the data located at the boundary of data blocks, leading to significant PSNR improvement, as shown in Figure~\ref{fig:sle}. 
Furthermore, the use of SHE reduces the number of Huffman trees needed for the data (since we do not need separate Huffman trees for different block shapes), thereby improving encoding efficiency and compression ratio.

\textcolor{black}{
Note that SZ's interpolation does not benefit from SHE. Unlike Lor/Reg that uses only neighboring data for predictions, Interp also considers global points from the dataset. However, with SHE, each small data block produced by OpST or AKDTree is predicted independently. This means Interp loses all global spatial information. In contrast, without SHE, the original OpST and AKDTree methods linearize the data blocks into a large 4D array, preserving more global spatial information for improved prediction. Thus, we exclusively apply the SHE approach to the Lor/Reg predictor.
}

\subsection{Hybrid Compression Strategy} \label{sec:hybrid}

In this section, we propose a solution to adaptively choose the best-fit 
compression strategy for both Lor/Reg (with SLE) and Interp algorithms.

For Lor/Reg (with SLE), we will choose from our proposed \textit{OpST with SHE (OpST+)}, \textit{AKDTree with SHE (AKDTree+)}, and \textit{GSP} based on the data characteristics (i.e., data density).
According to Section~\ref{sec:opst} and \ref{sec:kd}, OpST+ is more suitable for sparse (i.e., low-density) data, while AKDTree+ is designed to address the high time overhead of OpST+ when the density of data increases. 
Thus, there should be a data-density threshold to determine when to use OpST+ or AKDTree+.

To decide the threshold $T_0$ for switching between OpST+ and AKDTree+, we perform a series of experiments using Lor/Reg predictor, as shown in Figure~\ref{fig:tt}. 
The figure shows that OpST+ and AKDTree+ have almost identical compression performance in terms of bit-rate and PSNR on all four datasets/levels (from different timesteps) with different densities. 
Moreover, Figure~\ref{fig:time} shows the time costs of OpST+ and AKDTree+ (excluding compression). The figure demonstrates that the time of AKDTree+ is relatively stable, while the time of OpST+ increases linearly with the increase of data density. 
Overall, the only criterion for selecting OpST+ or AKDTree+ is the time cost rather than the compression performance. 
This is consistent with our previous design aim, that is, AKDTree is mainly designed to address the high time overhead issue of OpST+. 
Since OpST+ and AKDTree+ have a similar speed when the density is around 50\%, we use $T_0$ = 50 for choosing OpST+ or AKDTree+.

According to Section~\ref{sec:gsp}, GSP is designed to effectively handle the AMR level with high data density to prevent the negative impact of data partitioning without the use of SHE. 
In contrast, the data partition methods such as AKDTree require small blocks to be compressed together without SHE, which can significantly decrease prediction accuracy, as shown in Section~\ref{sec:sle}.
However, the negative impact on prediction accuracy caused by the partition can be eliminated by using SHE to compress each small block produced by the partition, which incurs little overhead, while GSP introduces significant overhead to the data size. 
As shown in Figure~\ref{fig:tt}, OpST+ and AKDTree+ outperform GSP across all the densities. 
As a result, the improved partition strategies using SHE can be a viable alternative to GSP for all levels.

In summary, for Lor/Reg (with SLE), our proposed hybrid compression approach is described as follows.
\begin{enumerate}[topsep=0pt,partopsep=1ex,parsep=0pt]
    \item When the density is smaller than $T_0 = 50\%$, we use OpST+ to remove empty regions and then compress;
    \item When the density is larger than $T_0 = 50\%$, we use AKDTree+ to remove empty regions and then compress.
\end{enumerate}

\textcolor{black}{For the Interp algorithm, we will select from our proposed methods: \textit{OpST}, \textit{AKDTree}, and \textit{GSP}, based on the data density. It is worth noting that we do not employ the SHE approach for the Interp algorithm due to its negligible improvement, as discussed at the end of Section~\ref{sec:sle}.}

\textcolor{black}{To determine the appropriate threshold $T_1$ for transitioning between OpST and AKDTree, we conducted a series of experiments, as depicted in Figure~\ref{fig:tt2}. These figures reveal that both OpST and AKDTree also deliver nearly equivalent compression performance on the Interp algorithm across all six AMR levels, each from different timesteps and with varied densities. Based on these findings, we recommend setting $T_1 = 50\%$ when selecting between OpST and AKDTree, consistent with our approach for Lor/Reg, aiming to enhance processing speed.}

\textcolor{black}{Next, to determine the threshold $T_2$ for switching between AKDTree and GSP, we also evaluate them on different datasets with different densities. 
As shown in Figure~\ref{fig:tt2}, when the density is relatively low, AKDTree outperforms GSP with respect to both bit-rate and PSNR; when the density gets higher and higher, GSP gradually outperforms AKDTree. 
We also observe that AKDTree and GSP have similar compression performance when the density is around 85\%. Thus, we use $T_2$ = 85\% to choose AKDTree or GSP.}

\textcolor{black}{Therefore, our proposed hybrid compression approach for the Interp algorithm is:
\begin{enumerate}[topsep=0pt,partopsep=1ex,parsep=0pt]
    \item When the density is smaller than $T_1$, we will use OpST to remove empty regions before compression;
    \item When the density is between $T_1$ and $T_2$, we will use AKDTree to remove empty regions before compression;
    \item When the density is larger $T_2$, we will use GSP to pad appropriate values before compression.
\end{enumerate}}

\textcolor{black}{It is evident that, compared to the Lor/Reg with SHE, GSP once again proves beneficial for Interp. The core reason is that, for Lor/Reg, we can use SHE to fully mitigate the impact of partitioning as just mentioned. Interp's global spatial information/locality inevitably gets disrupted by partitioning. GSP can thus be applied to the AMR level with high data density to counteract the effects of partitioning, while also minimizing data size overhead.}

\subsection{\textcolor{black}{Discussion on Generality and Distribution}} \label{sec:add-dis}
\textcolor{black}{\textit{\textbf{Generality of our design.}}
\textsc{TAC} and \textsc{TAC+} (including the pre-process strategies and SHE on SZ's Lor/Reg) are designed to handle AMR data level-wise where the data has an irregular distribution, such as many empty/zero regions. \textsc{TAC/TAC+} can also be employed for sparse non-AMR data compression. However, for sparse non-AMR data with very high density (e.g., 99.9\% of the data is non-zero), our \textsc{TAC/TAC+}, designed for irregular data, might not offer many advantages, as this data closely resembles ordinary non-sparse data without empty regions. On the other hand, \textsc{TAC/TAC+} is more apt for AMR data because AMR data has multi-resolution levels, and the density sum from all levels equals one. This means there is at most one super-dense level where the \textsc{TAC/TAC+} may be less effective.}

\textcolor{black}{Furthermore, the SHE approach can be utilized to enhance the compression ratio while maintaining prediction performance, especially when compressing numerous small-sized datasets collectively. Take, for instance, the scenario of compressing 10,000 2D images of $100\times100$. Compressing them individually would result in a reduced overall compression ratio (CR). Conversely, stacking them into a 3D block isn't optimal either, as it could compromise prediction accuracy due to the unsmoothness between images.}

\textcolor{black}{\textit{\textbf{Distributed scenario.}}
"In a distributed environment, parallelizing the GSP would be relatively straightforward because its padding approach is primarily local. The only minor challenge arises when padding values must be computed from other processes. This necessitates a boundary value exchange, similar to the Stencil problem.}

\textcolor{black}{However, OpST and AKDTree would face more significant challenges in parallelization due to their global nature. Merely using an embarrassingly parallel approach for these algorithms would yield smaller data blocks and lower spatial locality compared to the serial version. It's clear that communication is needed for these algorithms to be more effective, but balancing their ability to extract larger sub-blocks from the dataset with the communication cost would be a challenge. Addressing this challenge will be a primary goal in our future work.}

%% file: tex/04_evaluation.tex
\section{Experimental Evaluation}
\label{sec:evaluation}

\subsection{Experimental Setup}\label{sec:setup}

\textit{Test data.} Our evaluation primarily focuses on the AMReX framework~\cite{zhang2019amrex}, especially the Nyx cosmology simulation~\cite{nyx}, the WarpX electromagnetic and electrostatic Particle-In-Cell (PIC) simulation~\cite{warpx}, and the IAMR~\cite{IAMR}, which solves the incompressible Navier-Stokes equation.

\begin{figure}[t]
         \centering
         \vspace{-2mm}
         \includegraphics[width=0.55\columnwidth]{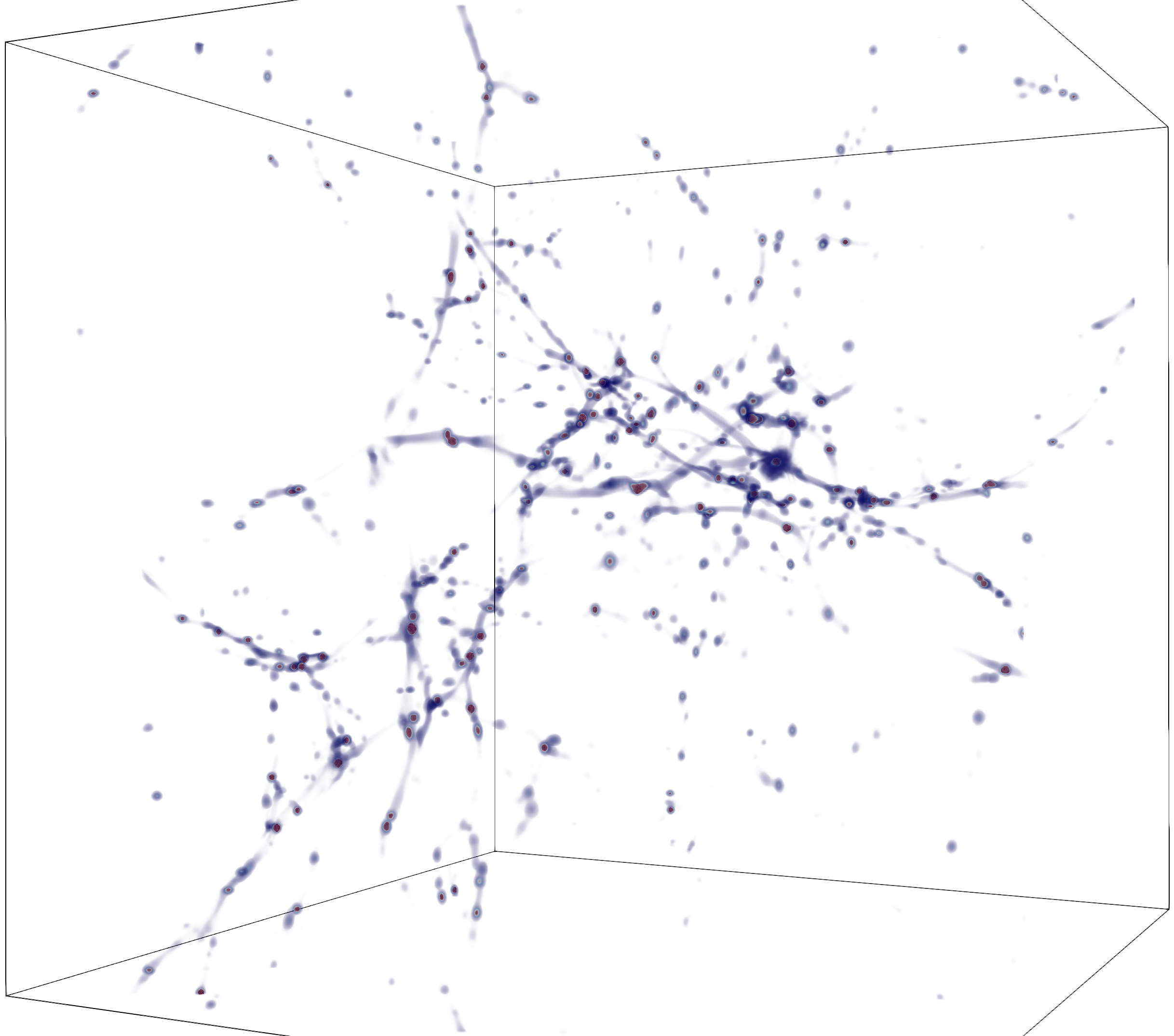}
        \caption[t]{\textcolor{black}{Visualization of Nyx's density field.}}
        \vspace{-4mm}
        \label{fig:nyx-example}
\end{figure}

Nyx, \textcolor{black}{as shown in Figure~\ref{fig:nyx-example}}, is a state-of-the-art, extreme-scale cosmology code that leverages AMReX. It produces six fields, which include baryon density, dark matter density, temperature, and velocities ($x$, $y$, and $z$). We analyzed six datasets produced by three real-world simulation runs, each having different numbers of AMR levels, and simulating a 64 megaparsec (Mpc) region. For these datasets, Z represents the redshift, i.e., the displacement of distant galaxies and celestial objects, as presented in Tab~\ref{tab:datasets1}.
Specifically:
the first run comprises two levels of refinement, with a coarse level of $256^3$ grids and a fine level of $512^3$ grids. We collected data from three timesteps, where the finest level density ranges from 23\% to 63\%.
The second run has up to four refinement levels. It began with a resolution of $128^3$ and refined to $1024^3$. From this run, we acquired data from two timesteps with the finest-level resolutions of $512^3$ (three levels) and $1024^3$ (four levels). The density at the finest level ranged from 0.2\% to 0.003\%.
The third run featured three refinement levels with grid sizes of $128^3$ at the coarsest level, $256^3$ at the intermediate level, and $512^3$ at the finest level. The density at the finest level is 0.90\%.

\begin{figure}[t]
         \centering
        \vspace{-2mm}
         \includegraphics[width=.9\columnwidth]{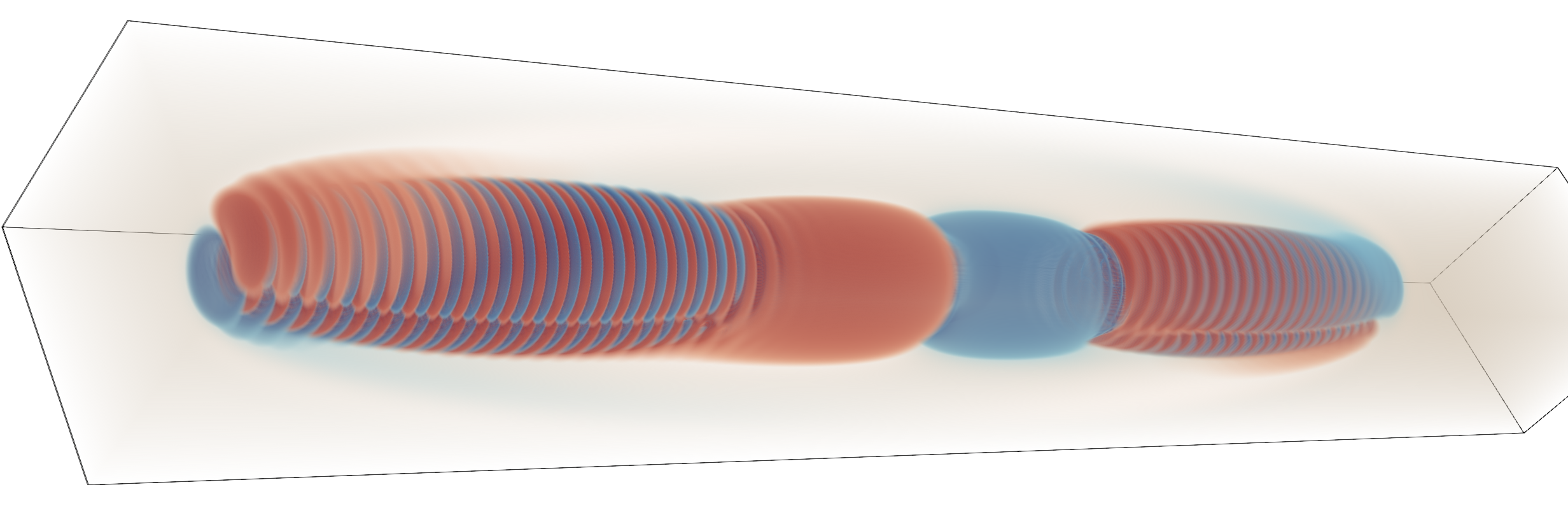}
        \caption[t]{\textcolor{black}{Visualization of WarpX's electric field (z-direction).}}
        \vspace{-2mm}
        \label{fig:wpx-example}
\end{figure}

\textcolor{black}{WarpX, as shown in Figure~\ref{fig:wpx-example}, is a highly parallel, optimized electromagnetic and electrostatic Particle-In-Cell (PIC) simulation that incorporates AMReX. It is designed to run on GPUs and multi-core CPUs and boasts load-balancing features. WarpX has the capability to scale up to the capacities of the world's most advanced supercomputers. Notably, it was awarded the 2022 ACM Gordon Bell Prize. We collected two timesteps from WarpX, as detailed in Tab~\ref{tab:datasets1}. Each timestep features two refinement levels, with dimensions of $128\times128\times1024$ and $256\times256\times2048$. The density at the finest level varies between 2\% and 8.6\%.}

\begin{figure}[b]
         \centering
         \vspace{-3mm}
         \includegraphics[width=0.5\columnwidth]{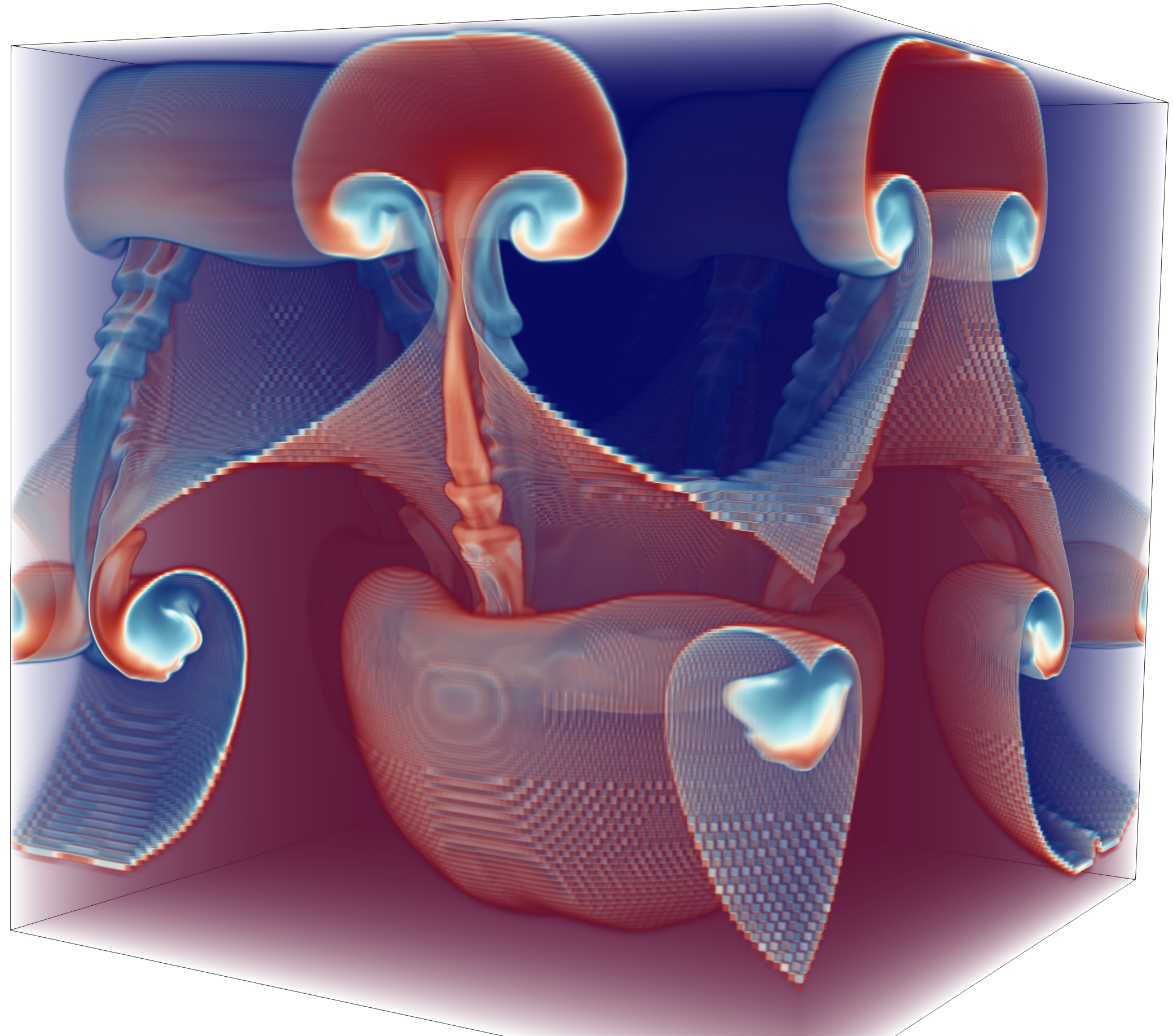}
        \caption[t]{\textcolor{black}{Visualization of IAMR's density field.}}
        \label{fig:iamr-example}
\end{figure}

\textcolor{black}{IAMR, as shown in Figure~\ref{fig:iamr-example}, is a highly parallel AMR code to solve the variable-density incompressible Navier-Stokes equations in either 2-D or 3-D. It also offers an embedded boundary (cut cell) representation for intricate geometries. We specifically executed the 3D Rayleigh-Taylor problem, which simulates a heavy fluid atop a light fluid under gravity's influence. We gathered data from two timesteps in IAMR. Each timestep comprises three refinement levels, beginning with dimensions of $128\times128\times128$ for the coarsest level and refining to $512\times512\times512$ for the finest level. The density at the finest level ranges from 0.6\% to 14.8\%.}

Note that the density of the finest level describes how much of the data in the dataset is at the highest resolution; a higher density of the finest level means that more data is refined to the highest resolution. Usually, the data density is gradually increasing at the finest level, within a single run.

\textit{Evaluation platform and compressor.}
The test platform is equipped with two 28-core Intel Xeon Gold 6238R processors and 384 GB of memory. \textcolor{black}{This work is based on two distinct SZ compression algorithms, namely Lor/Reg, which employs Lorenzo and linear regression predictors, and the Interp, which utilizes the spline interpolation approach as discussed in Section~\ref{sec:back-comps}. The Lor/Reg algorithm commences by partitioning the entire input data into 6$\times$6$\times$6 blocks, followed by the independent application of either Lorenzo predictor or high-dimensional linear regression on each individual block. Conversely, the Interp algorithm conducts interpolation across all three dimensions of the entire dataset.}

\textcolor{black}{A key distinction between Lor/Reg and Interp is that the former is block-based whereas the latter is global. Specifically, Lor/Reg algorithm divides data into blocks prior to compression, whereas the Interp algorithm applies global interpolation across the entire dataset.}

\begin{table}[t]
\caption{\textcolor{black}{Our tested AMR applications and datasets}.}
\vspace{-2mm}
\resizebox{\linewidth}{!}{
\input{Table/dataset}
 }
  \label{tab:dataset}
\end{table}

\textit{Comparison baselines.}
As outlined in Section~\ref{sec:background}, we compare against three baselines, either in 1D or 3D, and introduce two of our solutions.
Specifically, (1) the \textit{1D baseline (naive)}: each AMR level is compressed separately as a 1D array; (2) the \textit{1D baseline (zMesh)}~\cite{zMesh}: we refer readers to Section~\ref{sec:background} for more details about how the zMesh approach reorganize the AMR data for 1D compression; and (3) the \textit{3D baseline}: Different AMR levels are unified to the same resolution for 3D compression; (4) \textit{our \textsc{TAC}}: each AMR level is compressed using OpST (without SHE), AKDTree (without SHE), and GSP based on the data density. For more information, we refer readers to~\cite{tac} for more details. (5) \textit{our \textsc{TAC+}}: each AMR level is compressed using OpST+ or AKDTree+ based on the data density, with SHE using Lor/Reg algorithm. 

\textcolor{black}{As mentioned in Section~\ref{sec:sle}, \textsc{TAC+} is specifically paired with Lor/Reg algorithm. This is because TAC+ is developed to enhance the performance of Lor/Reg compression algorithm using SHE for AMR data. On the other hand, TAC serves as a pre-processing technique, making it compatible with both Lor/Reg and Interp algorithms.}

\begin{figure}[t]
     \centering
     \begin{subfigure}[t]{\columnwidth}
         \centering
         \includegraphics[width=0.82\columnwidth]{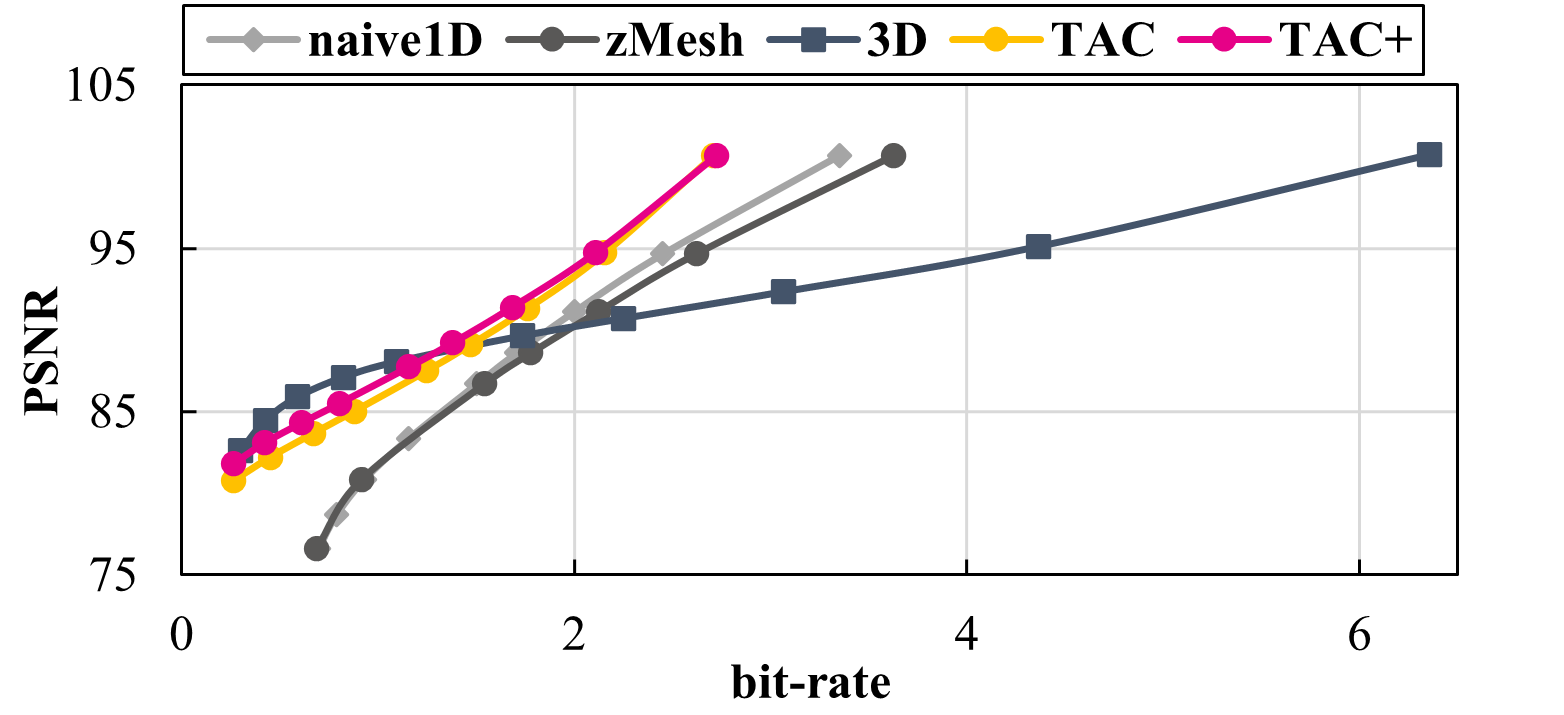} 
         \vspace{-2mm}
         \caption[t]{Run1\_Z10 (finest-level density = 23\%)}
         \label{fig:z10}
     \end{subfigure}
     \begin{subfigure}[t]{\columnwidth}
         \centering
         \includegraphics[width=0.82\columnwidth]{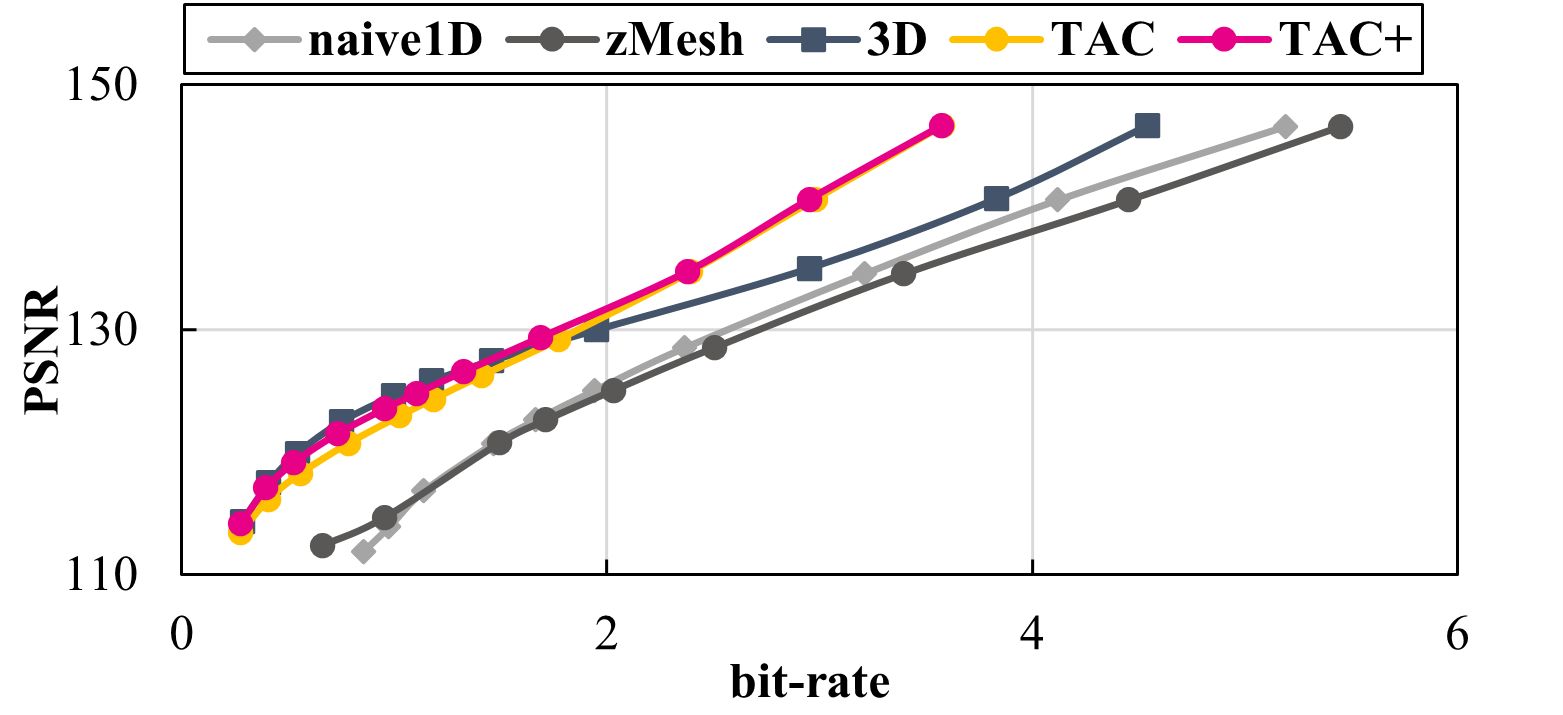}
         \vspace{-2mm}
         \caption{Run1\_Z5 (finest-level density = 58\%)}
         \label{fig:z5}
     \end{subfigure}

     \begin{subfigure}[t]{\columnwidth}
         \centering
         \includegraphics[width=0.82\columnwidth]{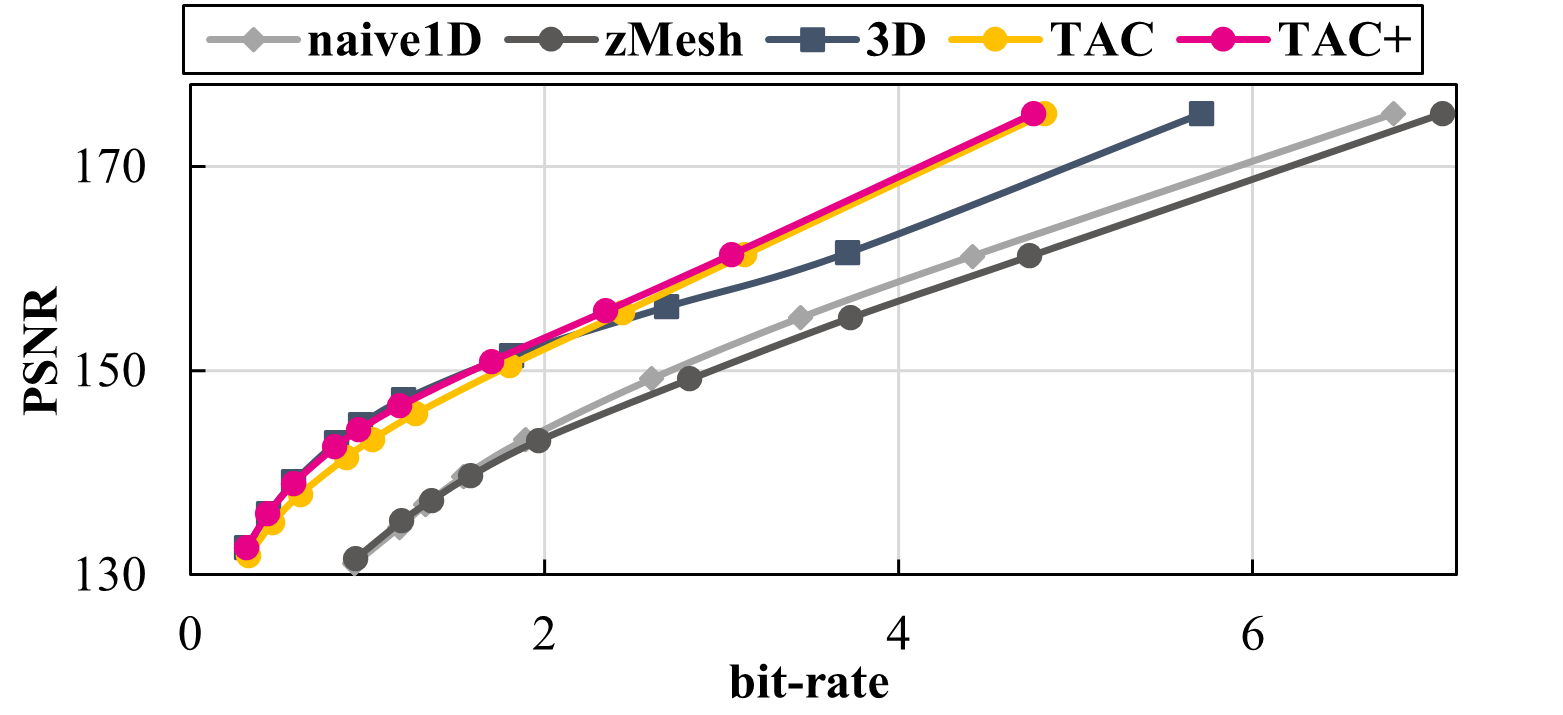}
         \vspace{-2mm}
         \caption{Run1\_Z2 (finest-level density = 63\%)}
         \label{fig:z2}
     \end{subfigure}
      \vspace{-3mm}
        \caption[t]{
        Rate-distortion comparison of our approaches (\textsc{TAC} and \textsc{TAC+}) and baselines on the early time-step (Z10) to the late time-step (Z2) from Nyx run1
        using Lor/Reg algorithm.}
        \vspace{-2mm}
        \label{fig:vsbs_1}
\end{figure}

\begin{figure}[t]
     \centering
     \begin{subfigure}[t]{\columnwidth}
         \centering
         \includegraphics[width=0.82\columnwidth]{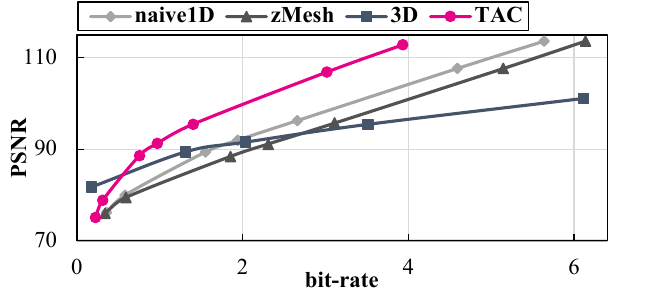} 
         \vspace{-2mm}
         \caption[t]{Run1\_Z10 (finest-level density = 23\%)}
         \label{fig:sz3-z10}
     \end{subfigure}
     \begin{subfigure}[t]{\columnwidth}
         \centering
         \includegraphics[width=0.82\columnwidth]{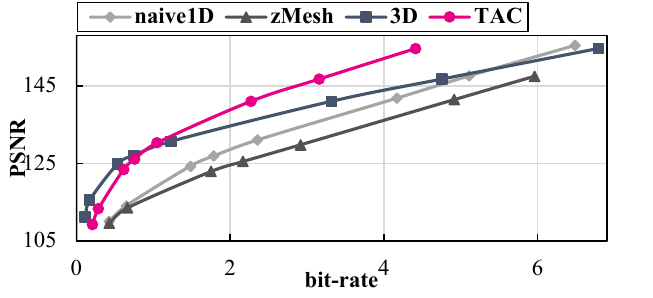}
         \vspace{-2mm}
         \caption{Run1\_Z5 (finest-level density = 58\%)}
         \label{fig:sz3-z5}
     \end{subfigure}

     \begin{subfigure}[t]{\columnwidth}
         \centering
         \includegraphics[width=0.82\columnwidth]{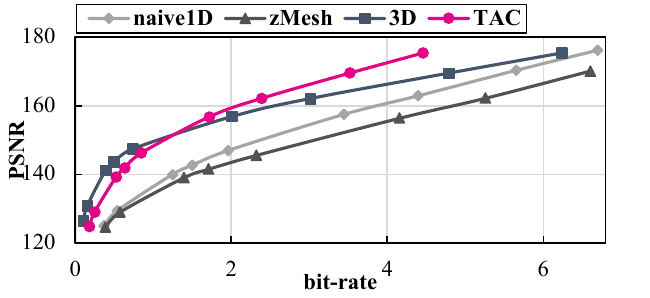}
         \vspace{-2mm}
         \caption{Run1\_Z2 (finest-level density = 63\%)}
         \label{fig:sz3-z2}
     \end{subfigure}
      \vspace{-3mm}
        \caption[t]{
        \textcolor{black}{Rate-distortion comparison of \textsc{TAC} and baselines on the Nyx run1 using Interp algorithm.}
        }
        \vspace{-2mm}
        \label{fig:nyx-sz3-1}
\end{figure}

\vspace{-2mm}
\subsection{Evaluation Metrics} \label{metric}
We will evaluate the compression performance based on the following metrics: 
(1) compression ratio or bit-rate (generic),
(2) distortion quality (generic),
(3) compression throughput (generic),
(4) rate-distortion (generic),
(5) power spectrum (cosmology specific),
(6) Halo finder (cosmology specific).

\textbf{Metric 1:} To evaluate the size reduction as a result of the compression, we use the compression ratio, defined as the original data size divided by the compressed data size, or bit-rate (bits/value), representing the amortized storage cost of each value. For single/double floating-point data, the bit-rate is 32/64 bits per value before compression. 
The compression ratio and bit-rate have a mathematical relationship as their multiplication is 32/64 so that a lower bit rate means a higher ratio.

\textbf{Metric 2:} Distortion is another important metric used to evaluate lossy compression quality. We use the peak signal-to-noise ratio (PSNR) to measure the distortion quality.
\begin{equation*}
\textstyle
\text{PSNR} = 20\cdot \log_{10}\left(R_X\right) - 10\cdot \log_{10}\left(\sum_{i=1}^{N} {e_i^2}/{N}\right),
\end{equation*}
where $e_i$ is the difference between the original and decompressed values for the point $i$, $N$ is the number of points, and $R_X$ is the value range of $X$. Higher PSNR means less error.

\textbf{Metric 3:} (De)compression throughputs are critical to improving the I/O performance. We calculate the throughput based on the original data size and (de)compression time. 

\textbf{Metric 4:} Similar to prior work~\cite{sz17,sz18,liang2018efficient,liang2021error,jin2020understanding,jin2021adaptive,zhao2020significantly}, we plot the rate-distortion curve to compare the distortion quality with the same bit-rate, for a fair comparison between different compression approaches, taking into account diverse compression algorithms.

\textbf{Metric 5:} 
Matter distribution in the Universe has evolved to form astrophysical structures on different physical scales, from planets to larger structures such as superclusters and galaxy filaments. The two-point correlation function $\xi(r)$, which gives the excess probability of finding a galaxy at a certain distance $r$ from another galaxy, statistically describes the amount of the Universe at each physical scale. The Fourier transform of $\xi(r)$ is called the matter power spectrum $P(k)$, where $k$ is the comoving wavenumber. 
The matter power spectrum describes how much structure exists at each physical scale. 
We run the power spectrum on the baryon density field by using a cosmology analysis tool called Gimlet. We compare the power spectrum $p'(k)$ of decompressed data with the original $p(k)$ and accept a maximum relative error within 1\% for all k < 10.

\textbf{Metric 6:} 
Halo finder aims to find the halos (over-densities) in the dark matter distribution and output the positions, the number of cells, and the mass for each halo it finds, respectively. Specifically, the halo-finder algorithm~\cite{Davis1985} searches for the halos from all the simulated data, with the following two criteria: (1) the mass of a data point must be greater than a threshold (e.g., 81.66$\times$ of the average mass of the whole dataset) to become a halo cell candidate~\cite{jin2020understanding, jin2021adaptive,ffis}, and (2) there must be enough halo cell candidates in a certain area to form a halo. For decompressed data, some of the information (mass and cells of halos) can be distorted from the original. 

\begin{figure}[t]
     \centering
     \begin{subfigure}[t]{\columnwidth}
         \centering
         \includegraphics[width=0.8\columnwidth]{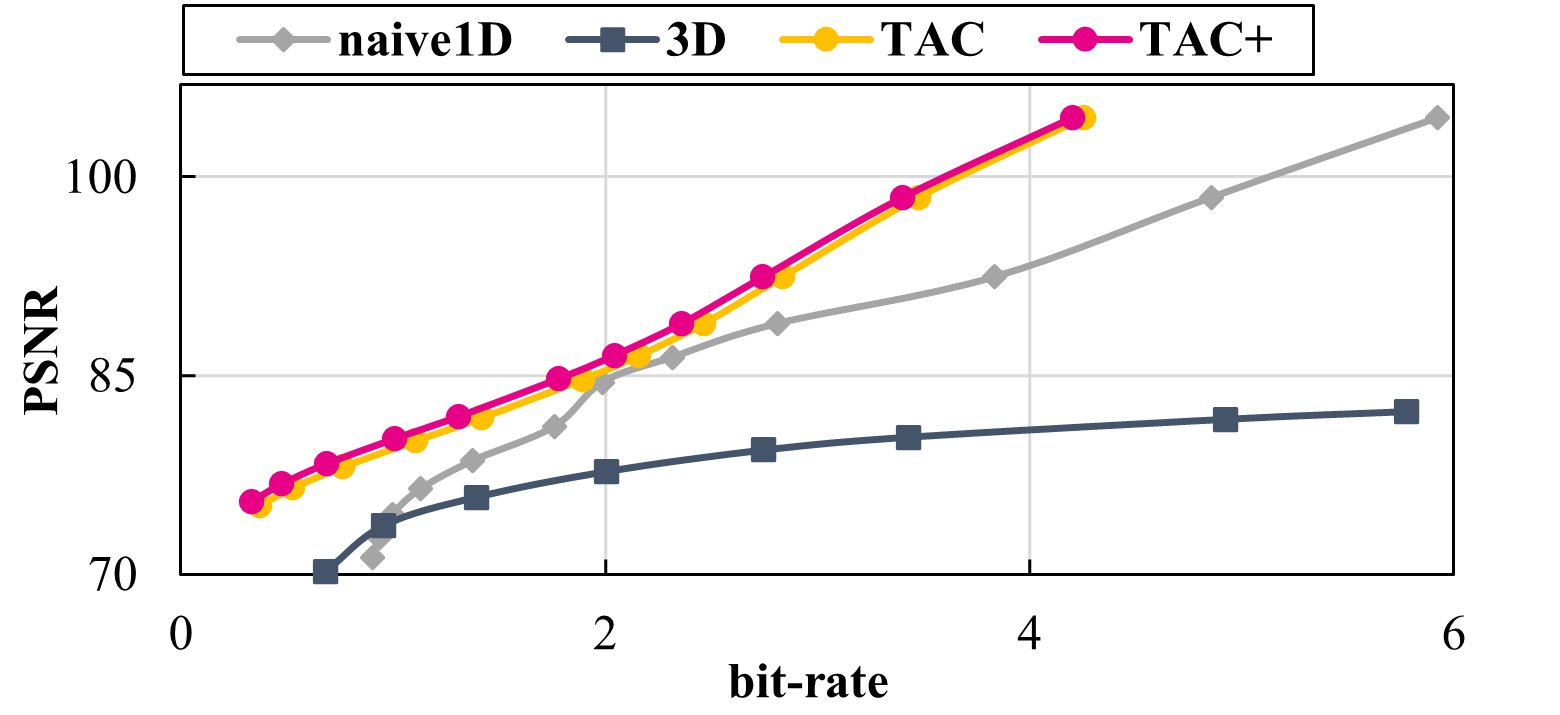}
         \vspace{-2mm}
         \caption{Run2\_T3 (finest-level density = 0.02\%)}
         \vspace{2mm}
         \label{fig:t3}
     \end{subfigure}
     
     \begin{subfigure}[t]{\columnwidth}
         \centering
         \includegraphics[width=0.8\columnwidth]{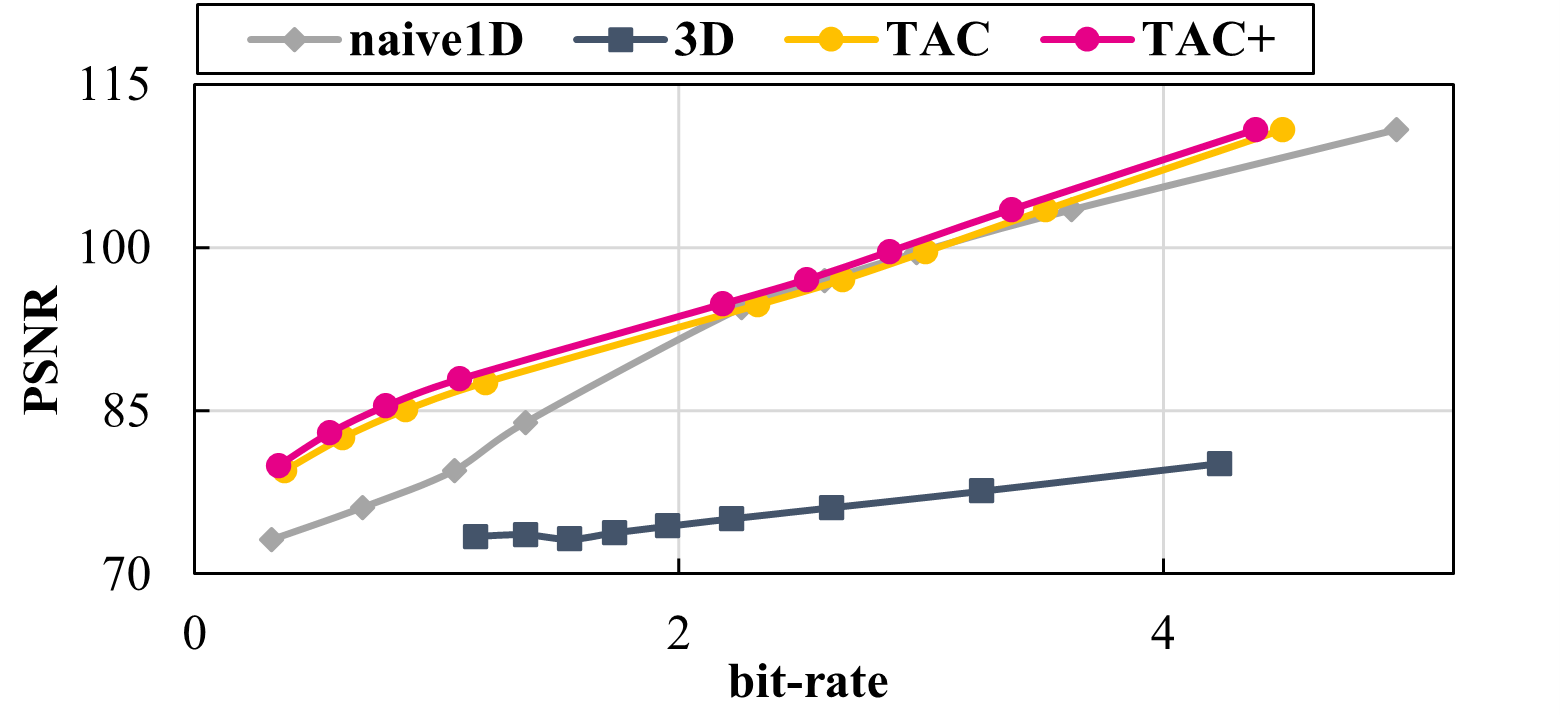}
         \vspace{-2mm}
         \caption{Run2\_T4 (finest-level density = 3E-5)}
         \label{fig:t4}
     \end{subfigure}
        \begin{subfigure}[t]{\columnwidth}
         \centering
         \includegraphics[width=0.8\columnwidth]{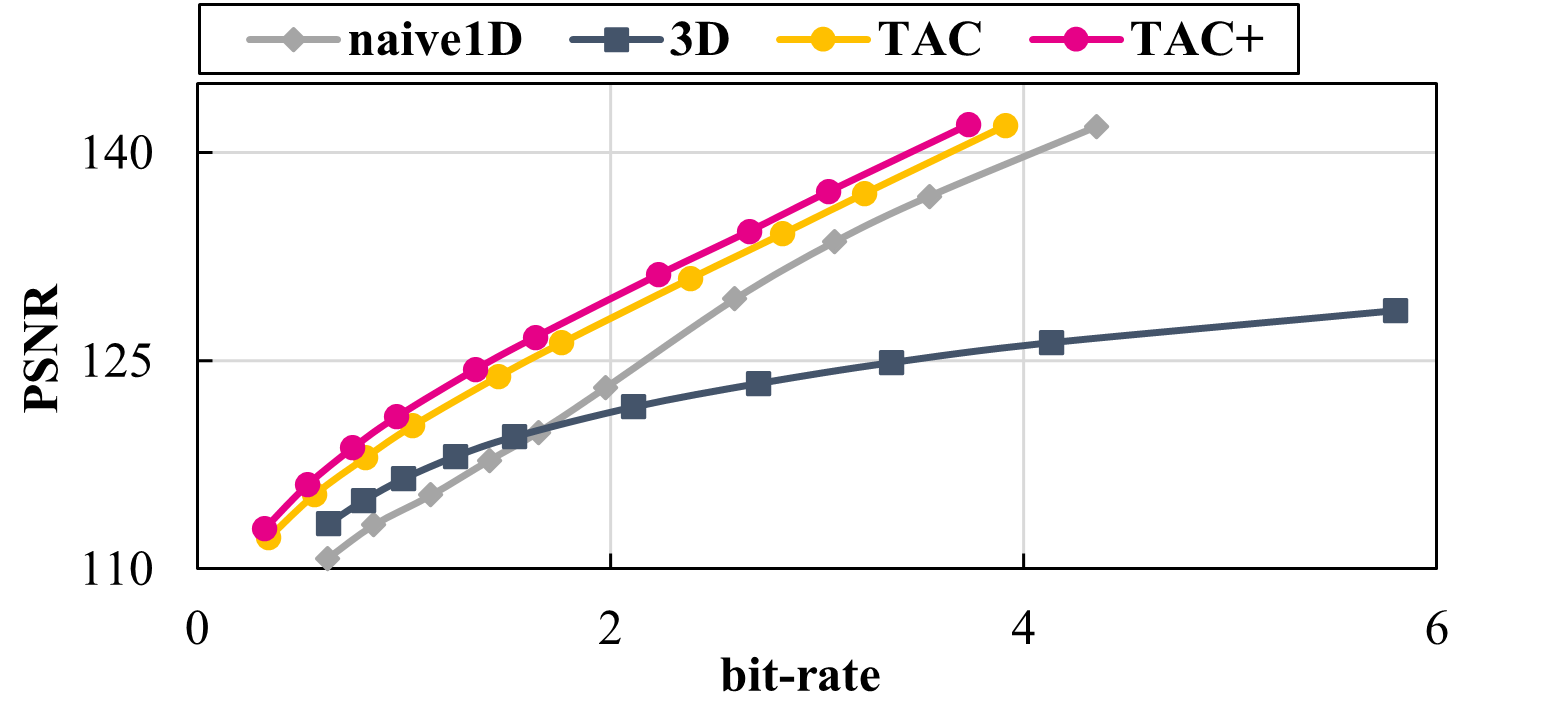}  
         \vspace{-2mm}
         \caption{Run3\_Z1.5 (density for each level: 0.87\%, 13.88\%, 85.25\% )}
         \label{fig:run3-1}
     \end{subfigure}
     \vspace{-3mm}
        \caption[t]{Rate-distortion comparison of our approaches and baselines on different time-steps from run2 and run3 using Lor/Reg algorithm.}
        \vspace{-2mm}
        \label{fig:vsbs_2}
\end{figure}

\begin{figure}[t]
     \centering
          \begin{subfigure}[t]{\columnwidth}
         \centering
         \includegraphics[width=0.82\columnwidth]{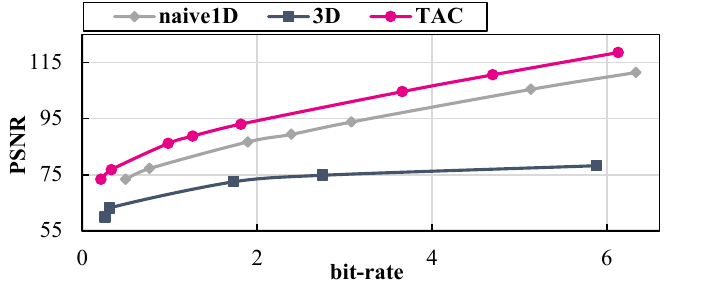}
         \vspace{-2mm}
         \caption{Run2\_T3 (finest-level density = 0.02\%)}
         \label{fig:t3-sz3}
     \end{subfigure}
     \begin{subfigure}[t]{\columnwidth}
         \centering
         \includegraphics[width=0.82\columnwidth]{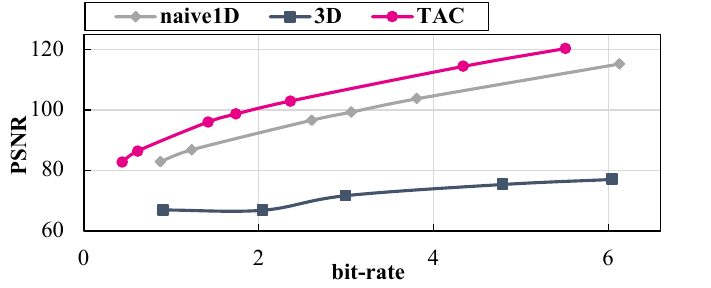}
         \vspace{-2mm}
         \caption{Run2\_T4 (finest-level density = 3E-5)}
         \label{fig:t4-sz3}
     \end{subfigure}
        \begin{subfigure}[t]{\columnwidth}
         \centering
         \includegraphics[width=0.82\columnwidth]{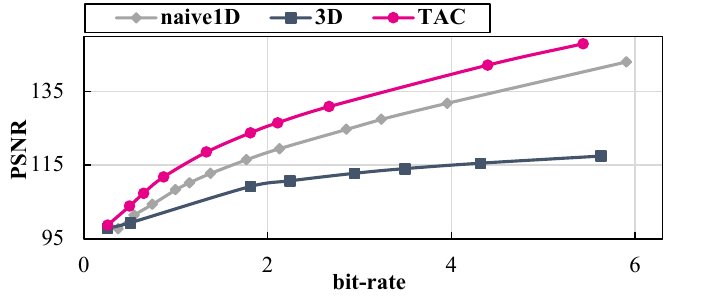}  
         \vspace{-2mm}
         \caption{Run3\_Z1.5 (density for each level: 0.87\%, 13.88\%, 85.25\% )}
         \label{fig:run3-1-sz3}
     \end{subfigure}
     \vspace{-3mm}
        \caption[t]{\textcolor{black}{Rate-distortion comparison of \textsc{TAC} and baselines on Nyx run2 and run3 using Interp algorithm.}
        }
        \vspace{-4mm}
        \label{fig:nyx-sz3-2}
\end{figure}

\begin{figure}[t]
     \centering   
     \begin{subfigure}[t]{\columnwidth}
         \centering
         \includegraphics[width=0.82\columnwidth]{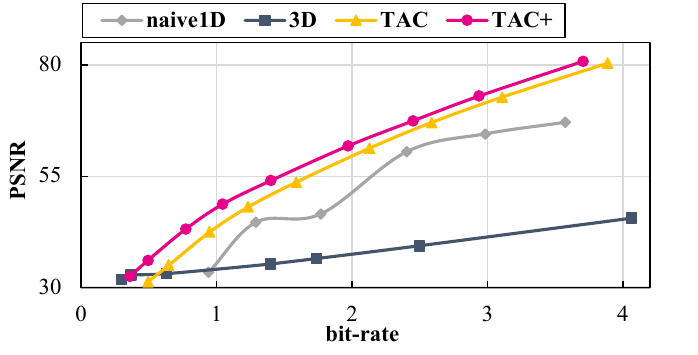}
         \vspace{-2mm}
         \caption{IAMR-90}
         \label{fig:rt90-sz2}
     \end{subfigure}
        \begin{subfigure}[t]{\columnwidth}
         \centering
         \includegraphics[width=0.82\columnwidth]{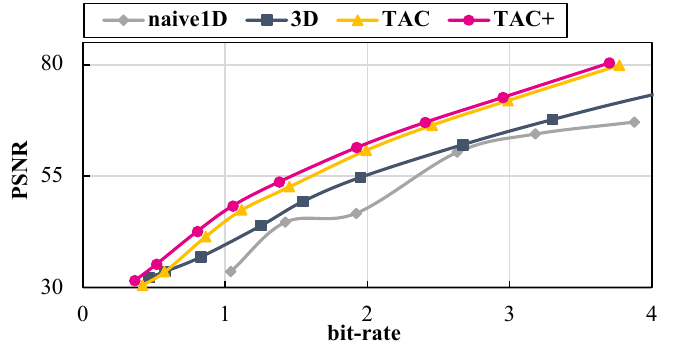}  
         \vspace{-2mm}
         \caption{IAMR-150}
         \label{fig:rt150-sz2}
     \end{subfigure}
     \vspace{-3mm}
        \caption[t]{\textcolor{black}{Rate-distortion comparison of our approaches and baselines on different time-steps from Rayleigh Taylor using Lor/Reg algorithm.}
        }
        \vspace{-2mm}
        \label{fig:rt-sz2}
\end{figure}

\begin{figure}[t]
     \centering   
     \begin{subfigure}[t]{\columnwidth}
         \centering
         \includegraphics[width=0.82\columnwidth]{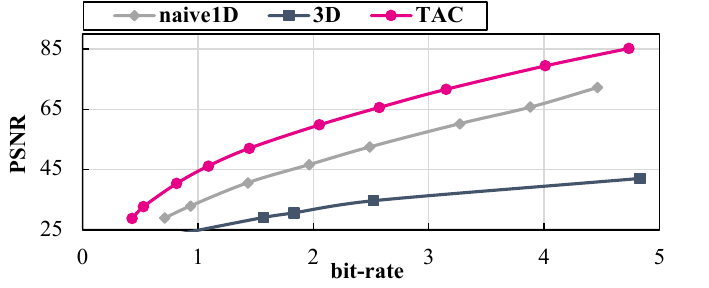}
         \vspace{-2mm}
         \caption{IAMR-90}
         \label{fig:rt90-sz3}
     \end{subfigure}
        \begin{subfigure}[t]{\columnwidth}
         \centering
         \includegraphics[width=0.82\columnwidth]{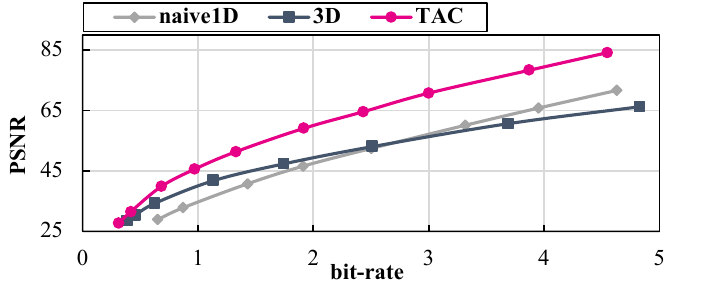}  
         \vspace{-2mm}
         \caption{IAMR-150}
         \label{fig:rt150-sz3}
     \end{subfigure}
     \vspace{-3mm}
        \caption[t]{\textcolor{black}{Rate-distortion comparison of \textsc{TAC} (top-left) and baselines on different time-steps from Rayleigh Taylor using Interp algorithm.}
        }
        \vspace{-2mm}
        \label{fig:rt-sz3}
\end{figure}

\subsection{Evaluation on Rate-distortion} 
\label{subsec:evaRateDis}
We first evaluate the rate-distortion of our proposed compression approaches and compare them with the baselines on different simulations and compression algorithms.

\begin{figure}[h]
     \centering   
     \begin{subfigure}[t]{\columnwidth}
         \centering
         \includegraphics[width=0.82\columnwidth]{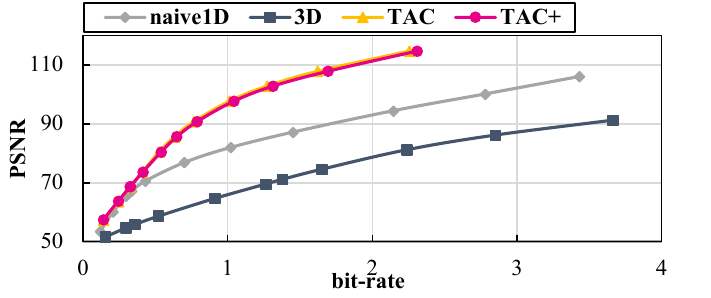}
         \vspace{-2mm}
         \caption{WarpX-800}
         \label{fig:wpx800-sz2}
     \end{subfigure}
        \begin{subfigure}[t]{\columnwidth}
         \centering
         \includegraphics[width=0.82\columnwidth]{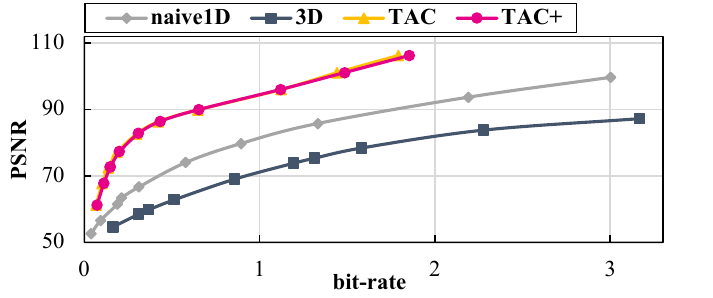}  
         \vspace{-2mm}
         \caption{WarpX-1600}
         \label{fig:wpx1600-sz2}
     \end{subfigure}
     \vspace{-3mm}
        \caption[t]{\textcolor{black}{Rate-distortion comparison of our approaches and baselines on different time-steps from WarpX using Lor/Reg algorithm.}
        }
        \vspace{-2mm}
        \label{fig:wpx-sz2}
\end{figure}

\textcolor{black}{For Lor/Reg algorithm, as shown in Figure~\ref{fig:vsbs_1},~\ref{fig:vsbs_2} and~\ref{fig:rt-sz2}, our new approach with SHE, \textsc{TAC+} (represented by the pink curve) yields better performance than the original \textsc{TAC} (the yellow curve) without SHE for all the datasets from Nyx and IAMR. 
However, in WarpX, \textsc{TAC+} does not surpass \textsc{TAC}, as depicted in Figure~\ref{fig:wpx-sz2}. This can be attributed to the simpler refinement structure of WarpX. Unlike Nyx and IAMR, which refine hundreds of regions (as shown in Figure~\ref{fig:dis_fine}), WarpX refines only one large rectangular area. Consequently, the unit block size in WarpX is four times larger than in Nyx and IAMR. After the \textsc{TAC+} partitioning process, this results in a substantially reduced number of unit blocks. This reduction negates the advantages of \textsc{TAC+}, which was specifically developed to handle numerous small AMR data blocks efficiently.}

\begin{figure}[h]
     \centering   
     \begin{subfigure}[t]{\columnwidth}
         \centering
         \includegraphics[width=0.82\columnwidth]{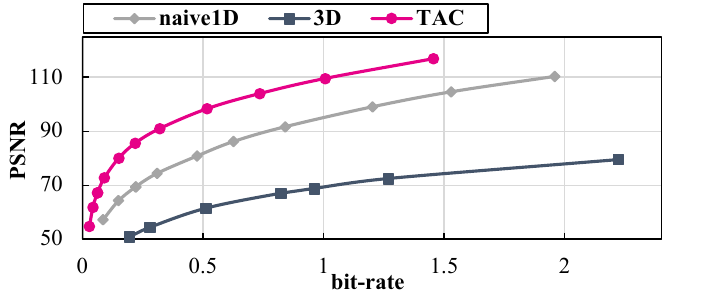}
         \vspace{-2mm}
         \caption{WarpX-800}
         \label{fig:wpx800-sz3}
     \end{subfigure}
        \begin{subfigure}[t]{\columnwidth}
         \centering
         \includegraphics[width=0.82\columnwidth]{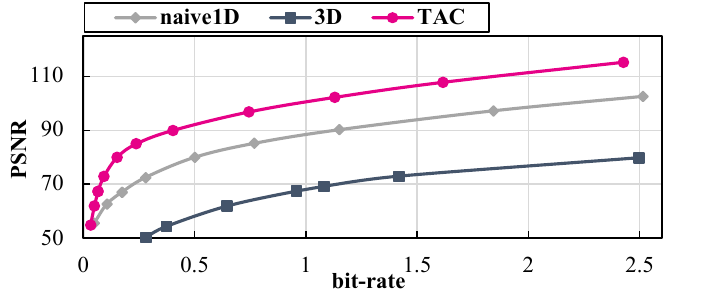}  
         \vspace{-2mm}
         \caption{WarpX-1600}
         \label{fig:wpx1600-sz3}
     \end{subfigure}
     \vspace{-3mm}
        \caption[t]{\textcolor{black}{Rate-distortion comparison of \textsc{TAC+} and baselines on different time-steps from WarpX using Interp algorithm.}
        }
        \vspace{-2mm}
        \label{fig:wpx-sz3}
\end{figure}

\textcolor{black}{Also, for the 1D baseline, our approaches including the \textsc{TAC+} and \textsc{TAC} outperform the 1D baseline across all the ten datasets from the three simulations and all two compression algorithms (i.e., Lor/Reg and Interp), as shown in 
Figure~\ref{fig:vsbs_1} to~\ref{fig:wpx-sz3}.}
Furthermore, the performance of our approach is more stable (i.e., smoother curve) than the 1D baseline.
We can also find that zMesh is slightly worse than the 1D baseline on our tested data as shown in Figure~\ref{fig:vsbs_1} and~\ref{fig:nyx-sz3-1}, which will be explained in the next section.

For the 3D baseline, we observe that our solutions has much better performance when the finest level has a relatively low density or the decompressed data has a high PSNR for all the datasets and compression algorithms, as shown in Figure~\ref{fig:vsbs_1} to~\ref{fig:wpx-sz3}. However, our approach cannot dominate the 3D baseline as shown in Figure~\ref{fig:vsbs_1} and~\ref{fig:nyx-sz3-1}, when the following criteria are satisfied: (1) the AMR data has only two levels of refinement, (2) the finest level has a relatively high density, and (3) the decompressed data has low PSNR/bitrate.

In the next section, we will discuss how the number of AMR levels, the density of the finest level, and the bit-rate/PSNR of decompressed data affect the performance of the 3D baseline and \textsc{TAC+} and \textsc{TAC} in detail.

\vspace{-4mm}
\subsection{Comparison with Baselines} \label{sec:dis}
\begin{figure}[t]
     \centering
     \begin{subfigure}[t]{0.37\linewidth}
         \centering
         \includegraphics[width=\linewidth]{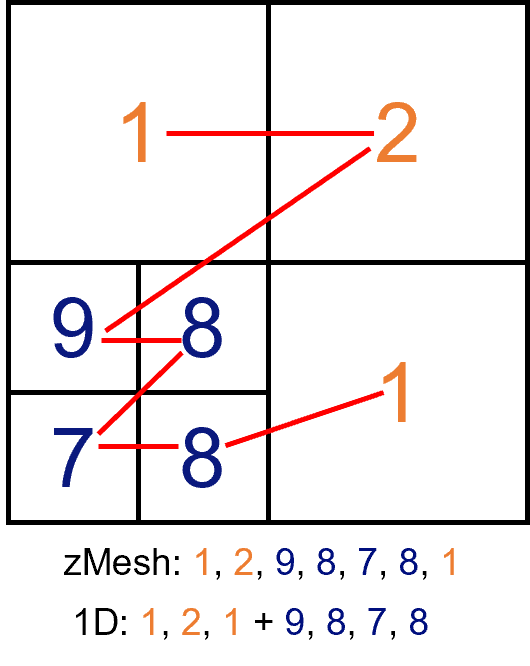} 
        \vspace{-6mm}
         \caption[t]{Tree-based AMR data}
         \label{fig:ourdata}
     \end{subfigure}
     \begin{subfigure}[t]{0.37\linewidth}
         \centering
         \includegraphics[width=\linewidth]{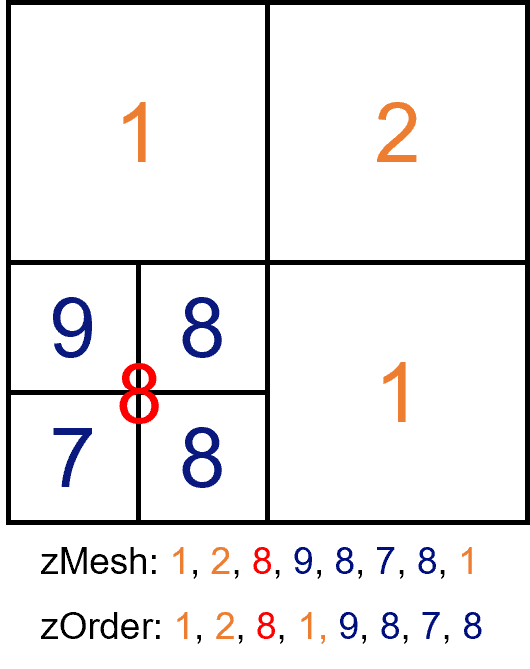}
        \vspace{-6mm}
         \caption{Patch-based AMR data}
         \label{fig:zdata}
     \end{subfigure}
         \vspace{-3mm}
        \caption[t]{An example of how the 1D baseline, zMesh, and original z-order reorder a simple 2D AMR data without and with redundancy. Orange: coarse level, blue: fine level, red: redundant data. 
        }
        \vspace{-2mm}
        \label{fig:zmesh}
\end{figure}

On compression, zMesh is meant to improve the smoothness of the patch-based AMR datasets by taking advantage of the data redundancy between each AMR level (as described in the introduction). 
Thus, zMesh cannot improve the smoothness if there is no data redundancy in the tree-structured AMR datasets (i.e., our tested datasets). 
A simple example is used to illustrate this in Figure \ref{fig:zdata}, where the finer-level data has higher values because a grid will be refined only if its value is larger than a certain threshold. For block-based AMR, when a grid needs to be refined because of its high value, the value will still remain in the level, resulting in a redundant value saved (i.e., the red 8). 
If one uses the original z-ordering to traverse the data level-by-level (shown in Figure \ref{fig:zdata}), the reordered data will have three significant value changes (i.e., from 2 to 8, from 8 to 1, and from 1 to 9). To solve this issue, zMesh traverses the two AMR levels together based on the layout of the 2D array. The reordered data are ``1-2-8-9-8-7-8-1'', which only has two significant value changes (from 2 to 8 and from 8 to 1). Thus, zMesh can improve the smoothness of patch-based AMR data. 

However, as shown in Figure \ref{fig:ourdata}, for tree-structured AMR data (without saving a redundant ``8''), compared to the 1D baseline that compresses each level separately, zMesh introduces two significant data changes (i.e., from 2 to 9 and from 8 to 1) as it traverses between two AMR levels. This explains why zMesh is slightly worse than the 1D baseline. 

When considering the 3D baseline, we observe that it works slightly better than \textsc{TAC+} in the following circumstances: (1) the AMR data has only two levels of refinement, (2) the finest level of the data has a relatively high density, and (3) the decompressed data has a low PSNR/bit-rate.

We now explain each observation point individually.
As outlined in Section~\ref{sec:baseline}, a primary drawback of the 3D baseline is the redundant data produced by the up-sampling process. This redundancy becomes even more pronounced for AMR data with more than two refinement levels, especially when the finest level of data is sparsely populated. Such extra data increases the data size, leading to sub-optimal compression performance for the 3D baseline (points 1 \& 2).

Conversely, the 3D baseline yields superior data locality and smoothness compared to \textsc{TAC+} and \textsc{TAC}. This is because it processes the entire dataset as a unified block, without segmenting it. Consequently, the 3D baseline exhibits enhanced compressibility, resulting in a quicker bitrate reduction and a slower reduction in PSNR as the error bound increases, compared to \textsc{TAC+} and \textsc{TAC}. As a result, the 3D baseline performs better at a low bit-rate (point 3).

\subsection{\textcolor{black}{\textsc{TAC/TAC+} for Different Compression Algorithms}} \label{sec:algovs}
\textcolor{black}{Note that the Lor/Reg compression algorithm can use the enhanced \textsc{TAC+}, resulting in improved compression performance. Conversely, with the Interp, we did not use \textsc{TAC+} because it cannot bring a performance advantage over \textsc{TAC} as mentioned in Section~\ref{sec:sle}.
The underlying reason for this difference lies in the nature of the AMR data and the Lor/Reg algorithm. Both AMR data and Lor/Reg are block-based, whereas Interp is global.}

\textcolor{black}{Specifically, a significant challenge in compressing high-dimensional AMR data is the need for partitioning, which can compromise spatial information and data smoothness. Given that Lor/Reg also segments data into blocks, it aligns naturally with the partitioning approach inherent to AMR data.
By leveraging our pre-process and optimization technique, Lor/Reg combined with \textsc{TAC+} using SLE can effectively address the issues of local data smoothness and smoothness caused by partitioning, making it especially compatible with AMR data.
In contrast, the Interp method implements a global interpolation approach on partitioned AMR data. Due to partitioning, small-sized blocks inevitably disrupt global spatial information for global interpolation. Although we can use the OpST and AKDTree to obtain larger partitioned blocks and improve spatial locality, we still cannot achieve perfect compatibility between interpolation prediction and AMR data.}


\subsection{Post-analysis Quality with Adaptive Error Bound}
\begin{table}[ht]
\caption{Halo finder analysis with different methods.}
\vspace{-2mm}
\centering
\resizebox{.95\linewidth}{!}{
\begin{tabular}{|c|c|c|c|}
\hline
              & \textbf{CR}    & \textbf{Avg Rel Mass Diff} & \textbf{Avg Rel Cells Diff} \\ \hline
3D baseline   & 188.7          & 2.7E-04               & 2.2E-03                   \\ \hline
\textsc{TAC+} (1:1) & 189.1         & 2.2E-04               & 2.1E-03                   \\ \hline
\textsc{TAC+} (2:1) & \textbf{192.5} & \textbf{1.1E-04}      & \textbf{9.0E-04}          \\ \hline
\end{tabular}
}
\label{tab:hf}
\vspace{-2mm}
\end{table}
When factoring level-wise compression, our approach can apply different error bounds to different AMR levels based on (1) the post-analysis metrics, (2) the up-sampling rates of coarse levels, and (3) the rate-distortion trade-off between different AMR levels.
We now evaluate our approach with the two cosmology-specific post-analysis metrics (i.e., power spectrum and halo finder) to demonstrate the benefit of the adaptive error bound method.
We chose the dataset run1-Z2 for evaluation because \textsc{TAC+} has a similar performance as the 3D baseline on this dataset.

\begin{table*}
\label{tab:nyxtp}
\centering
\caption{\textcolor{black}{Overall compression/decompression throughputs (MB/s) of different approaches with different absolute error bounds on Nyx.}}
\vspace{-2mm}
\resizebox{0.85\linewidth}{!}{
\textcolor{black}{\input{Table/nyxspeed}}
}
\vspace{-4mm}
\end{table*}

\begin{table}
\label{tab:rttp}
\centering
\caption{\textcolor{black}{Overall compression/decompression throughputs (MB/s) of different approaches with different absolute error bounds on IAMR.}}
\vspace{-2mm}
\resizebox{0.85\linewidth}{!}{
\textcolor{black}{\input{Table/rtspeed}}
}
\vspace{-2mm}
\end{table}

\begin{figure}[ht]
    \centering 
    \includegraphics[width=0.82\columnwidth]{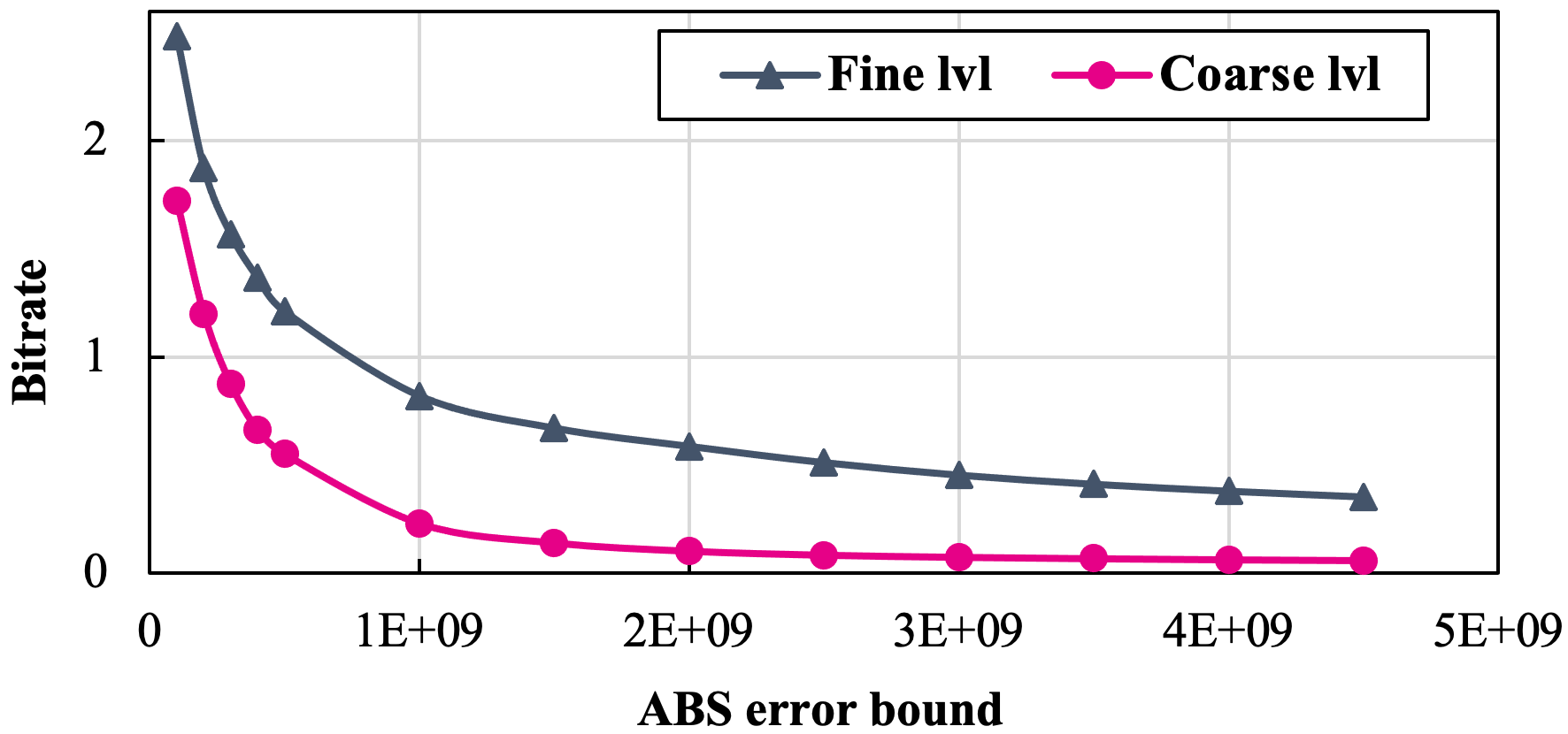}
    \vspace{-2mm}
    \caption{Bit-rates with different error bounds using SZ lossy compression for fine and coarse levels on Run1\_Z2 dataset.}
    \vspace{-4mm}
    \label{fig:db}
\end{figure}

Figure~\ref{fig:db} shows the motivation of performing rate-distortion trade-off between different AMR levels.
As the error bounds for the fine and coarse levels increase, their bit rates will converge to a similar value. This means that when the error bound is relatively large, the reduction in data size will be insignificant compared to the compression error increment (i.e., the slopes of both curves are small). Thus, we conclude that when the error is large, trading data quality for size reduction is not worth.

\begin{table}
\label{tab:wpxtp}
\centering
\caption{\textcolor{black}{Overall compression/decompression throughput (MB/s) of different approaches with different absolute error bounds on WarpX.}}
\vspace{-2mm}
\resizebox{0.85\linewidth}{!}{
\textcolor{black}{\input{Table/wpxspeed}}
}
\vspace{-2mm}
\end{table}

\begin{figure}[ht]
     \centering
     \begin{subfigure}[t]{\columnwidth}
         \centering
         \includegraphics[width=0.87\columnwidth]{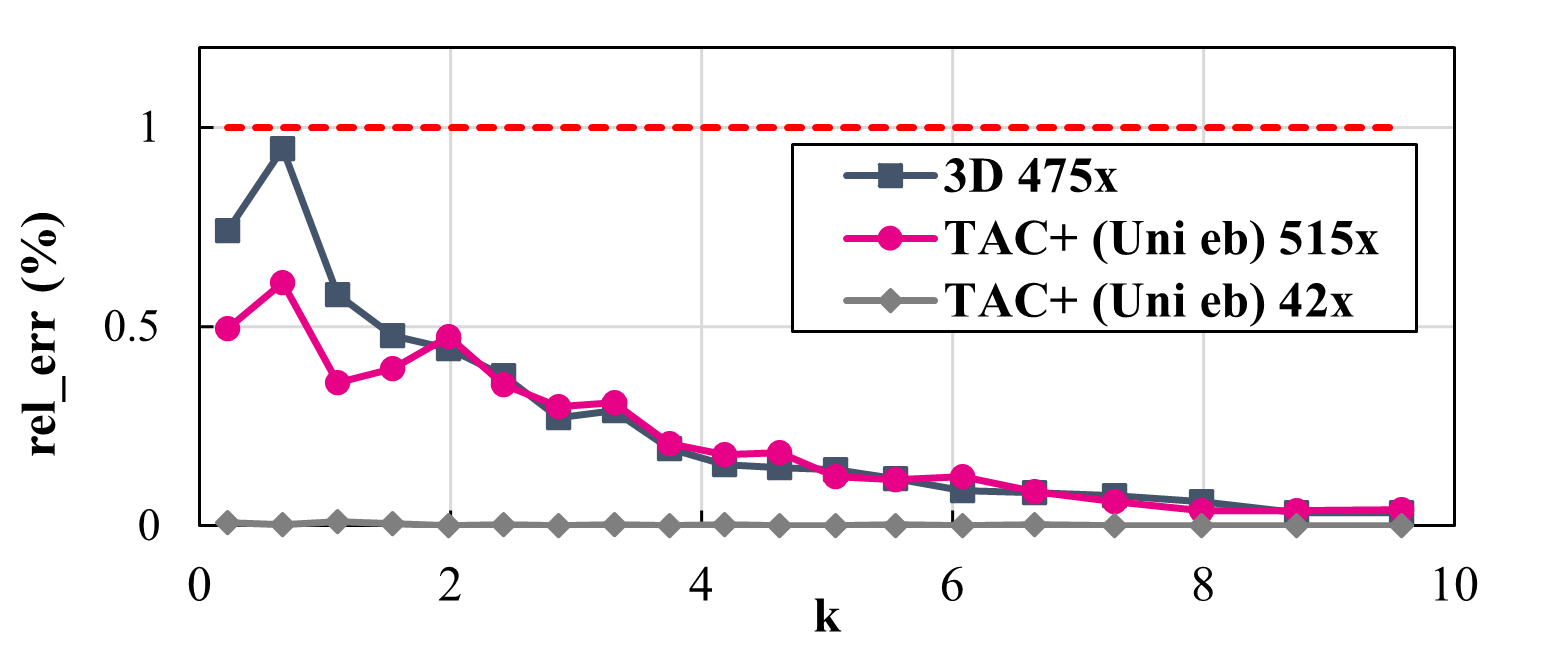}  
         \vspace{-3mm}
         \caption{Power spectrum error under 1\% limit }
         \label{fig:ps1}
     \end{subfigure}
     \begin{subfigure}[t]{\columnwidth}
         \centering
         \includegraphics[width=0.89\columnwidth]{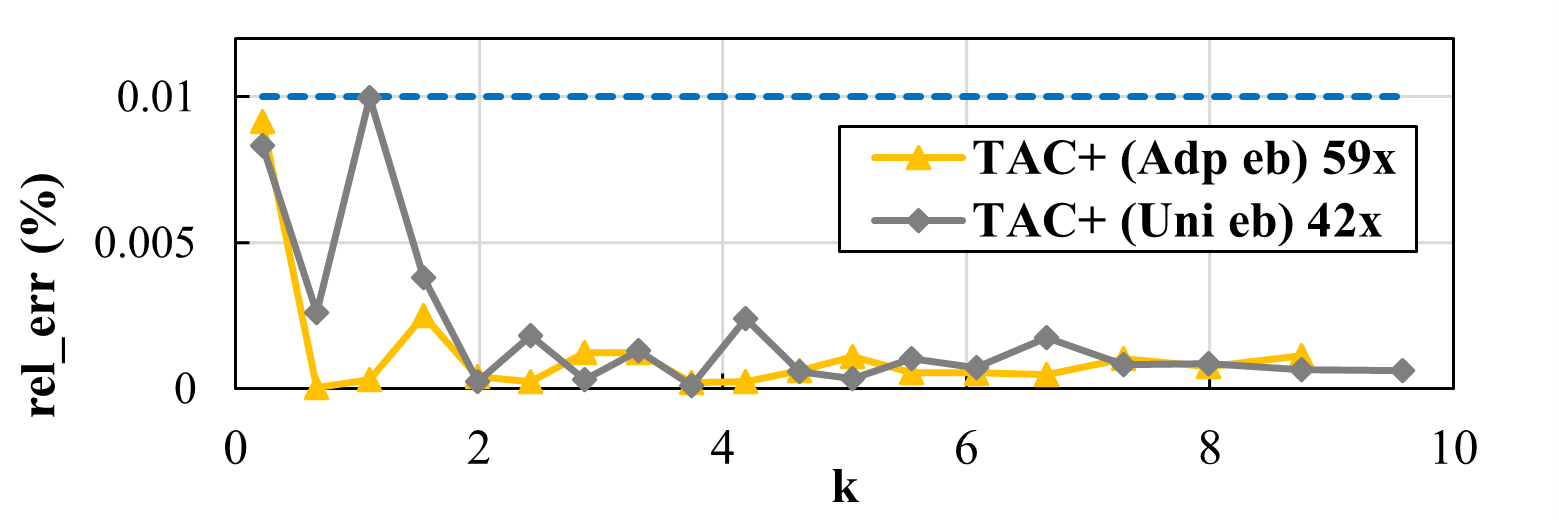}
         \vspace{-3mm}
         \caption{Power spectrum error under 0.01\% limit }
         \label{fig:ps2}
     \end{subfigure}
      \vspace{-4mm}
        \caption{Power spectrum error (in relative) of the 3D baseline and \textsc{TAC+} (uniform error bound) and \textsc{TAC+} (adaptive error bound) on baryon density field on run1-Z2. The red and blue dashed line is the 1\% and 0.01\% limit of acceptable power spectrum error.}
        \vspace{-2mm}
        \label{fig:psall}
\end{figure}

\textbf{Power Spectrum} Figure~\ref{fig:ps1} shows that under the (almost) same compression ratio, \textsc{TAC+} (with the uniform error bound) has a better power-spectrum error compared to the 3D baseline.
Also note that \textsc{TAC+} yields nearly lossless power-spectrum distortion (less than 0.01\%) under the compression ratio of 42$\times$, as shown in Figure~\ref{fig:ps1}'s gray curve.

Now, let us follow the three steps mentioned at the beginning of this section to adjust the error bound for each AMR level.
First, the post-analysis metric--power spectrum---needs to be run on the uniform-resolution data and focuses on the global quality of data. Thus, the ideal error-bound configuration/ratio for the fine and coarse levels on the uniform-resolution data would be 1:1.

as mentioned before, the coarse level of the AMR dataset needs to be up-sampled to uniform the resolution. As a result, the compression error of the coarse level will be up-sampled as well, resulting in more error in the post-analysis. 
Thus, we then need to give the coarse level a smaller error bound based on the up-sample rate. Here the up-sample rate for Z2's coarse level is $2^3$, leading to an ideal error-bound ratio of the fine and coarse levels changed to 8:1.

Finally, this 8:1 ratio needs to be adjusted based on the rate-distortion trade-off as mentioned before.
As shown in Figure~\ref{fig:db}, when using the error-bound ratio of 8:1 (e.g., 4E+9 for the fine level and 5E+8 for the coarse level), the error bound of the fine level is too large, resulting in an ineffective rate-distortion trade-off.
Thus, we can balance two levels by increasing the error bound for the coarse level (to gain compression ratio) and decreasing the error bound for the fine level (to add compression error), which can achieve an overall rate-distortion benefit. Based on our experiments, we adjust the error-bound ratio from 8:1 to 3:1,
As shown in Figure~\ref{fig:ps2}, \textsc{TAC+} with adaptive error bound can significantly improve the compression ratio compared with uniform error bound under similar power spectrum error.

\textbf{Halo finer} 
We evaluate the mass change, and the number of cells change for the three largest halos identified using the 3D baseline, \textsc{TAC+} (with uniform error bound), and \textsc{TAC+} (with adaptive error bound), as shown in Table~\ref{tab:hf}.
We can see that \textsc{TAC+} with uniform error bound produces better halo-finer analysis quality than the 3D baseline.

Similar to the error-bound configuration analysis done for the power spectrum, let us now adjust the error-bound ratio between the fine and coarse levels for halo finder.
The halo-finer analysis also requires uniform-resolution data as input. However, different from the power-spectrum analysis, the halo-finder analysis focuses more on high-value points at the fine level, since only high-value data points qualify as halo candidates, as described in Section~\ref{metric}. Note that this does not mean we can directly discard the coarse-level data with small values as they still contribute to the average value of the dataset, which is also an important parameter for the halo finder~\cite{Davis1985}. 
Therefore, we set the ideal error-bound ratio to 1:2 (i.e., fine level vs coarse level) for the uniform-resolution data based on our massive experiments.
After that, considering the up-sampling rate of $2^3$, the error-bounded ratio is changed to 4:1.  
Finally, we adjust the ratio to 2:1 based on the rate-distortion trade-off. Table~\ref{tab:hf} shows that \textsc{TAC+} with adaptive error bound obtains the minimal differences of the mass and cell numbers. 

\subsection{Evaluation on Time Overhead}
We evaluate the overall throughput (including pre-process and (de)compression) on the datasets with different error bounds. 
\textcolor{black}{As shown in Table~\ref{tab:nyxtp},~\ref{tab:rttp}, and~\ref{tab:wpxtp}, compared to the 3D baseline, the throughput of \textsc{TAC+} is up to $2\times$ higher on the Nyx Run1 datasets, $75\times$ higher on the Nyx Run2 dataset, $11.8\times$ higher on the Nyx Run1 dataset, $10\times$ higher on IAMR dataset, and $14\times$ higher on WarpX dataset.
This is because the WarpX, IAMR, Nyx Run2, and Nyx Run3 datasets have a lower density than the Run1 datasets at the finest level, resulting in a higher overhead of redundant data for the 3D baseline, which is consistent with our discussion in Section \ref{sec:dis}.}
We note that \textsc{TAC+} and performs better than the 1D baseline on all the tested datasets except the Nyx Run2\_T4 datasets, due to its long data-partition time (relative to the total time) on the small-sized datasets. \textcolor{black}{Also, in the three tables, the symbol ``\textbackslash{}'' denotes that \textsc{TAC+} is specifically paired with the Lor/Reg algorithm as mentioned in Section~\ref{sec:setup}.}

\textcolor{black}{The overall higher throughput of both \textsc{TAC+} and \textsc{TAC} can be attributed to the high efficiency of the three pre-process strategies.
Firstly, the GSP pre-process operates in linear time, as only a subset of the boundary data needs to be processed once to calculate the padding value. Additionally, padding is restricted to specific empty regions and is executed just once. Next, the time complexities for OpSt(+) and AKDTree(+) are $O(N^2 \cdot d)$ and $O(\frac{1}{3} N \cdot \log N)$ respectively, where $N$ represent the unit block number and $d$ indicates the density. Given that each unit block comprises a substantial number of data points, both OpSt(+) and AKDTree(+) exhibit high efficiency. Moreover, by employing our hybrid compression strategy detailed in Section~\ref{sec:hybrid}, we can further enhance the performance of \textsc{TAC(+)} by opting for the faster algorithm between OpSt(+) and AKDTree(+).
}

In addition, \textsc{TAC+} has almost the same throughput as \textsc{TAC}. The very slight decrease is because \textsc{TAC+} compresses the data in a more fine-grained manner.
We exclude zMesh in this evaluation as it is theoretically slower than the 1D baseline due to the extra z-ordering and provides worse rate-distortion according to our evaluation.

%% file: Table/dataset.tex
\newcommand\alignmiddle[2]{
\makebox[3em][r]{$(#1$}\makebox[.8em]{$,\ $}\makebox[2em][l]{$#2)$}}

\begin{tabular}{|l|c|c|c|c|}
\hline
\textbf{Dataset} &
  \textbf{\# Levels} &
  \textbf{\begin{tabular}[c]{@{}c@{}}Grid Size of Each Level\\  (Fine to Coarse)\end{tabular}} &
  \textbf{\begin{tabular}[c]{@{}c@{}}Density of Each Level\\ (Fine to Coarse)\end{tabular}}  & 
  \textbf{\begin{tabular}[c]{@{}c@{}}\textcolor{black}{Data Size}\\  \textcolor{black}{(Per timestep)}\end{tabular}} \\ \hline
Nyx Run1\_Z10        & 2 & 512, 256             & 23\%, 77\%    & \textcolor{black}{3.3 GB}               \\ \hline
Nyx Run1\_Z5        & 2 & 512, 256            & 58\%, 42\%  & \textcolor{black}{6.4 GB}                \\ \hline
Nyx Run1\_Z2        & 2 & 512, 256            & 63\%, 37\% & \textcolor{black}{6.7 GB}                 \\ \hline
Nyx Run2\_T3 & 3 & 512, 256, 128       & 0.02\%, 0.56\%, 99.42\% & \textcolor{black}{169 MB}      \\ \hline
Nyx Run2\_T4 & 4 & 1024, 512, 256, 128 & 3E-5, 0.02\%, 2.2\%, 97.7\% & \textcolor{black}{190 MB} \\ \hline
Nyx Run3\_Z1 & 3 & 512, 256, 128 & 0.90\%, 14.70\%, 84.40\% & \textcolor{black}{416 MB} \\ \hline
\textcolor{black}{WarpX\_800} & \textcolor{black}{2} & \textcolor{black}{$128^2$*1024; $256^2$*2048} &  \textcolor{black}{8.6\%,91.4\%}& \textcolor{black}{2 GB}\\ \hline
\textcolor{black}{WarpX\_1600}  & \textcolor{black}{2} & \textcolor{black}{$128^2$*1024; $256^2$*2048} & \textcolor{black}{2.0\%, 98.0\%} & \textcolor{black}{1.4 GB} \\ \hline
\textcolor{black}{IAMR\_90} & \textcolor{black}{3} & \textcolor{black}{512, 256, 128} & \textcolor{black}{0.6\%, 10.5\%, 88.9\%} & \textcolor{black}{336 MB} \\ \hline
\textcolor{black}{IAMR\_150}  & \textcolor{black}{3} & \textcolor{black}{512, 256, 128} & \textcolor{black}{14.8\%, 30.9\%, 54.3\%} & \textcolor{black}{2 GB}\\ \hline
\end{tabular}

\label{tab:datasets1}

%% file: Table/nyxspeed.tex

\begin{tabular}{|c|c|cccc|cccc|cccc|cccc|}
\hline
\multirow{2}{*}{\begin{tabular}[c]{@{}c@{}}Comp\\ Algo\end{tabular}} & \multirow{2}{*}{EB\_abs} & \multicolumn{4}{c|}{Run1\_Z10}                                                                  & \multicolumn{4}{c|}{Run1\_Z2}                                                                    & \multicolumn{4}{c|}{Run2\_T4}                                                                     & \multicolumn{4}{c|}{Run3\_Z1}                                                                    \\ \cline{3-18} 
                                                                     &                          & \multicolumn{1}{c|}{1D} & \multicolumn{1}{c|}{3D} & \multicolumn{1}{c|}{TAC} & TAC+             & \multicolumn{1}{c|}{1D} & \multicolumn{1}{c|}{3D}  & \multicolumn{1}{c|}{TAC} & TAC+             & \multicolumn{1}{c|}{1D} & \multicolumn{1}{c|}{3D}   & \multicolumn{1}{c|}{TAC} & TAC+             & \multicolumn{1}{c|}{1D} & \multicolumn{1}{c|}{3D}  & \multicolumn{1}{c|}{TAC} & TAC+             \\ \hline
\multicolumn{1}{|l|}{\multirow{3}{*}{Lor/Reg}}                       & 1E+08                    & \multicolumn{1}{c|}{51} & \multicolumn{1}{c|}{37} & \multicolumn{1}{c|}{79}  & 77               & \multicolumn{1}{c|}{51} & \multicolumn{1}{c|}{79}  & \multicolumn{1}{c|}{85}  & 84               & \multicolumn{1}{c|}{49} & \multicolumn{1}{c|}{0.32} & \multicolumn{1}{c|}{24}  & 24               & \multicolumn{1}{c|}{49} & \multicolumn{1}{c|}{5.6} & \multicolumn{1}{c|}{63}  & 63               \\ \cline{2-18} 
\multicolumn{1}{|l|}{}                                               & 1E+09                    & \multicolumn{1}{c|}{81} & \multicolumn{1}{c|}{48} & \multicolumn{1}{c|}{94}  & 93               & \multicolumn{1}{c|}{55} & \multicolumn{1}{c|}{92}  & \multicolumn{1}{c|}{103} & 97               & \multicolumn{1}{c|}{53} & \multicolumn{1}{c|}{0.39} & \multicolumn{1}{c|}{26}  & 26               & \multicolumn{1}{c|}{53} & \multicolumn{1}{c|}{6.5} & \multicolumn{1}{c|}{78}  & 75               \\ \cline{2-18} 
\multicolumn{1}{|l|}{}                                               & 1E+10                    & \multicolumn{1}{c|}{89} & \multicolumn{1}{c|}{53} & \multicolumn{1}{c|}{107} & 102              & \multicolumn{1}{c|}{58} & \multicolumn{1}{c|}{100} & \multicolumn{1}{c|}{111} & 104              & \multicolumn{1}{c|}{84} & \multicolumn{1}{c|}{0.41} & \multicolumn{1}{c|}{28}  & 27               & \multicolumn{1}{c|}{57} & \multicolumn{1}{c|}{7.1} & \multicolumn{1}{c|}{91}  & 84               \\ \hline
\multirow{3}{*}{Interp}                                              & 1E+08                    & \multicolumn{1}{c|}{74} & \multicolumn{1}{c|}{35} & \multicolumn{1}{c|}{80}  & \textbackslash{} & \multicolumn{1}{c|}{73} & \multicolumn{1}{c|}{73}  & \multicolumn{1}{c|}{90}  & \textbackslash{} & \multicolumn{1}{c|}{67} & \multicolumn{1}{c|}{0.27} & \multicolumn{1}{c|}{24}  & \textbackslash{} & \multicolumn{1}{c|}{68} & \multicolumn{1}{c|}{4.9} & \multicolumn{1}{c|}{70}  & \textbackslash{} \\ \cline{2-18} 
                                                                     & 1E+09                    & \multicolumn{1}{c|}{82} & \multicolumn{1}{c|}{42} & \multicolumn{1}{c|}{91}  & \textbackslash{} & \multicolumn{1}{c|}{82} & \multicolumn{1}{c|}{81}  & \multicolumn{1}{c|}{95}  & \textbackslash{} & \multicolumn{1}{c|}{77} & \multicolumn{1}{c|}{0.32} & \multicolumn{1}{c|}{25}  & \textbackslash{} & \multicolumn{1}{c|}{78} & \multicolumn{1}{c|}{5.8} & \multicolumn{1}{c|}{77}  & \textbackslash{} \\ \cline{2-18} 
                                                                     & 1E+10                    & \multicolumn{1}{c|}{90} & \multicolumn{1}{c|}{46} & \multicolumn{1}{c|}{97}  & \textbackslash{} & \multicolumn{1}{c|}{87} & \multicolumn{1}{c|}{85}  & \multicolumn{1}{c|}{104} & \textbackslash{} & \multicolumn{1}{c|}{84} & \multicolumn{1}{c|}{0.33} & \multicolumn{1}{c|}{26}  & \textbackslash{} & \multicolumn{1}{c|}{87} & \multicolumn{1}{c|}{6.3} & \multicolumn{1}{c|}{87}  & \textbackslash{} \\ \hline
\end{tabular}
\label{tab:nyxtp}

%% file: Table/rtspeed.tex
\begin{tabular}{|c|c|cccc|cccc|}
\hline
\multirow{2}{*}{\begin{tabular}[c]{@{}c@{}}Comp\\ Algo\end{tabular}} & \multirow{2}{*}{EB\_abs} & \multicolumn{4}{c|}{IAMR\_90}                                                                      & \multicolumn{4}{c|}{IAMR\_150}                                                                    \\ \cline{3-10} 
                                                                     &                          & \multicolumn{1}{c|}{1D} & \multicolumn{1}{c|}{3D}  & \multicolumn{1}{c|}{TAC} & TAC+             & \multicolumn{1}{c|}{1D} & \multicolumn{1}{c|}{3D} & \multicolumn{1}{c|}{TAC} & TAC+             \\ \hline
\multirow{3}{*}{Lor/Reg}                                             & 2E+03                    & \multicolumn{1}{c|}{50} & \multicolumn{1}{c|}{5.0} & \multicolumn{1}{c|}{67}  & 66               & \multicolumn{1}{c|}{51} & \multicolumn{1}{c|}{28} & \multicolumn{1}{c|}{76}  & 78               \\ \cline{2-10} 
                                                                     & 2E+04                    & \multicolumn{1}{c|}{52} & \multicolumn{1}{c|}{5.2} & \multicolumn{1}{c|}{73}  & 71               & \multicolumn{1}{c|}{50} & \multicolumn{1}{c|}{30} & \multicolumn{1}{c|}{85}  & 86               \\ \cline{2-10} 
                                                                     & 2E+05                    & \multicolumn{1}{c|}{54} & \multicolumn{1}{c|}{5.6} & \multicolumn{1}{c|}{79}  & 78               & \multicolumn{1}{c|}{53} & \multicolumn{1}{c|}{31} & \multicolumn{1}{c|}{94}  & 94               \\ \hline
\multirow{3}{*}{Interp}                                              & 2E+03                    & \multicolumn{1}{c|}{72} & \multicolumn{1}{c|}{4.6} & \multicolumn{1}{c|}{68}  & \textbackslash{} & \multicolumn{1}{c|}{68} & \multicolumn{1}{c|}{25} & \multicolumn{1}{c|}{81}  & \textbackslash{} \\ \cline{2-10} 
                                                                     & 2E+04                    & \multicolumn{1}{c|}{80} & \multicolumn{1}{c|}{4.8} & \multicolumn{1}{c|}{75}  & \textbackslash{} & \multicolumn{1}{c|}{78} & \multicolumn{1}{c|}{27} & \multicolumn{1}{c|}{92}  & \textbackslash{} \\ \cline{2-10} 
                                                                     & 2E+05                    & \multicolumn{1}{c|}{86} & \multicolumn{1}{c|}{5.1} & \multicolumn{1}{c|}{80}  & \textbackslash{} & \multicolumn{1}{c|}{81} & \multicolumn{1}{c|}{28} & \multicolumn{1}{c|}{96}  & \textbackslash{} \\ \hline
\end{tabular}
\label{tab:rttp}

%% file: Table/wpxspeed.tex
\begin{tabular}{|c|l|cccc|cccc|}
\hline
\multirow{2}{*}{\begin{tabular}[c]{@{}c@{}}Comp\\ Algo\end{tabular}} & \multicolumn{1}{c|}{\multirow{2}{*}{EB\_abs}} & \multicolumn{4}{c|}{WarpX\_800}                                                                 & \multicolumn{4}{c|}{WarpX\_1600}                                                                 \\ \cline{3-10} 
                                                                     & \multicolumn{1}{c|}{}                         & \multicolumn{1}{l|}{1D} & \multicolumn{1}{c|}{3D} & \multicolumn{1}{c|}{TAC} & TAC+             & \multicolumn{1}{c|}{1D} & \multicolumn{1}{c|}{3D}  & \multicolumn{1}{c|}{TAC} & TAC+             \\ \hline
\multirow{3}{*}{Lor/Reg}                                             & 1E+06                                         & \multicolumn{1}{l|}{49} & \multicolumn{1}{c|}{15} & \multicolumn{1}{c|}{97}  & 85               & \multicolumn{1}{c|}{47} & \multicolumn{1}{c|}{8.9} & \multicolumn{1}{c|}{102} & 89               \\ \cline{2-10} 
                                                                     & 1E+07                                         & \multicolumn{1}{c|}{52} & \multicolumn{1}{c|}{18} & \multicolumn{1}{c|}{107} & 92               & \multicolumn{1}{c|}{50} & \multicolumn{1}{c|}{11}  & \multicolumn{1}{c|}{126} & 103              \\ \cline{2-10} 
                                                                     & 1E+08                                         & \multicolumn{1}{c|}{52} & \multicolumn{1}{c|}{19} & \multicolumn{1}{c|}{115} & 98               & \multicolumn{1}{c|}{53} & \multicolumn{1}{c|}{14}  & \multicolumn{1}{c|}{159} & 121              \\ \hline
\multirow{3}{*}{Interp}                                              & 1E+06                                         & \multicolumn{1}{c|}{73} & \multicolumn{1}{c|}{14} & \multicolumn{1}{c|}{128} & \textbackslash{} & \multicolumn{1}{c|}{68} & \multicolumn{1}{c|}{8.7} & \multicolumn{1}{c|}{123} & \textbackslash{} \\ \cline{2-10} 
                                                                     & 1E+07                                         & \multicolumn{1}{c|}{79} & \multicolumn{1}{c|}{16} & \multicolumn{1}{c|}{140} & \textbackslash{} & \multicolumn{1}{c|}{75} & \multicolumn{1}{c|}{11}  & \multicolumn{1}{c|}{140} & \textbackslash{} \\ \cline{2-10} 
                                                                     & 1E+08                                         & \multicolumn{1}{c|}{83} & \multicolumn{1}{c|}{17} & \multicolumn{1}{c|}{147} & \textbackslash{} & \multicolumn{1}{c|}{79} & \multicolumn{1}{c|}{12}  & \multicolumn{1}{c|}{159} & \textbackslash{} \\ \hline
\end{tabular}
\label{tab:wpxtp}

%% file: tex/06_conclusion.tex
\section{Conclusion and Future Work} 
\label{sec:conclusion}

In conclusion, this paper leverages 3D compression for AMR data on a systemic level. We propose three pre-process strategies that can adapt based on the density of each AMR level.
Our approach improves the compression ratio compared to the state-of-the-art approach by up to 4.9$\times$ under the same data quality loss.
With our level-wised compression approach, we are able to tune the error-bound ratio of fine and coarse levels to be 3:1 and 2:1 for better power-spectrum and halo-finder analyses, respectively, under the same compression ratio.
In future work, we will apply \textsc{TAC+} to more AMR simulations and improve its throughput on multi-core CPUs using OpenMP and GPUs using CUDA. 

%% file: tex/99_acknowledge.tex
\section*{Acknowledgement}
\small This work has been authored by employees of Triad National Security, LLC, which operates Los Alamos National Laboratory under Contract No. 89233218CNA000001 with the U.S. Department of Energy (DOE) and the National Nuclear Security Administration (NNSA). 
This research was supported by the Exascale Computing Project (ECP), Project Number: 17-SC-20-SC, a collaborative effort of the DOE SC and NNSA.
This work was also supported by NSF awards 2303064, 2247080, 2311876, and 2312673.
This research used resources of the National Energy Research Scientific Computing Center, a DOE SC User Facility located at Lawrence Berkeley National Laboratory, operated under Contract No. DE-AC02-05CH11231. We would like to thank Dr. Zarija Lukić and Dr. Jean Sexton from the NYX team at Lawrence Berkeley National Laboratory for granting us access to cosmology datasets. 
This research used the open-source particle-in-cell code WarpX, primarily funded by the US DOE Exascale Computing Project. 